\documentclass[12pt,letterpaper]{article}

\usepackage{epsfig}
\usepackage{amsmath}
\usepackage{amssymb}
\usepackage{amsthm}
\usepackage{mathrsfs}
\usepackage{bbm}
\usepackage{bm}
\usepackage[T1]{fontenc}

\usepackage[utf8]{inputenc}

\usepackage{graphicx}
\usepackage[percent]{overpic}
\usepackage{indentfirst}
\usepackage{xspace}
\usepackage{multirow}
\usepackage{hyperref}
\usepackage{xcolor}
\definecolor{darkred}{rgb}{0.5,0.15,0.15}
\hypersetup{colorlinks=true,urlcolor=darkred,linkcolor=darkred,citecolor=darkred}
\usepackage{verbatim}
\usepackage[letterpaper,margin=1in,headheight=15pt]{geometry}
\usepackage{mathpazo}
\usepackage{authblk}
\usepackage{tikz-cd}
\usepackage{capt-of}
\usepackage{silence}
\usepackage{enumerate}

\usepackage{tikz}
\usetikzlibrary{calc}   
\usetikzlibrary{topaths}
\usetikzlibrary{decorations}
\usetikzlibrary{decorations.pathmorphing}

\usepackage{csquotes}
\PassOptionsToPackage{%
backend=bibtex8,bibencoding=ascii,%
language=auto,%
style=alphabetic, 
sorting=nyt, 
maxbibnames=10, 
natbib=true 
}{biblatex}
\usepackage{biblatex}
\numberwithin{equation}{section}

\theoremstyle{definition}

\newtheorem{theorem*}{Theorem}

\newtheorem{example}[equation]{Example}
\newtheorem{notation}[equation]{Notation}

\newcommand{\tops}{\texorpdfstring}

\newcommand{\nid}{\noindent}

\newcommand{\cS}{\ensuremath{\mathcal S}}
\newcommand{\cF}{\ensuremath{\mathcal F}}
\newcommand{\cK}{\ensuremath{\mathcal K}}

\newcommand{\cX}{\ensuremath{\mathcal X}}

\newcommand{\cW}{\ensuremath{\mathcal W}}

\newcommand{\scrW}{\ensuremath{\mathscr W}}
\newcommand{\scrT}{\ensuremath{\mathscr T}}
\newcommand{\scrM}{\ensuremath{\mathscr M}}
\newcommand{\scrB}{\ensuremath{\mathscr B}}

\newcommand{\scrE}{\ensuremath{\mathscr E}}
\newcommand{\scrR}{\ensuremath{\mathscr R}}
\newcommand{\scrG}{\ensuremath{\mathscr G}}

\newcommand{\sfS}{\ensuremath{\mathsf S}}
\newcommand{\sfT}{\ensuremath{\mathsf T}}

\newcommand{\sfa}{\ensuremath{\mathsf a}}
\newcommand{\sfb}{\ensuremath{\mathsf b}}
\newcommand{\sff}{\ensuremath{\mathsf f}}

\newcommand{\R}{\ensuremath{\mathbb R}}
\newcommand{\C}{\ensuremath{\mathbb C}}
\newcommand{\PP}{\ensuremath{\mathbb P}}
\newcommand{\Z}{\ensuremath{\mathbb Z}}

\newcommand{\bbS}{\ensuremath{\mathbb S}}

\newcommand{\N}{{\mathcal N}}

\newcommand{\I}{{\mathrm i}}
\newcommand{\E}{{\mathrm e}}
\newcommand{\de}{\mathrm{d}}

\DeclareMathOperator{\Tr}{Tr}

\newcommand{\Mflat}{\mathscr{M}_{\text{flat}}}

\newcommand{\Tdual}{\hat{\sfT}}

\newcommand{\tgamma}{\tilde{\gamma}{}}
\newcommand{\hgamma}{\hat{\gamma}{}}
\newcommand{\gammavf}{\gamma_1+\gamma_2+\gamma_{\rm f}}

\newcommand{\bps}{\textrm{\tiny BPS}}

\definecolor{Blue}{RGB}{50,50,255}
\definecolor{Orange}{RGB}{255,153,0} 
\definecolor{Red}{RGB}{255,0,0}
\definecolor{Green}{RGB}{0,175,0}
\definecolor{ToDo}{RGB}{150,0,0}

\begin{document}

\title{{\bf Novel wall-crossing behaviour \\[1mm] in rank one $\boldsymbol{\N=2^*}$ gauge theory}}
\date{{{\tiny \color{gray} \tt}} 
{{\tiny \color{gray} \tt}}} 
\author[1]{Philipp R\"uter}
\author[2]{Richard J. Szabo}
\affil[1,2]{\small Department of Mathematics, Heriot-Watt
  University}
\affil[1,2] {\small Maxwell Institute for Mathematical Sciences, Edinburgh}
\affil[2]{\small Higgs Centre for Theoretical Physics, Edinburgh}

\maketitle

\begin{abstract}
\noindent
We study the BPS spectrum of four-dimensional $\N=2$ supersymmetric Yang-Mills theory with gauge group $SU(2)$ and a massive adjoint hypermultiplet, which has an extremely intricate structure with infinite spectrum in all chambers of its Coulomb moduli space, and is not well understood. We build on previous results by employing the BPS quiver description of the spectrum, and explore the qualitative features in detail using numerical techniques. We find novel and unexpected behaviour in the form of wall-crossings involving interactions between BPS particles with negative electric-magnetic pairings, which we interpret in terms of the reverse orderings of the usual wall-crossing formulas for rank one $\N=2$ field theories. This identifies new {\it a priori} unrelated states in the spectrum.
\end{abstract}

\vspace{1cm}

\begin{flushright}
{\sf\small EMPG--21--11}
\end{flushright}


\newpage
{\baselineskip=12pt
\tableofcontents
}
\newpage

\bigskip

\section{Introduction}
\nid
The $\N=2^*$ gauge theory is a deformation of $\N=4$ supersymmetric Yang-Mills theory in four dimensions by giving the adjoint-valued hypermultiplet a mass $m$, which breaks the supersymmetry from $\N=4$ to $\N=2$ while preserving S-duality and superconformal symmetry~\cite{Seiberg:1994aj}. Alternatively, it can be regarded as pure $\N=2$ gauge theory extended by an adjoint hypermultiplet with a mass term that preserves $\N=2$ supersymmetry. For gauge group $SU(2)$, which is the focus of the present paper, its Seiberg-Witten curve $\Sigma$ may be realized as a branched two-sheeted covering of a once-punctured torus $C$~\cite{Donagi:1995cf,Witten:1997sc}, which can be used to geometrically engineer the $\N=2^*$ theory as a class $\cS$ theory obtained from compactification of the six-dimensional $(2,0)$-theory on~$C$~\cite{gaiotto2012n}.

In this paper we are concerned with the spectrum of BPS particles in the $\N=2^*$ theory, which was originally explored in~\cite{schulze1997bps,ferrari1997dyon} and more recently in~\cite{alim2013bps,alim2014mathcal,Cecotti:2015qha,longhi2015structure,Longhi:2016wtv}. The BPS spectrum exhibits very complicated behaviour at generic points of the Coulomb branch $\scrB_m$, and in particular it has no chambers containing only finitely many BPS states~\cite{alim2013bps}. Results thus far have therefore been limited to either specific loci in $\scrB_m$ or  subsets of the full spectrum. This is in marked constrast with the two natural mass limits at $m=0$ and $m\to\infty$, in which the $\N=2^*$ theory becomes $\N=4$ supersymmetric Yang-Mills theory and pure $\N=2$ gauge theory respectively, where the spectra are completely understood. The purpose of this paper is to combine algebraic and geometric constructions to obtain a more extensive and explicit description of the $\N=2^*$ spectrum.

In our study of the $\N=2^*$ spectrum, we represent BPS states as elements of a charge lattice $\gamma\in\Gamma$ which is equipped with an {electric-magnetic pairing} $\langle\cdot,\cdot\rangle\colon \Gamma\times\Gamma\rightarrow\mathbb{Z}$ and a {central charge} $Z\colon\Gamma\rightarrow\mathbb{C}$. The spectrum can be studied as one varies over the Coulomb branch $\scrB_m$, which is parametrized by a complex variable $u\in\C$, with $Z(\gamma)=Z(\gamma;u)$ a holomorphic function of $u$. In $\scrB_m$ there are real codimension one {walls} where two BPS states have parallel central charges, and on crossing these walls the spectrum undergoes discontinuous changes dependent on the pairing of the charges involved, which are computed by the Kontsevich-Soibelman wall-crossing formula. For the $\N=2^*$ theory, every state in the spectrum can be written as a linear combination of three charges $\gamma_i$, $i=1,2,3$, with pairings $\langle\gamma_i,\gamma_{i+1}\rangle=2$, which correspond respectively to the primitive monopole, dyon and quark.\footnote{This naming is simply for convenience, as it depends on a choice of electromagnetic duality frame. However, there is no canonical such choice, and indeed there is a $\Z_3$ symmetry which cyclically permutes the charges $\gamma_i$.} In particular, the adjoint hypermultiplet of mass $m$ is represented by the flavour charge $\gamma_{\rm f} = \gamma_1+\gamma_2+\gamma_3$.

In \cite{longhi2015structure} the BPS spectrum was considered on three particular walls $\scrE_i\subset\scrB_m$, $i=1,2,3$ (see \eqref{longhiwall} below for their precise definitions).
On these walls, the spectrum then agrees with a semi-classical ansatz. For example, on $\scrE_3$ it consists of hypermultiplets $\gamma_1,\gamma_2,\gamma_3,\gammavf$, a vector multiplet $\gamma_1+\gamma_2$ (the $W$-boson), and two infinite families of hypermultiplets which are bound states of $\gamma_1$ and $\gamma_2$ \cite{longhi2015structure} whose central charges accumulate near the ray $Z(\gamma_1+\gamma_2)$.
This spectrum can be confirmed by comparing it to the spectrum generator, which is also constructed in \cite{longhi2015structure}, as explained in more detail in Section~\ref{longhispectrum}. Generic points in $\scrB_m$ can then be explored by a perturbation away from this wall, leading to new vector multiplets that are again accompanied by infinite families of hypermultiplets, which in turn can form further bound states \cite{longhi2015structure}.

In this paper we begin by briefly reviewing the class $\cS$ geometry of the $\N=2^*$ theory in Section~\ref{sec:SWgeometry}, and relevant general aspects of BPS spectra as well as wall-crossing in $\N=2$ theories in Section~\ref{bpsspectra} together with geometric techniques for computing them; in particular, we review the current state of understanding of the $\N=2^*$ spectrum in Section~\ref{longhispectrum}, emphasising its complexities from various points of view. In Section~\ref{sec:BPSquivers} we reproduce the results of \cite{longhi2015structure} on the wall $\scrE_3$ using algebraic techniques based on representations and mutations of BPS quivers to explicitly construct the spectrum on $\scrE_3$, and map the construction to the geometric engineering of Section~\ref{bpsspectra}. In Section~\ref{perturbedspectrum} we discuss the perturbation away from the wall $\scrE_3$. This perturbation leads to multiple new vector multiplets which renders the mutation method problematic, as infinitely many quiver mutations occur on either side of a vector multiplet. In Section~\ref{PrimaryStates} we propose an ansatz that quells the situation and obtain the BPS quiver descriptions of the $\N=2^*$ theory on either side of each of the accumulation rays in the central charge plane.

Beyond these primary bound states, there are potentially arbitrary numbers of ``secondary'' BPS states. These are difficult to systematically investigate with analytic methods. We therefore implement numerical techniques to gain further insight; this is described in Section~\ref{Pythonspectrum}. This allows us to study much of the qualitative behaviour of the $\N=2^*$ spectrum for different perturbations. The details of the precise implementation of the Python algorithm~\cite{Rueter} used to explore the spectrum are discussed in Appendix~\ref{PythonAlgorithm} at the end of the paper, which we hope will also be useful for further numerical investigations.

In Section~\ref{SecondaryStates} we observe that it is possible for wall-crossings to occur that involve interactions  among BPS particles with negative electric-magnetic pairing $\langle\gamma,\gamma'\rangle$. We  confirm the existence of these wall-crossings in our numerical exploration of the spectrum in Section~\ref{Pythonspectrum}. Such wall-crossings are not expected in $\N=2$ field theories with gauge group $SU(2)$ as they lead to {exotic} BPS states, which are not singlets of the $\mathfrak{su}(2)_R$ R-symmetry and violate the {no-exotics theorem}~\cite{gaiotto2013framed,galakhov2013wild}.

In Section~\ref{missingstates} we interpret these negative pairings in terms of the reverse orderings of the primitive $\langle\gamma,\gamma'\rangle=2$ wall-crossing formula. This interpretation is motivated by the Denef formula and its four-dimensional field theory limit as discussed in~\cite{galakhov2013wild}, which indicates that the bound states between ${\gamma}$ and ${\gamma'}$ in the Kontsevich-Soibelman wall-crossing formula are necessarily in the spectrum as well. Indeed we find that at least some of these bound states can be found by reinterpreting other bound states in the spectrum which were generated by previous pairing two wall-crossings. This is novel behaviour which has not been observed previously in other rank one $\N=2$ theories. It also hints at an intricate structure of the spectrum which connects {\it a priori} unrelated states, some of which result from the complicated formation of secondary BPS states.

\section{\tops{Geometry of Rank One $\boldsymbol{\N=2^*}$ Gauge Theory}{Geometry of the Rank One N=2* Gauge Theory with Gauge Group SU(2)}}
\label{sec:SWgeometry}
\nid
The $\N=2^*$ gauge theory is defined as $\N=2$ supersymmetric Yang-Mills theory on $\R^{3,1}$ coupled to a hypermultiplet of mass $m$ in the adjoint representation of the gauge group. In the massless limit $m=0$, the $\N=2$ supersymmetry is enhanced and the field theory becomes the  $\N=4$ supersymmetric Yang-Mills theory. On the other hand, in the infinite mass limit $m\to\infty$ the hypermultiplet decouples and the field theory becomes the asymptotically free pure $\N=2$ gauge theory consisting solely of a vector multiplet in the adjoint representation of the gauge group, whose bosonic field content consists of a gauge field $A$, a complex scalar Higgs field $\phi$ valued in the complexified gauge algebra, and an auxiliary scalar field $D$. In this paper we consider only the gauge group $SU(2)$ of rank one. Like its massless limit, the $\N=2^*$ theory then enjoys manifest $SL(2,\Z)$ S-duality and superconformal symmetry.

The Coulomb branch $\scrB_m$ of the $\N=2^*$ gauge theory is the base of a fibration whose fibre over $u\in\scrB_m$ is the Seiberg-Witten curve $\Sigma=\Sigma_{u;m}$. For gauge group $SU(2)$ this is the family of elliptic curves~\cite{Seiberg:1994aj}
\begin{equation}\label{eq:SWcurvemassive}
\Sigma_{u;m} : \quad y^2 = \big(x-\varepsilon_1(\tau_0)_{u;m}\big)\, \big(x-\varepsilon_2(\tau_0)_{u;m}\big)\, \big(x-\varepsilon_3(\tau_0)_{u;m}\big) \ ,
\end{equation}
where
\begin{equation}
\varepsilon_i(\tau_0)_{u;m} =e_i(\tau_0) \,u+e_i(\tau_0)^2 \, m^2 \ .
\end{equation}
Here $\tau_0$ is the complexified (exactly marginal) microscopic gauge coupling, while the Coulomb branch modulus $u\in\scrB_m$ is given by
\begin{equation}
u = u_0-\frac12\,e_1\, m^2 + m^2\,\sum_{n=1}^\infty\, c_n\, q_0^n \ ,
\end{equation}
where \smash{$u_0=\frac1{16\pi^2} \, \langle{\rm Tr}\,\phi^2\rangle$} is the Coulomb order parameter, with the trace in the fundamental representation of $\mathfrak{su}(2)$. The series coefficients $c_n$ correspond to instanton corrections~\cite{Dorey:1996ez} with $q_0=\E^{\,2\pi\,\I\,\tau_0}$.
The half-periods $(e_1,e_2,e_3)$ form a modular vector of weight two under $SL(2,\Z)$. An explicit closed expression for $u=u(\tau,\tau_0)$ can be found in~\cite{Huang:2011qx} as a function of both the ultraviolet coupling $\tau_0$ and the complex structure modulus $\tau$ of the curve $\Sigma$ which is identified as the infrared gauge coupling of the $\N=2^*$ field theory. This defines a bimodular form of weight $(0,2)$, and consequently $\varepsilon_i(\tau,\tau_0)$ form a bimodular vector of weight $(0,4)$.

The Coulomb moduli space $\scrB_m$ is topologically the four-punctured sphere \smash{$\PP^1_{\infty,u_1,u_2,u_3}$}, where the elliptic curve \eqref{eq:SWcurvemassive} degenerates at the singularities $ u_i=e_i(\tau_0)\, m^2$ in the $u$-plane. The singularities for $i=1,2,3$ are respectively called electric, magnetic and dyonic according to the $U(1)$ charges of the massless particles which appear at the singularity; they meet at equal distances from each other because of the cyclic $\Z_3$ symmetry acting on $\scrB_m$, and hence only one ultraviolet scale $\Lambda_m$ is generated. These massless particles can be used to give a unique weak coupling description of the gauge theory near each singularity. At the singularity $u=\infty$, the $W^\pm$-bosons become infinitely massive and $\tau=\tau_0$, so that the low energy effective field theory is an almost superconformal theory.

In this paper we are interested in the microscopic description of the $\N=2^*$ theory as a field theory of class $\cS$~\cite{Witten:1997sc,gaiotto2012n}. For this, we exhibit the Seiberg-Witten curve $\Sigma$ as a branched covering of an ultraviolet curve $C$. There is an $SL(2,\C)$ M\"obius transformation sending the four branch points at $x=\infty,\varepsilon_1,\varepsilon_2,\varepsilon_3$ to the points $ -u/m^2, e_1,e_2,e_3$ on $\PP^1$, which maps~\cite{Donagi:1995cf} the elliptic curve  \eqref{eq:SWcurvemassive} to the double cover $w^2 = m^2\,x+u$ of the curve
\begin{align}
y^2=\big(x-e_1(\tau_0)\big)\, \big(x-e_2(\tau_0)\big)\, (x-e_3(\tau_0)\big) \ .
\end{align}
This is the Weierstrass canonical form of the elliptic curve which describes a torus with complex structure modulus $\tau_0$. It can be uniformized by taking $x=\wp(z)$ and $y=\frac12\,\wp'(z)$, where $\wp(z|\tau_0)$ is the Weierstrass $\wp$-function which is a doubly periodic even function of a local holomorphic coordinate $z\in\C$ on this torus, with a single double pole $\wp(z|\tau_0)\simeq\frac1{z^2}$ for $z\to0$. Then the equation for $\Sigma$ becomes
\begin{equation}\label{eq:SW2sheeted}
w^2=m^2\,\wp(z|\tau_0) + u \ .
\end{equation}
This exhibits the Seiberg-Witten curve as a branched two-sheeted covering $\Sigma\to C$ of genus two over the torus $C$ of modulus $\tau_0$ with a single regular puncture at $z=0$. 

The Seiberg-Witten differential $\lambda=\lambda_{u;m}$ on $\Sigma$ is the canonical one-form
\begin{equation}\label{eq:quaddiff}
\lambda = w \, {\rm d}z
\end{equation}
on the holomorphic cotangent bundle $T^*C\cong C\times\C$, with coordinates $(z,w)$, restricted to the covering $\Sigma\subset T^*C$. The meromorphic differential $\lambda$ has two simple poles at the corresponding preimages of $z=0$ on $\Sigma$, with residues $\pm\,m$, and two simple zeroes which correspond to branch points of $\Sigma\rightarrow C$. In this realization, the Seiberg-Witten curve $\Sigma$ is equivalently the spectral curve for the moduli space of the $SU(2)$ Hitchin system on $C$, interpreted as the space of quantum vacua when the four-dimensional theory is compactified on a circle. This space is biholomorphic (in a certain complex structure) to the moduli space of stable Higgs bundles on $C$, which is a torus fibration over the Coulomb branch whose fiber over $u\in\scrB_m$ is an abelian variety of line bundles on $\Sigma_{u;m}$~\cite{hitchin1987stable},\footnote{This structure makes the moduli space into an algebraic completely integrable system. In fact, $\Sigma_{u;m}$ is the spectral curve of the elliptic Calogero-Moser integrable system~\cite{DHoker:1997hut}.} and also (in another complex structure) to the moduli space $\Mflat(C)$ which parameterizes flat $SL(2,\mathbb{C})$ connections on $C$.

It is instructive to consider the two natural mass limits of this class $\cS$ geometry:
\begin{itemize}
\item[$\diamond$] In the massless limit $m=0$, the elliptic curve \eqref{eq:SWcurvemassive} becomes
\begin{align}
\Sigma_{u_0;0} : \quad y^2 = \big(x-e_1(\tau_0)\, u_0\big)\,\big(x-e_2(\tau_0)\, u_0\big)\,\big(x-e_3(\tau_0)\, u_0\big) \ .
\end{align}
This is the Seiberg-Witten curve of $\N=4$ supersymmetric Yang-Mills theory with gauge group $SU(2)$~\cite{Seiberg:1994aj}. In this limit the infrared and ultraviolet moduli coincide, $\tau=\tau_0$, and the Coulomb moduli space is zero-dimensional, $\scrB_0=\{u_0\}$. The Seiberg-Witten curve $\Sigma$ coincides with the ultraviolet curve $C$, and the Seiberg-Witten differential is given by the canonical $(1,0)$-differential as $\lambda_{u_0;0} = \sqrt{u_0} \ \de z$.
\item[$\diamond$] In the asymptotically free limit $m\to\infty$, we double scale with the weak coupling limit $\tau_0\to \I\,\infty$ and $8\,q_0^{1/2}\, m^2 = \Lambda_\infty^2$ fixed. After a shift $x\to x-\frac12\,e_1\,u+\frac12\,e_1^2\,m^2$ in the $x$-plane and an infinite additive renormalization \smash{$u_\infty=u+\frac12\,e_1\,m^2$}, the elliptic curve \eqref{eq:SWcurvemassive} becomes
\begin{align}
\Sigma_{u_\infty;\infty} : \quad y^2 = (x-u_\infty)\,(x^2-\Lambda_\infty^4) \ .
\end{align}
This is the Seiberg-Witten curve for the pure $\N=2$ gauge theory with ultraviolet scale $\Lambda_\infty$~\cite{Seiberg:1994rs}. The Coulomb  branch $\scrB_\infty$ is topologically a three-punctured sphere \smash{$\PP^1_{\infty,\Lambda_\infty^2,-\Lambda_\infty^2}$}, where the quark point has disappeared, and the duality group is reduced to the congruence subgroup $\varGamma(2)\subset SL(2,\Z)$~\cite{Seiberg:1994rs}. In this limit, the covering equation \eqref{eq:SW2sheeted} describes $\Sigma_{u_\infty;\infty}$ as a genus one double cover and becomes~\cite{Itoyama:1995uj}
\begin{align}
w^2 = -\pi^2\,\Lambda_\infty^2\,\cos(2\pi\,z)+u_\infty \ ,
\end{align}
where the ultraviolet curve $C$ is a now a cylinder with irregular punctures at its ends.
\end{itemize}

\section{BPS Spectra and Wall-Crossing}\label{bpsspectra}
\noindent
Our object of study in this paper is the BPS spectrum of the $\N=2^*$ theory. In this section we shall review some generalities about BPS spectra and wall-crossing in $\N=2$ theories, focusing on the rank one case and highlighting essential features relevant to the $\N=2^*$ gauge theory. Most of the material of this section is based on~\cite{gaiotto2013wall}, where further details may be found.

\subsection{BPS States and Walls}
\label{subsec:BPSStatesWalls}
\noindent
A BPS state of a quantum field theory on $\R^{3,1}$ with rigid $\N=2$ supersymmetry is a one-particle state in the Hilbert space whose mass $M$ and central charge $Z$ saturate the BPS bound
\begin{equation}
M \ge |Z| \ .
\end{equation}
To each state we assign an electromagnetic charge $\gamma$ in the charge lattice $\Gamma$, which for our purposes can be identified geometrically with homology classes of cycles on the Seiberg-Witten curve $\Sigma$ in $H_1(\Sigma,\mathbb{Z})$.\footnote{More precisely, $\Gamma$ should be identified with a sublattice of $H_1(\Sigma,\mathbb{Z})$, since we require charges to be invariant under a combined operation of exchanging the sheets of $\Sigma$ and reversing the orientation of~$\gamma$~\cite{gaiotto2013wall}.} The central charge $Z$ is a homomorphism of abelian groups
\begin{equation}
Z\,:\,\Gamma\longrightarrow\mathbb{C} \ ,
\end{equation}
whose value on $[\gamma]\in H_1(\Sigma,\mathbb{Z})$ is given by the period of the Seiberg-Witten differential $\lambda$:
\begin{equation}
Z(\gamma)=\oint_{\gamma} \, \lambda \ .
\end{equation}
It therefore inherits the dependence on the Coulomb moduli $u\in\scrB$ from $\Sigma_u$ and $\lambda_u$, with $\Gamma$ forming a local system of abelian groups over $\scrB$. 

Another important piece of data is the covariantly constant antisymmetric electric-magnetic pairing
\begin{align}
\langle\cdot,\cdot\rangle\,:\,\Gamma\times\Gamma\longrightarrow\mathbb{Z} \ , 
\end{align}
which can be taken to be the oriented intersection pairing of $H_1(\Sigma,\mathbb{Z})$. Pairs of charges with non-zero electric-magnetic pairing are said to be mutually non-local, otherwise they are mutually local. Charges $\gamma_{\rm f}$ in the radical of $\langle\cdot,\cdot\rangle$ are called {flavour charges}.\footnote{We shall often abuse terminology in the following, referring to an electromagnetic or flavour charge $\gamma$ and its central charge $Z(\gamma)$ synonymously. We will also frequently refer to a BPS particle by its charge $\gamma$; the corresponding antiparticle in the BPS spectrum has opposite charge $-\gamma$.}

The spectrum of BPS states is piecewise constant on the Coulomb branch $\scrB$: there are real codimension one {walls of marginal stability} which separate different regions of $\scrB$. Such a wall $\scrW(\gamma,\gamma')$ occurs when the central charges of two states $\gamma$ and $\gamma'$ in the spectrum align:
\begin{align}
\scrW(\gamma,\gamma') = \big\{u\in\scrB \ \big| \ {\rm arg}\big(Z(\gamma;u)\big)={\rm arg}\big(Z(\gamma';u)\big)\big\} \ .
\end{align}
The spectrum on the wall is ill-defined since BPS bound states $n\,\gamma + n'\,\gamma'$ (with $n,n'\in\Z$) can freely form and decay, as
\begin{equation}
|Z(n\,\gamma+n'\,\gamma')| = |n\,Z(\gamma)+n'\,Z(\gamma')| = n\,|Z(\gamma)|+n'\,|Z(\gamma')|
\end{equation}
for electromagnetic charges $\gamma$ and $\gamma'$ with parallel central charges $Z(\gamma)$ and $ Z(\gamma')$.
As we vary $u\in\scrB$, the spectrum changes on crossing such a wall; in particular, a BPS particle with charge $\gamma+\gamma'$ decays into $\gamma$ and $\gamma'$ on crossing $\scrW(\gamma,\gamma')$. 

\begin{example}\label{ex:puresu2walls}
For the pure $SU(2)$ gauge theory, there are two regions of the Coulomb moduli space which are separated by a wall which is roughly an ellipse around $0\in\scrB_\infty$ passing through the singularities $u=\pm\,\Lambda_\infty^2$. For small $u$ (the strong coupling region), the spectrum consists of two BPS hypermultiplets $\gamma_1$ and $\gamma_2$ (the primitive monopole and dyon). On crossing the wall the spectrum changes: it still contains $\gamma_1$ and $\gamma_2$, but now additionally a vector state $\gamma_1+\gamma_2$ (the $W$-boson), along with two infinite families of hypermultiplets $\gamma_i+n\,(\gamma_1+\gamma_2)$ with $i=1,2$ and $n\in\Z$ that accumulate towards the $W$-boson as $n\to\pm\,\infty$.
\end{example}

\subsection{Ideal Triangulations}\label{BPSgeometry}
\noindent
One method for computing the spectrum of BPS states at a given point of the Coulomb branch $\scrB$ is through a family of ideal triangulations $\scrT_\vartheta$ of the ultraviolet curve $C$, which depends on an angle $\vartheta\in\R/2\pi\,\Z$. These triangulations are constructed from a singular foliation of $C$ which is defined as follows~\cite{gaiotto2013wall,hollands2016spectral}.

With the Seiberg-Witten differential $\lambda$ from \eqref{eq:quaddiff}, we call a $\vartheta${-trajectory} a path $\varrho$ on $C$ such that 
\begin{equation}\label{trajectory}
\E^{-\I\,\vartheta}\,\lambda(v) \ \in \ \R^\times=\mathbb{R}\setminus\{0\} \ ,
\end{equation}
for any tangent vector $v$ to $\varrho$.
These trajectories are the leaves of a singular foliation $\cF_\vartheta$ of $C$, where the singular points are the zeroes of $\lambda$, which are also the branch points of the covering $\Sigma\rightarrow C$. The types of curves $\varrho$ which are relevant for us are:
\begin{itemize}
\item[$\diamond$] {Generic curves} which have both of their endpoints on a puncture.
\item[$\diamond$]{Separating curves} which have one endpoint on a puncture and one endpoint on a branch point.
\item[$\diamond$] {Finite curves} which have both endpoints on a branch point.
\end{itemize}
The separating and finite curves are also called {Stokes curves} in direction $\vartheta$, and they form the {Stokes graph} in direction $\vartheta$ with vertices at the branch points on $C$. The Stokes graph is also called a WKB spectral network, while its oriented edges (the Stokes curves) are also called S-walls and double walls for separating and finite curves, respectively (see~\cite{recipe} for an up to date review); we will use both nomenclatures interchangeably in this paper.

\begin{figure}[h!]
\small
\centering
\begin{overpic}
[width=0.60\textwidth]{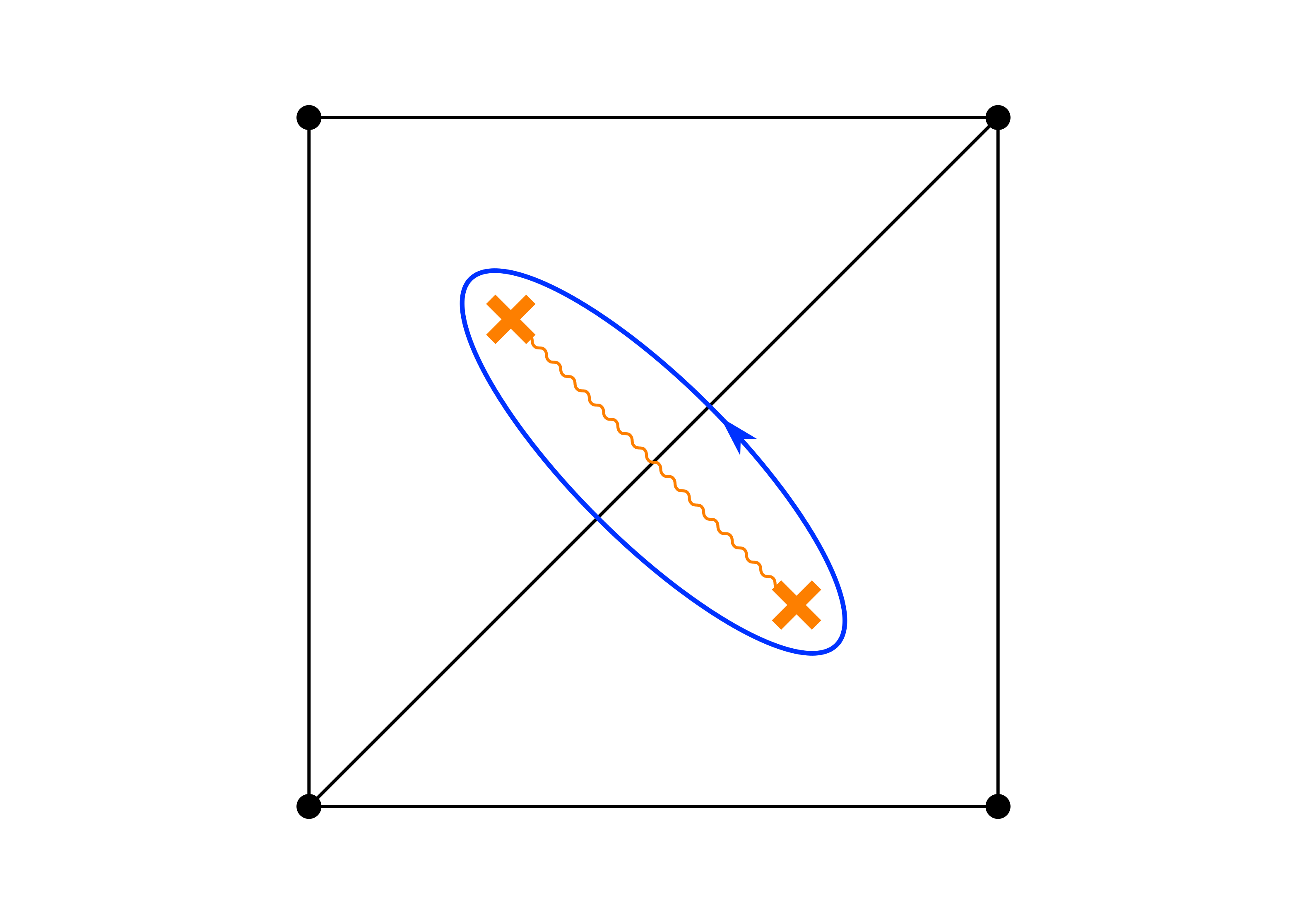}
\put(66,21){\textcolor{Blue}{$\gamma$}}
\put(41,22){$E_{\gamma}$}
\end{overpic}
\caption{\small To each edge of $\scrT_\vartheta$ we can assign a cycle $\gamma$ on the Seiberg-Witten curve $\Sigma$, and denote the corresponding edge by $E_\gamma$. Punctures are indicated by solid circles and branch points by orange crosses. The orange wavy line represents a choice of branch cut.}
\label{triangulationcycle}
\normalsize
\end{figure}

At generic values of $\vartheta$ there are no finite trajectories and the Stokes graph separates $C$ into simply connected cells of the foliation $\cF_\vartheta$. Each such cell is bounded by separating curves and contains an infinite family of generic curves.
We can define an ideal triangulation $\scrT_\vartheta$ of $C$ by taking one representative of the generic curves in each cell to be an edge and the punctures of $C$ to be the vertices. 
A neighbourhood of each simple branch point has exactly three Stokes curves emanating to or from it, so that each triangle of $\scrT_\vartheta$ contains one branch point. A basis of $H_1(\Sigma,\mathbb{Z})$ is obtained by fixing a local trivialization of the covering $\Sigma\to C$, via a choice of branch cuts on the base $C$, and from curves which encircle two branch points each. We can use this to pair each edge with a cycle $\gamma$, and denote it by $E_{\gamma}$ as shown in Figure~\ref{triangulationcycle}.\footnote{There is some subtlety involved in this to get the orientations of curves right. See \cite{gaiotto2013wall} for details.}

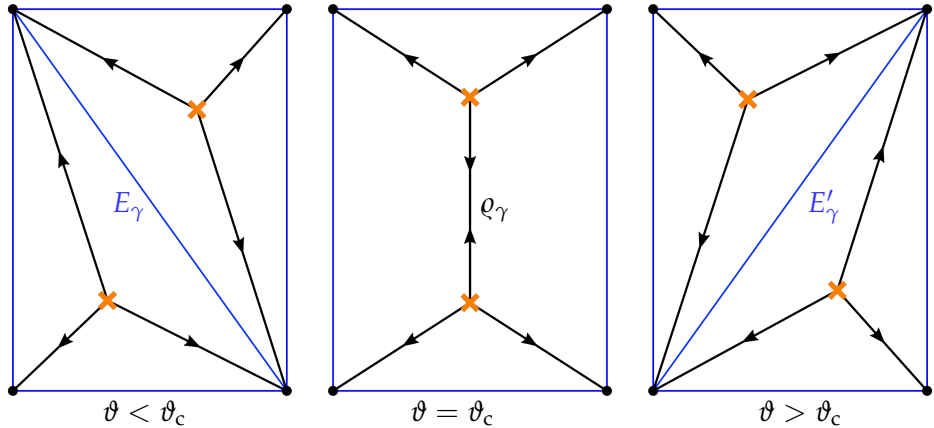
\begin{figure}[h!]
\small
\centering
\begin{overpic}
[width=0.80\textwidth]{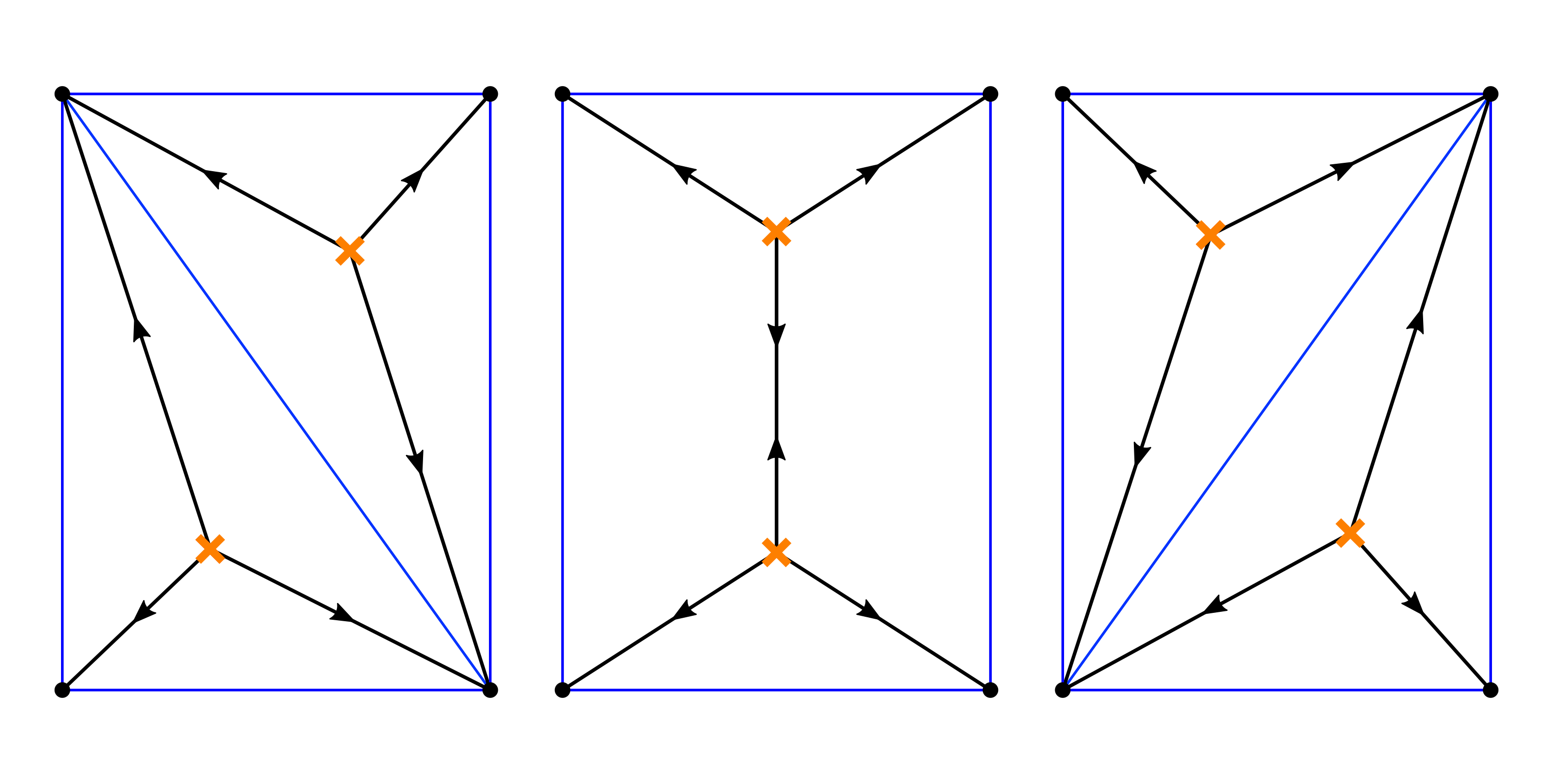}
\put(13,3){$\vartheta<\vartheta_{\rm c}$}
\put(44,3){$\vartheta=\vartheta_{\rm c}$}
\put(79,3){$\vartheta>\vartheta_{\rm c}$}
\put(14,24){\textcolor{Blue}{$E_{\gamma}$}}
\put(84,24){\textcolor{Blue}{$E'_{\gamma}$}}
\put(51,24){$\varrho_{\gamma}$}
\end{overpic}
\caption{\small Changes of the Stokes graph and the triangulation $\scrT_\vartheta$ as a critical value $\vartheta_{\rm c}$ is crossed. The edges of the triangulation are indicated with blue dashed lines. The edge $E_{\gamma}$ undergoes a flip when $\vartheta_{\rm c}$ is crossed. At $\vartheta=\vartheta_{\rm c}$ the edge $E_{\gamma}$ cannot be defined due to the presence of $\varrho_{\gamma}$ which lifts to $[\gamma]\in H_1(\Sigma,\Z)$.}
\label{jumpstandw}
\normalsize
\end{figure}

As we vary $\vartheta$ continuously, the topology of the triangulation $\scrT_\vartheta$ is piecewise constant and undergoes discontinuous jumps at critical values $\vartheta_{\rm c}$. These jumps give rise to {flips} of the edge $E_{\gamma}$ when going from $\vartheta<\vartheta_{\rm c}$ to $\vartheta>\vartheta_{\rm c}$. At $\vartheta=\vartheta_{\rm c}$ there is a finite curve $\varrho_{\gamma}$ that makes it impossible to define the edge $E_\gamma$. This process is shown in Figure~\ref{jumpstandw}. 
The path $\varrho_{\gamma}$ can be lifted to the Seiberg-Witten curve $\Sigma$ where it supports the cycle $\gamma$.

The flip of $E_{\gamma}$ can be identified with a BPS state of charge $\gamma$ in the spectrum, and $\vartheta_{\rm c}$ as the phase of its central charge $Z(\gamma)$. At the critical phase $\vartheta=\vartheta_{\rm c}$, the period 
\begin{align}
\E^{-\I\,\vartheta_{\rm c}}\,\oint_\gamma\,\lambda \ \in \ \R
\end{align}
is real and there is a Lagrangian disk in $T^*C$ bounded by the one-cycle $\gamma$ of $H_1(\Sigma,\Z)$. By wrapping an M2-brane around this disk, whose end lies on an M5-brane wrapped on $\R^{3,1}\times\Sigma$,
the state can be understood geometrically as a self-dual string in the M-theory background $\R^{3,1}\times T^*C\times\R^3$ which wraps the cycle $\gamma$, generating a BPS particle in the four-dimensional field theory upon (twisted) dimensional reduction over $C$~\cite{Klemm:1996bj}. The entire spectrum of the theory can be found by varying $\vartheta$ over a range of $\pi$ and keeping track of the changes of the triangulation $\scrT_\vartheta$; varying through a further angle $\pi$ gives the corresponding antiparticles. From this perspective the different regions of the Coulomb moduli space $\scrB$ correspond to different orderings of the central charges within an angular wedge of opening $\pi$.

For $SU(2)$ theories there are two possible topologies for double walls $\varrho_{\gamma}$, which lead to BPS hypermultiplets and vector multiplets respectively; their BPS degeneracies $\Omega(\gamma)=\Omega(-\gamma)$ is given by an index (second helicity supertrace) that {counts} BPS states of charge $\gamma$ in a precise sense. The state shown in Figure~\ref{jumpstandw} is called a {saddle} and corresponds to a BPS hypermultiplet $\gamma$ with degeneracy $\Omega(\gamma)=1$. The other topology is a closed curve on $C$. 

\begin{example}\label{ex:puresu2triang}
Going back to Example~\ref{ex:puresu2walls}, consider the large $u$ region of the Coulomb moduli space $\scrB_\infty$ of the pure $SU(2)$ gauge theory, where the ultraviolet curve $C$ is topologically a cylinder with irregular punctures at each end. These punctures can be replaced by boundary edges with one puncture on each of them. Figure~\ref{puresu2flips} shows the triangulation $\scrT_\vartheta$ of $C$ and its first flip as $\vartheta$ is varied.
\begin{figure}[h!]
\small
\centering
\begin{overpic}
[width=0.60\textwidth]{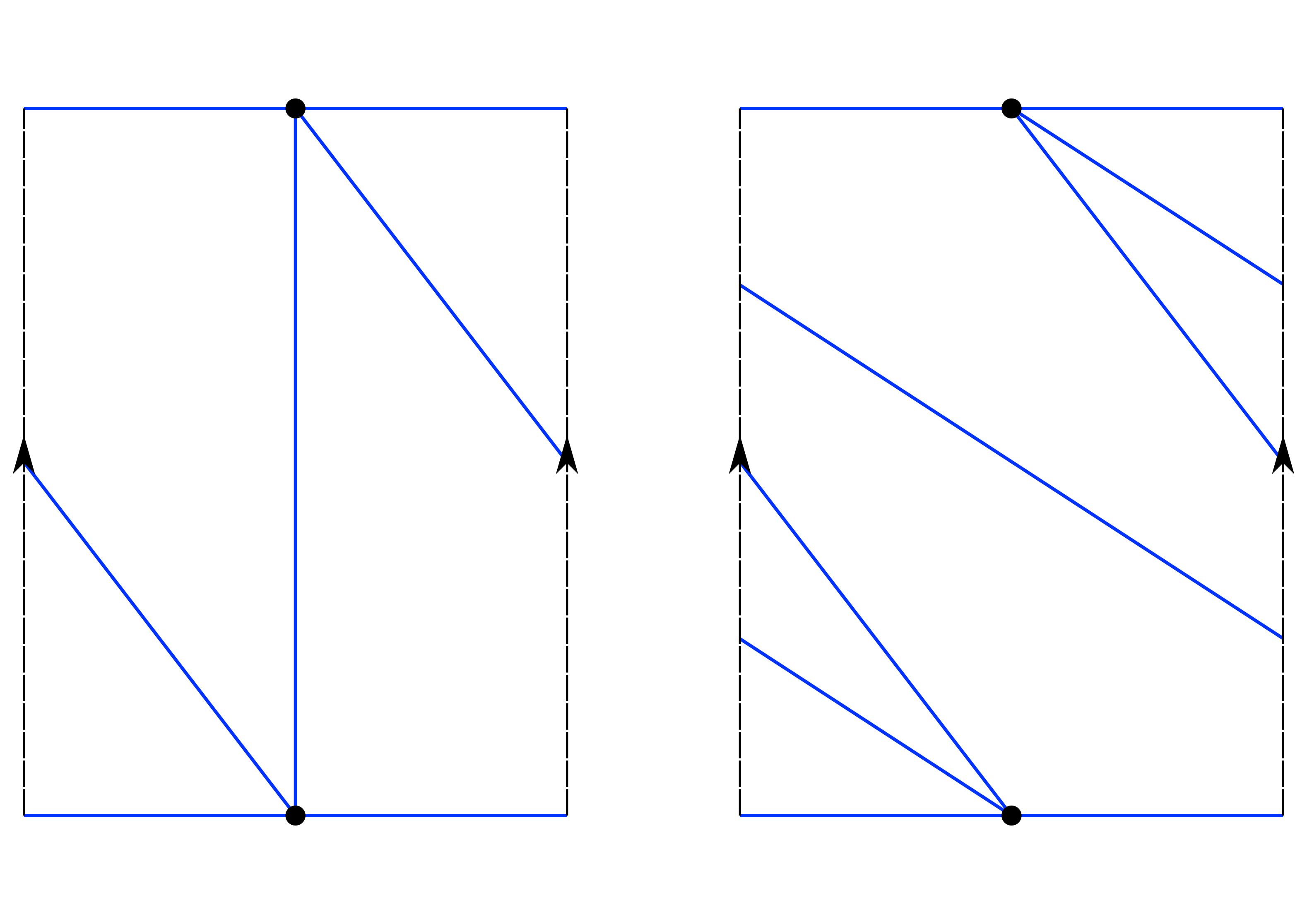}
\put(23.5,34){\textcolor{Blue}{$\gamma_1$}}
\put(33,40){\textcolor{Blue}{$\gamma_2$}}
\put(76,40){\textcolor{Blue}{$2\gamma_1+\gamma_2$}}
\put(69,34){\textcolor{Blue}{$-\gamma_1$}}
\end{overpic}
\caption{\small Triangulation $\scrT_\vartheta$ for the pure $SU(2)$ gauge theory and its first flip as $\vartheta$ is varied. The ultraviolet curve $C$ is a cylinder in this case, and its boundaries provide boundary edges which do not change and do not correspond to any electromagnetic charge $\gamma$. We draw it as a plane with opposite sides identified as indicated by the arrows.}
\label{puresu2flips}
\normalsize
\end{figure}
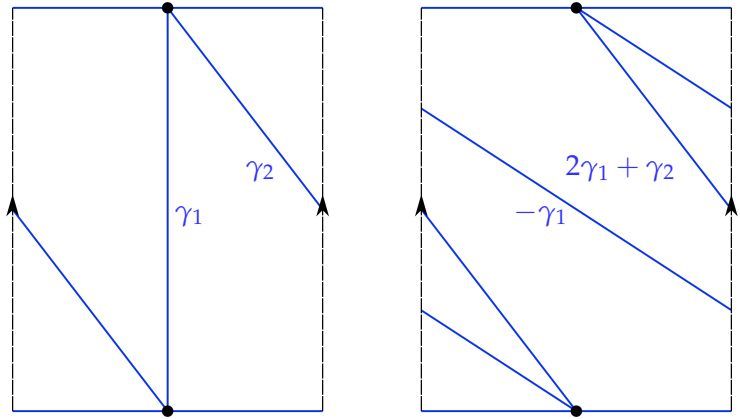
As we increase $\vartheta$, the edges wind the cycle of the cylinder. There is a critical value $\vartheta_{\rm c}$ where two finite curves $\varrho_{\gamma_1+\gamma_2}$ go around the cylinder and can both be lifted to $\gamma_1+\gamma_2$.\footnote{Likewise the finite curves which can be lifted to support the BPS states $\gamma_1$ and $\gamma_2$ also wind the cylinder.} This corresponds to a BPS vector multiplet $\gamma$ with degeneracy $\Omega(\gamma)=-2$.\footnote{There is actually a one-parameter family of closed trajectories and the two finite trajectories with ends on the branch point are just the boundaries of this family. The Hilbert space of one-particle states associated to this family is the cohomology of its moduli space, and the prescription of~\cite{gaiotto2012spectral} for the invariant trace over this Hilbert space yields the degeneracy $\Omega(\gamma)=-2$.} Increasing $\vartheta$ further beyond $\vartheta_{\rm c}$ leads to triangulations with edges winding the cylinder in the opposite direction. This change from infinite positive winding to infinite negative winding is called a {juggle} and is illustrated in Figure~\ref{juggletriangulations}. 
\begin{figure}[h!]
\small
\centering
\begin{overpic}
[width=0.80\textwidth]{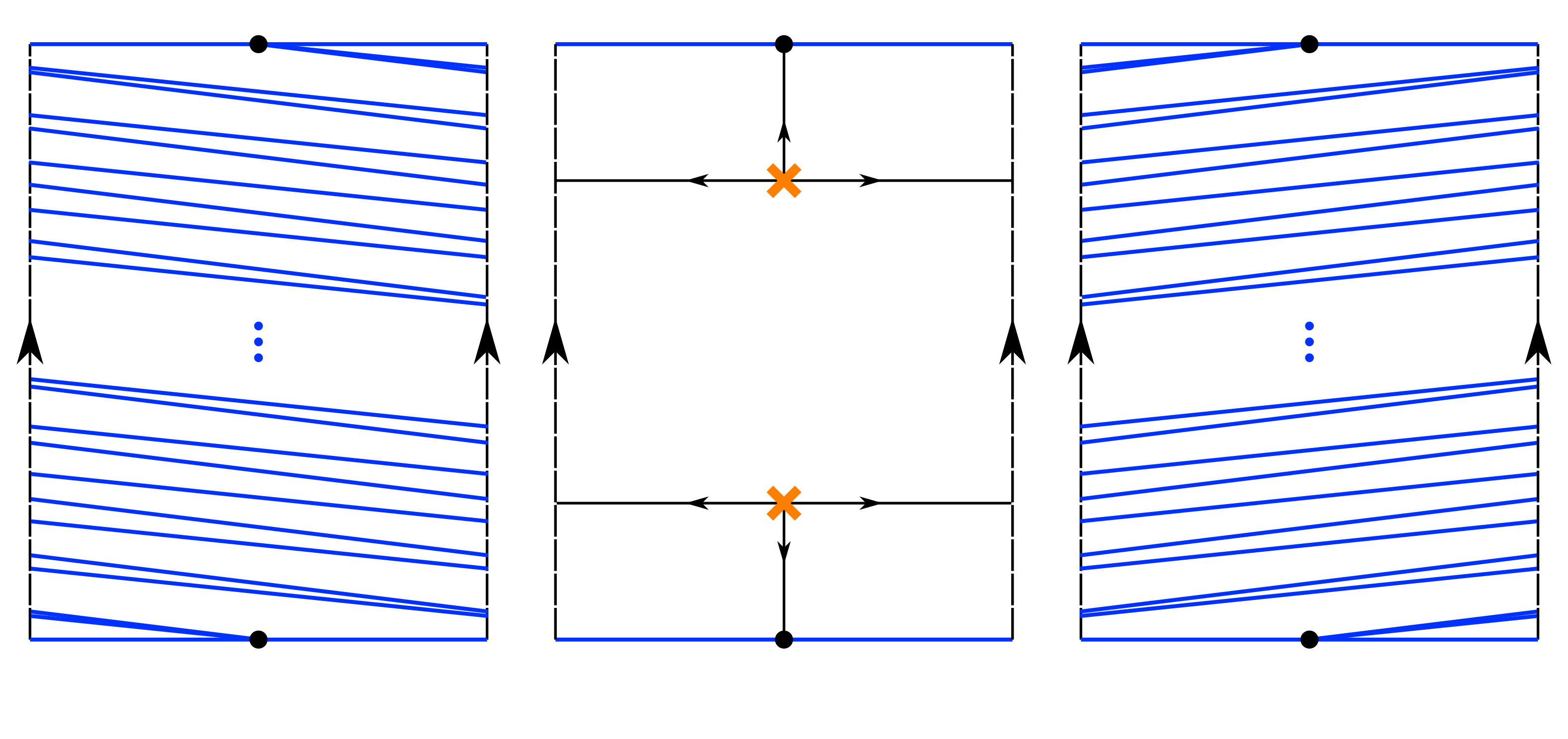}
\put(12,0){$\vartheta<\vartheta_{\rm c}$}
\put(46,0){$\vartheta=\vartheta_{\rm c}$}
\put(80,0){$\vartheta>\vartheta_{\rm c}$}
\end{overpic}
\caption{\small Triangulations $\scrT_\vartheta$ for the pure $SU(2)$ gauge theory for $\vartheta$ below and above $\vartheta_{\rm c}$ have edges winding the cylinder. At $\vartheta=\vartheta_{\rm c}$ the triangulation cannot be defined and we instead draw the Stokes graph, which contains two closed finite curves that lift to $\gamma_1+\gamma_2$.}
\label{juggletriangulations}
\normalsize
\end{figure}
The resulting spectrum consists of hypermultiplets $\gamma_i + k\,(\gamma_1+\gamma_2)$, for $i=1,2$ and $k\ge 0$, and the vector state $\gamma_1+\gamma_2$.
\end{example}

\subsection{Fock-Goncharov Coordinates}
\noindent
The triangulations $\scrT_\vartheta$ can also be used to define a system of holomorphic Darboux coordinates \smash{$\cX_\gamma=\cX_{\gamma}^{u,\vartheta}$} for the moduli space $\Mflat(C)$ of flat $SL(2,\mathbb{C})$ connections on $C$~\cite{gaiotto2013wall,gaiotto2012spectral,hollands2016spectral}. They are indexed by $[\gamma]\in H_1(\Sigma,\Z)$ and depend on the parameters $(u,\vartheta)\in\scrB\times[0,2\pi)$. Defining $\cX_{\gamma}$ requires a choice of some extra discrete data called a {framing}: given an $SL(2,\C)$ local system on $C$, this assigns to every puncture $p_i\in C$ a flat section $s_i$ of the associated projective line bundle over a neighbourhood of $p_i$ with monodromy eigenvalue $\mu_i$ when going around $p_i$. For the triangulations defined by the WKB foliations $\cF_\vartheta$ in Section~\ref{BPSgeometry}, these can be obtained as {small flat sections} by using WKB analysis \cite{gaiotto2013wall}, and the triangulation $\scrT_\vartheta$ with this framing is called a {WKB triangulation}. In addition to the flip which changes edges of $\scrT_\vartheta$, there is now also an operation that acts on the framing: the {pop} sends $\mu_i$ to \smash{$\mu_i^{-1}$}. For an edge $E_{\gamma}$ we can then define the coordinate $\cX_{\gamma}$ by assigning single-valued smooth sections $s_i$, $i=1,2,3,4$, to the corners of the quadrilateral shown in Figure~\ref{triangulationcycle}, ordered clockwise starting from the top right corner.

The Fock-Goncharov coordinate is then given by the cross-ratio
\begin{equation}\label{Xgammadef}
\cX_{\gamma}=-\frac{s_1\wedge s_2}{s_2\wedge s_3} \, \frac{s_3 \wedge s_4}{s_4\wedge s_1} \ ,
\end{equation}
which is invariant under complex rescaling of the sections $s_i$ and lies in $\mathbb{C}^{\times}$ for generic points in $\Mflat(C)$~\cite{gaiotto2013wall}.\footnote{There are loci in $\Mflat(C)$ where coordinates defined like this vanish as two sections align and their exterior product is equal to $0$. The Fock-Goncharov coordinates are therefore defined on a dense open subset of $\Mflat(C)$ where this does not happen.} These coordinates span the group ring $\C[\Gamma]$ of the charge lattice $\Gamma$,
\begin{equation}
\C[\Gamma] = \bigoplus_{\gamma\in\Gamma} \, \C \, \cX_\gamma \qquad \mbox{with} \quad \cX_{\gamma} \, \cX_{\gamma'} = \cX_{\gamma+\gamma'} \ ,
\end{equation}
and they are (exponentiated) Darboux coordinates with respect to the natural holomorphic Poisson structure on the moduli space $\scrM_{\rm flat}(C)$ with Poisson bracket $\{\cdot,\cdot\}$ given by
\begin{equation}
\left\{\cX_{\gamma},\cX_{\gamma'}\right\} = \langle\gamma,\gamma'\rangle \, \cX_{\gamma+\gamma'} \ .
\end{equation}
Thus $\cX_\gamma$ generate the coordinate ring of the complex algebraic torus with character lattice $\Gamma$, called the complex Poisson torus.

We have seen that as $\vartheta$ varies continuously there are critical values $\vartheta_{\rm c}$ where the topology of the triangulation undergoes a discontinuous change due to the presence of a BPS state $\gamma_{\bps}$ with arg$(Z(\gamma_{\bps}))=\vartheta_{\rm c}$. This results in a discontinuity in the coordinates \smash{$\big\{\cX_{\gamma}^{u,\vartheta}\big\}_{\gamma\in\Gamma}$} given by \smash{$\cX_\gamma\mapsto\cK_{\gamma_\bps}^{\Omega(\gamma_\bps)}(\cX_\gamma)$}, where $\cK_{\gamma_{\bps}}$ is the  Kontsevich-Soibelman symplectomorphism  acting on the complex Poisson torus as the birational cluster transformation~\cite{kontsevich2008stability}
\begin{equation}\label{Kwall}
\cK_{\gamma_{\bps}}:\cX_{\gamma}\longmapsto\cX_{\gamma}\,\big(1-\sigma(\gamma_{\bps})\,\cX_{\gamma_{\bps}}\big)^{\langle\gamma,\gamma_{\bps}\rangle} \ ,
\end{equation}
and the quadratic refinement $\sigma(\gamma_{\bps})$ is $-1$ for BPS hypermultiplets and $+1$ for BPS vector multiplets. This transformation of the coordinates is also based on a relabelling of edges which changes the label $\gamma_{\bps}$ as well as any other label $\gamma$ as
\begin{align}\label{fliprules}
\gamma_{\bps}\longmapsto - \gamma_{\bps}\qquad \mbox{and} \qquad
\gamma \longmapsto \gamma + \langle\gamma,\gamma_{\bps}\rangle_+ \, \gamma_{\bps} \ ,
\end{align}
where $\langle\cdot,\cdot\rangle_+=\frac12\,\big(\langle\cdot,\cdot\rangle+|\langle\cdot,\cdot\rangle|\big)$ is the positive part of the electric-magnetic pairing $\langle\cdot,\cdot\rangle$.

As we vary $\vartheta$ over an angle $\pi$, we get a symplectomorphism $\cK_{\gamma_\bps}$ for each BPS state $\gamma_\bps$ in the spectrum and the resulting ordered product of symplectomorphisms \smash{$\cK_{\gamma_\bps}^{\Omega(\gamma_\bps)}$} is called the {spectrum generator} $\bbS$. The symplectomorphism $\bbS$ is constant on the Coulomb branch $\scrB$, but its decomposition in terms of $\cK_{\gamma_\bps}$ can change throughout $\scrB$. The general form of $\bbS$ is the ordered product
\begin{equation}\label{Sdef}
\bbS(\vartheta;u)=\prod^{\curvearrowleft}_{\gamma_{\bps}\in\Gamma_\vartheta} \, \cK_{\gamma_{\bps}}^{\Omega(\gamma_{\bps};u)} \ ,
\end{equation}
where $\Gamma_\vartheta=\{\gamma\in\Gamma\ |\ \vartheta<\text{arg}(-Z(\gamma))<\vartheta+\pi\}$ and the product is taken in increasing order of $\text{arg}(-Z(\gamma_{\bps}))$ from right to left. The  $u$-dependence of $\bbS(\vartheta;u)$ enters through reordering of charges and jumps of the BPS degeneracies $\Omega(\gamma_{\bps};u)$ when walls of marginal stability are crossed. Recall that in $SU(2)$ theories one should only encounter charges lifted from saddle trajectories or closed loops on the ultraviolet curve $C$, which correspond to BPS hypermultiplets with $\Omega(\gamma)=1$ and BPS vector multiplets with $\Omega(\gamma)=-2$ respectively.

\subsection{Wall-Crossing Formulas}\label{wallcrossing}
\nid
As discussed in Section~\ref{subsec:BPSStatesWalls}, there are real codimension one walls in the Coulomb branch $\scrB$ at which the spectrum undergoes discontinuous changes. One way to keep track of the changes in the BPS spectrum on crossing these walls of marginal stability is through the {Kontsevich-Soibelman wall-crossing formula}~\cite{kontsevich2008stability}, which governs permutations of the factors $\cK_{\gamma_\bps}$ which make up the spectrum generator $\bbS$ in~(\ref{Sdef}). 

\begin{example}\label{ex:AD3wallcrossing}
The AD$_3$ theory is the rank one superconformal field theory at the Argyres-Douglas point in the Coulomb moduli space of $\N=2$ supersymmetric Yang-Mills theory with gauge group $SU(3)$~\cite{Argyres:1995jj}, whose Seiberg-Witten curve is $w^2=z^3-3\,\Lambda^2\,z+u$ with ultraviolet curve $C=\C=\PP^1\setminus\{\infty\}$. The Coulomb branch $\scrB$ in this case has two regions separated by a wall of marginal stability which is roughly an ellipse around $u=0$, passing through the two singular points at $u=\pm\,2\,\Lambda^3$. Near $u=0$ the spectrum consists of two hypermultiplets $\gamma_1$ and $\gamma_2$, with pairing $\langle\gamma_1,\gamma_2\rangle=1$ and phases arg$(Z(\gamma_1))>\text{arg}(Z(\gamma_2))$. The spectrum generator then reads
\begin{equation}
\bbS=\cK_{\gamma_1}\,\cK_{\gamma_2} \ .
\end{equation}
At the wall of marginal stability these two states align and in the outside region their ordering is changed to arg$(Z(\gamma_1))<\text{arg}(Z(\gamma_2))$. We therefore have to permute the symplectomorphisms $\cK_{\gamma_i}$ in the spectrum generator and for $\langle\gamma_1,\gamma_2\rangle=1$ this permutation is given by the primitive wall-crossing formula
\begin{equation}\label{KSWCFone}
\bbS=\cK_{\gamma_1}\,\cK_{\gamma_2} = \cK_{\gamma_2}\,\cK_{\gamma_1+\gamma_2}\,\cK_{\gamma_1} \ .
\end{equation}
We therefore find an additional bound state $\gamma_1+\gamma_2$ in the region outside of the wall, which is a BPS hypermultiplet. 
\end{example}

\begin{example}\label{ex:puresu2wallcrossing}
Recall from Example~\ref{ex:puresu2walls} that the Coulomb branch $\scrB_\infty$ of the pure $SU(2)$ gauge theory also has one wall of marginal stability around $u=0$ with two hypermultiplets $\gamma_1$ and $\gamma_2$ in the interior region. However, compared to Example~\ref{ex:AD3wallcrossing}, now their pairing is $\langle\gamma_1,\gamma_2\rangle=2$. Crossing the wall of marginal stability then gives the formula
\begin{equation}\label{KSWCFtwo}
\bbS=\cK_{\gamma_1}\,\cK_{\gamma_2} = \varPi_2^{\gamma_1,\gamma_2} \, \cK_{\gamma_1+\gamma_2}^{-2} \, \varPi_1^{\gamma_1,\gamma_2} \ ,
\end{equation}
where 
\begin{align}
\varPi_1^{\gamma_1,\gamma_2} = \prod_{k\geq0}^\curvearrowleft \, \cK_{(k+1)\,\gamma_1+k\,\gamma_2} \qquad \mbox{and} \qquad \varPi_2^{\gamma_1,\gamma_2} = \prod_{k\geq0}^\curvearrowright \, \cK_{k\,\gamma_1 + (k+1)\,\gamma_2} \ .
\end{align}
This shows that, in addition to the state $\gamma_1+\gamma_2$ which is a BPS vector multiplet in this case, there are two infinite families given by $(k+1)\,\gamma_1+k\, \gamma_2$ and $k\,\gamma_1+(k+1)\,\gamma_2$ for $k\geq0$ which are hypermultiplets. 
\end{example}

For $SU(2)$ theories, besides the pairings one and two encountered in Examples~\ref{ex:AD3wallcrossing} and~\ref{ex:puresu2wallcrossing}, the only other electric-magnetic pairing that should occur in wall-crossings is between mutually local charges, i.e. $\langle\gamma,\gamma'\rangle=0$, in which case no new bound states are formed, and the symplectomorphisms $\cK_\gamma$ and $\cK_{\gamma'}$ commute. This is particularly relevant for flavour states $\gamma_{\rm f}$, which are mutally local with all other BPS states and cannot be detected by the wall-crossing formula. Pairings larger than two are not expected to occur, as they would lead to {``wild'' spectra}~\cite{galakhov2013wild}. 

Negative pairings should only occur in terms of the reverse orderings of wall-crossing formulas such as (\ref{KSWCFone}) and (\ref{KSWCFtwo}), as they would otherwise violate the no-exotics theorem~\cite{gaiotto2013framed,galakhov2013wild}. In fact, it can be argued from {Denef's multicentre equations} that primitive wall-crossings with negative pairings cannot occur~\cite{galakhov2013wild}. These equations describe BPS bound states of D-branes in supergravity and string theory settings~\cite{denef2002quantum,denef2011split}, but as shown in \cite{galakhov2013wild} a four-dimensional field theory limit of them should be applicable to BPS states in $\N=2$ field theories.

Consider a set of $N$  BPS states $\gamma_1,\dots,\gamma_N$ placed at points $\boldsymbol{r}_1,\dots,\boldsymbol{r}_N\in\mathbb{R}^3$. Denef's equations give a sufficient condition for this collection to again form a BPS configuration, namely
\begin{equation}\label{denefeqn}
\sum_{\stackrel{\scriptstyle j=1}{\scriptstyle j\neq i}}^N\, \frac{\langle\gamma_i,\gamma_j\rangle}{|\boldsymbol{r}_i-\boldsymbol{r}_j|}=2\,\text{Im}\big(\E^{-\I\,\vartheta}\,Z(\gamma_i)\big) \ ,
\end{equation}
for $i=1,\dots,N$, where $\vartheta=\text{arg}\big(\sum_{i=1}^N\, Z(\gamma_i)\big)$.
Now consider two  BPS states with pairing $\langle\gamma,\gamma'\rangle<0$, where $\text{arg}(Z(\gamma))>\text{arg}(Z(\gamma'))$ and no other symplectomorphisms lie between $\cK_{\gamma}$ and $\cK_{\gamma'}$ in the spectrum generator. By approaching the wall where the central charges of $\gamma$ and $\gamma'$ align, this would lead to a wall-crossing with negative pairing. Then using (\ref{denefeqn}) it can be shown that $\gamma+\gamma'$ is a BPS state as well, thereby violating the initial assumption of having no further symplectomorphism factors in between $\cK_{\gamma}\,\cK_{\gamma'}$. In fact, from this argument we expect to see exactly the states that would be created by crossing the wall from the other side with pairing $\langle\gamma',\gamma\rangle>0$.

In Section~\ref{missingstates} we will discuss how negative pairing wall-crossings in the $\N=2^*$ gauge theory can be interpreted in terms of the reverse orderings of the pairing two wall-crossing formula, and we identify explicitly some of the bound states that we need for this interpretation. 

\subsection{\tops{$\N=2^*$ Spectrum}{N=2* Spectrum via Spectrum Generator}}\label{longhispectrum}
\nid
Let us now summarise some of the known features of the BPS spectrum of the rank one $\N=2^*$ gauge theory, following~\cite{schulze1997bps,ferrari1997dyon,alim2014mathcal,longhi2015structure}. Recall that the $\N=2^*$ theory can be regarded as an interpolation between the $\N=4$ and the pure $\N=2$ supersymmetric Yang-Mills theories, to which it respectively reduces by taking the limit in which the adjoint hypermultiplet mass $m$ is sent to zero or decoupled. For $m\to\infty$, the complete spectrum of the pure $\N=2$ theory was described in Example~\ref{ex:puresu2walls}. For $m=0$, the BPS spectrum of $\N=4$ supersymmetric Yang-Mills theory consists of any dyon of primitive electromagnetic charge, i.e. all charges 
\begin{align}
\gamma_{[p,q]}=p\,A+q\,B \ ,
\end{align}
where $(p,q)$ are relatively prime integers and $(A,B)$ is the canonical symplectic homology basis for the torus of modulus $\tau_0$ (cf.~Section~\ref{sec:SWgeometry}); in particular,  the spectrum is not finitely generated. This is a consequence of S-duality: by B\'ezout's identity, any such dyon $\gamma_{[p,q]}$ is generated by an application of $SL(2,\Z)$ to the primitive monopole state with charge $\gamma_{[0,1]}$. These BPS states have mass \smash{$M\big(\gamma_{[p,q]}\big)=\big|Z\big(\gamma_{[p,q]}\big)\big|=\sqrt{u_0} \, |p+\tau_0\,q|$}.

One might  expect that S-duality of the $\N=2^*$ theory could be similarly exploited to analyse the spectrum, but this is not so: whereas in the massless limit there are no walls of marginal stability where BPS bound states can form or decay, there are plenty of walls in the Coulomb branch $\scrB_m$ for $m\neq0$ and the action of S-duality on the modulus $u$ (cf.~Section~\ref{sec:SWgeometry}) generally maps between different chambers of $\scrB_m$; later on we shall encounter many explicit instances of such walls and chambers. As one continuously varies the bare mass $m$ and coupling $\tau_0$, and uses the renormalization group flow, the change in spectrum from $m=\infty$ to $m=0$ is continuous~\cite{ferrari1997dyon}. Hence all BPS states of the pure $\N=2$ theory are also in the spectrum of the $\N=2^*$ theory, whereas dyons $\gamma_{[p,q]}$ are unstable for $m\neq0$ when $(p,q)$ have a non-trivial common divisor~\cite{schulze1997bps}. The latter feature restricts the number of walls of marginal stability that one must consider to study the eventual decays of BPS states; for example, the primitive monopole cannot decay for $m\neq0$ outside the wall that is inherited from the pure $\N=2$ theory. In~\cite{schulze1997bps,ferrari1997dyon} it is shown that BPS states with \emph{any} magnetic (or electric) charges exist in the $\N=2^*$ spectrum in the region of the Coulomb branch $\scrB_m$ outside the wall of marginal stability of the $W$-boson, which however have mass larger than $m$ and so decouple in $\scrB_\infty$. The problem then is to determine the extent to which $\N=4$ BPS states destabilize and disappear from the spectrum as the mass increases from $m=0$. 

The complexity of the $\N=2^*$ spectrum was exemplified in~\cite{alim2014mathcal} where it was shown that there exists two vector particles, which are mutually non-local, for any choice of central charges. These lead to stable vector states that could form a highly complicated spectrum of bound states, as multiple accumulation rays (one at each vector state) in the region of the central charge plane between the two vector multiplets could be arbitrarily ``wild'': the additional wall-crossings of vector multiplets will produce some highly complicated spectrum with infinitely many vector states. We shall encounter precisely this behaviour in our numerical explorations of the $\N=2^*$ spectrum in Section~\ref{Pythonspectrum1}, as well as further complexities throughout Section~\ref{perturbedspectrum}. Geometrically, this is the phenomenon of ``accumulation of accumulation rays" (and higher iterations thereof) in the central charge plane (see~\cite[Section 5.6.1]{Coman:2020qgf}), which occurs for ultraviolet curves $C$ such as the once-punctured torus or Riemann surfaces without punctures, and is measured by the Cantor-Bendixson rank of the quadratic differential $\lambda^{\otimes 2}$~\cite{Aulicino2015}. In Example~\ref{ex:Fterm} we shall mention an algebraic interpretation of this ``wild'' behaviour in terms of the representation theory of the associated BPS quiver.

Generally, the spectrum generator for $SU(2)$ theories of class $\cS$ can also be constructed by using just one framed ideal triangulation $\scrT_\vartheta$ at fixed generic angle $\vartheta$. This was done by Longhi in~\cite{longhi2015structure} for the $\N=2^*$ theory following the prescription in \cite{gaiotto2013wall}. Choose a symplectic basis of one-cycles $\{\gamma_i=(\gamma_{\rm e}^i,\gamma_{\rm m}^i)\}_{i=1,2,3}$ for the homology group $H_1(\Sigma_{u;m},\Z)=\Z^2\oplus\Z^2\oplus\Z^2$ of the $\N=2^*$ Seiberg-Witten curve. The idea then is to determine the spectrum generator by its action on the basis of Fock-Goncharov coordinates $\cX_{\gamma_i}$ through
\begin{equation}
\cX_{\gamma_i}^{\vartheta+\pi} = \bbS \cX_{\gamma_i}^{\vartheta} = \bbS_i\, \cX_{\gamma_i}^{\vartheta} \ .
\end{equation}
The spectrum generator can be obtained from $\scrT_\vartheta$ by using the fact that the resulting triangulation at $\vartheta+\pi$ is the original triangulation but with the entire framing popped. The action of this `omnipop' can then be computed directly.

To find the spectrum, we need to find a decomposition in terms of symplectomorphisms $\cK_\gamma$ as given in (\ref{Sdef}). This can be arbitrarily difficult to do, but sometimes an ansatz for the spectrum can be found and then checked with the spectrum generator. For the $\N=2^*$ theory this is justified by the observation that, even at strong coupling, no new BPS states appear other than the ones expected from perturbation theory around all of its phases~\cite{schulze1997bps}. In this case Longhi uses in~\cite{longhi2015structure} a semi-classical ansatz on a very specific subset of the $\N=2^*$ Coulomb branch $\scrB_m$, namely on the walls of marginal stability defined by 
\begin{equation}\label{longhiwall}
\scrE_i = \Big\{u\in\scrB_m \ \Big| \ \frac{Z(\gamma_i;u)}{Z(\gamma_{\rm f};u)}\in\R_{>0} \ \ , \ \ \text{arg}\big(Z(\gamma_{i+1};u)\big)<\text{arg}\big(Z(\gamma_{i-1};u)\big)\Big\} \ ,
\end{equation}
where the index $i=1,2,3$ is read cyclically modulo~$3$.
On these walls one of the basis charges aligns with the flavour charge $\gamma_{\rm f}=\gamma_1+\gamma_2+\gamma_3$, which corresponds to the adjoint hypermultiplet of the $\N=2^*$ gauge theory with bare mass $m$ and constant central charge $Z(\gamma_{\rm f};u)=m$ throughout $\scrB_m$. This does not lead to any wall-crossing that the Kontsevich-Soibelman wall-crossing formula could detect, since $\gamma_{\rm f}$ has zero pairing with any other state. Hence the BPS spectrum is unambiguously defined on the walls $\scrE_i$.

A semi-classical analysis of the field theory yields hypermultiplets $\gamma_{i\pm1}$, as well as a vector state $\gamma_{i-1}+\gamma_{i+1}$ with the usual families of hypermultiplets $\gamma_{i\pm1}+n\,(\gamma_{i-1}+\gamma_{i+1})$ in the $m\to\infty$ spectrum from Example~\ref{ex:puresu2wallcrossing}. In addition to these, there are hypermultiplets with electromagnetic charges $\gamma_{\rm f}+\gamma_{i-1}+\gamma_{i+1}$ and $\gamma_{\rm f}-(\gamma_{i-1}+\gamma_{i+1})=\gamma_i$. The corresponding spectrum generator is~\cite{longhi2015structure,Longhi:2016wtv}
\begin{equation}\label{Spectrum}
\bbS=\varPi_2^{\gamma_{i+1},\gamma_{i-1}}\,\cK_{\gamma_{\rm f}-(\gamma_{i-1}+\gamma_{i+1})}\,\cK^{-2}_{\gamma_{i-1}+\gamma_{i+1}}\,\cK_{\gamma_{\rm f}+(\gamma_{i-1}+\gamma_{i+1})}\,\varPi_1^{\gamma_{i+1},\gamma_{i-1}} \ ,
\end{equation}
where the symplectomorphisms \smash{$\varPi_j^{\gamma_{i+1},\gamma_{i-1}}$} for $j=1,2$ correspond to infinite families of BPS hypermultiplets with charges \smash{$\gamma_j+k\,(\gamma_{i-1}+\gamma_{i+1})$, $\cK^{-2}_{\gamma_{i-1}+\gamma_{i+1}}$} to a BPS vector multiplet of charge $\gamma_{i-1}+\gamma_{i+1}$, and the remaining \smash{$\cK_{\gamma_{\rm f}\pm(\gamma_{i-1}+\gamma_{i+1})}$} to two more hypermultiplets.\footnote{The infinite mass limit of the spectrum on the walls of marginal stability $\scrE_i$ can be taken as follows. Since $\gamma_{\rm f}$ and $\gamma_i$ align, sending $m$ to infinity does not change the phase of $Z(\gamma_i)$, and similarly for the state $\gamma_{i-1}+\gamma_{i+1}+\gamma_{\rm f}$ which will also acquire infinite mass but is likewise aligned with $\gamma_{\rm f}$. Therefore the states $\gamma_i$ and $\gamma_{i-1}+\gamma_{i+1}+\gamma_{\rm f}$ decouple and the remaining states are the vector multiplet $\gamma_{i-1}+\gamma_{i+1}$ and the hypermultiplets $\gamma_{i\pm 1}+k\,(\gamma_{i-1}+\gamma_{i+1})$ for $k\ge 0$, which are  unaffected by the limit and constitute the spectrum of the pure $SU(2)$ gauge theory. For general regions of the Coulomb branch $\scrB_m$, however, taking this limit is more difficult as varying one of the central charges will in general lead to a reordering of symplectomorphisms $\cK_{\gamma}$ and therefore to wall-crossing. Notice also that the walls $\scrE_i$ are ill-defined in the massless limit to the $\N=4$ supersymmetric Yang-Mills theory.}

We will compute this spectrum through an algebraic method in Section \ref{spectrumonwall}, and then discuss a perturbation away from the walls \eqref{longhiwall} in Section \ref{perturbedspectrum}. 

\section{\tops{$\boldsymbol{\N=2^*}$ Spectrum from Representation Theory}{N=2* Spectrum from BPS Quivers}}
\label{sec:BPSquivers}
\nid
In this section we will construct the spectrum of the $\N=2^*$ theory on the walls defined in (\ref{longhiwall}) by using quiver representation theory methods, reproducing in this way the results of~\cite{longhi2015structure}, summarised in Section~\ref{longhispectrum}, from an algebraic perspective that is tailored to our later investigations. We will first briefly review the general theory of BPS quivers and the mutation method from~\cite{alim2013bps,alim2014mathcal,Cecotti:2014zga}, then construct the $\N=2^*$ spectrum, and finally discuss the underlying triangulations and the geometry of finite curves.

\subsection{BPS Quivers}
\nid
We can compute the BPS spectrum of the $\N=2^*$ theory from {BPS quivers}. A quiver is a directed graph, i.e. a set of {nodes} with directed {arrows} between them. A quiver can be associated to the BPS spectrum of an $\N=2$ theory, where the nodes $\{\bullet_i\}$ correspond to certain charges $\{\gamma_i\}$ and the arrows to their intersection pairings. A linear representation of a quiver is an assignment of a complex vector space to each node and a linear map between vector spaces for each arrow. Any BPS state will then define a space of quiver representations, characterized uniquely by the dimension vector $\vec n=(n_i)$ where $n_i$ is the dimension of the vector space at node $\bullet_i$, together with the moduli of linear maps \smash{${\sf a}^{\vec n}_{ji}$} represented by $n_j{\times}n_i$ complex matrices for each arrow $\bullet_i\longrightarrow \bullet_j$.

To assign a quiver to a BPS spectrum, we first fix the cone of particles, i.e. a choice of half-plane $\mathbb{H}_{\vartheta}\hookrightarrow\mathbb{C}$ centred on the ray $\{z\in\C~|~\text{arg}(z)=\vartheta\}$ such that no central charge lies on the boundary $\partial\mathbb{H}_\vartheta = \E^{\,\I\,\vartheta}\,\R$.\footnote{This is not always possible. In fact for some $\N=2$ theories the BPS rays are dense in the circle and any choice of upper half-plane will have charges on its boundary. The spectra of such theories are not finitely generated. An example is $\N=4$ supersymmetric Yang-Mills theory, and more generally any class~$\cS$ theory whose ultraviolet curve $C$ is a Riemann surface without punctures.} Any state $\gamma$ whose central charge $Z(\gamma)$ lies in this charge plane is considered to be a particle, while $-\gamma$ is considered as the corresponding antiparticle. We can then find a positive integral basis $\{\gamma_i\}$ for the BPS states and write the electromagnetic charge $\gamma_\bps$ of any BPS particle as
\begin{equation}\label{eq:gammabps}
\gamma_{\bps}=\sum_i\, n_i\,\gamma_i \qquad \mbox{with} \quad n_i\in\mathbb{Z}_{\geq0} \ ,
\end{equation}
with corresponding central charge
\begin{align}\label{eq:gammabpsZ}
Z(\gamma_\bps;u) = \sum_i \, n_i\,Z(\gamma_i;u) = \oint_{\gamma_\bps} \, \lambda
\end{align}
in $\mathbb{H}_\vartheta$. This choice of basis is in fact unique.

For a theory with gauge group of rank $r$ and with $n_{\rm f}$ flavours, the local system $\Gamma$ has rank $2r+n_{\rm f}$, and we thus need $2r+n_{\rm f}$ basis elements $\gamma_i$.\footnote{Note that we ask only for a basis for all BPS particles, not a $\Z$-basis that spans the lattice $\Gamma$.} From a physical perspective, we can think of the $\gamma_i$ as BPS hypermultiplets and of the particles formed by sums of these as composite objects or bound states. This point of view is reinforced by the fact that the basis states $\gamma_i$ are always BPS hypermultiplets of the theory (see Example~\ref{ex:nodes} below).

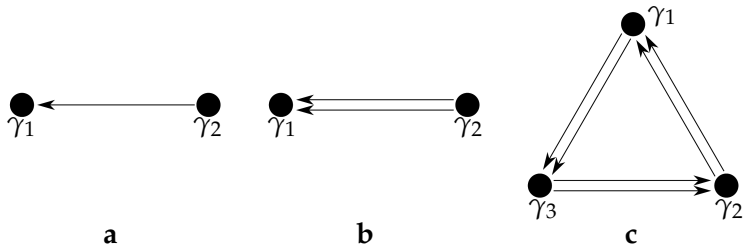
\begin{figure}[h!]
\small
\centering
\begin{overpic}
[width=0.60\textwidth]{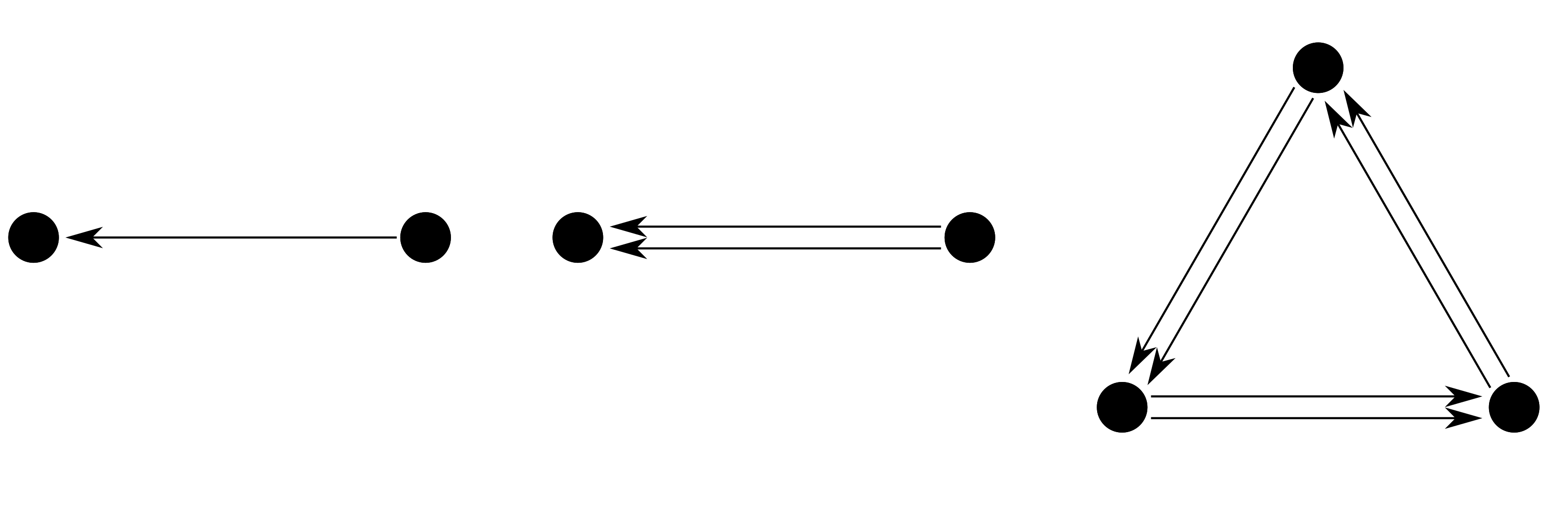}
\put(13,0){\textbf{a}}
\put(47,0){\textbf{b}}
\put(83,0){\textbf{c}}
\put(0,15){$\gamma_1$}
\put(25,15){$\gamma_2$}
\put(35,15){$\gamma_1$}
\put(60,15){$\gamma_2$}
\put(70,4){$\gamma_3$}
\put(95,4){$\gamma_2$}
\put(86,30){$\gamma_1$}
\end{overpic}
\caption{\small BPS quivers of \textbf{a:} AD$_3$ theory (Dynkin quiver of type $A_2$), \textbf{b:} pure $SU(2)$ theory (Kronecker quiver), \textbf{c:} $\N=2^*$ theory (Markov quiver).}
\label{quiverexamples}
\normalsize
\end{figure}

Having fixed a particle cone and hence obtained the basis $\{\gamma_i\}$, we define the {BPS quiver} as follows:
\begin{itemize}
\item[$\diamond$] To each $\gamma_i$ we assign a node $\bullet_i$.
\item[$\diamond$] For each pair $\gamma_i$ and $\gamma_j$ we compute the electric-magnetic pairing $\langle\gamma_j,\gamma_i\rangle$, and we draw $\langle\gamma_j,\gamma_i\rangle_+$ arrows from node $\bullet_i$ to node $\bullet_j$. 
\end{itemize}
There is a small subtlety with the second point. Since the direction of arrows is determined by the sign of the electric-magnetic pairing, one does not obtain quivers in this way where arrows between two nodes have opposite directions. This subtlety can be important for certain theories (see \cite{alim2014mathcal}), but it will be irrelevant for us. Figure~\ref{quiverexamples} shows the resulting BPS quivers of the AD$_3$ theory from Example~\ref{ex:AD3wallcrossing} and the pure $SU(2)$ gauge theory from Example~\ref{ex:puresu2wallcrossing}, as well as for the $\N=2^*$ theory.

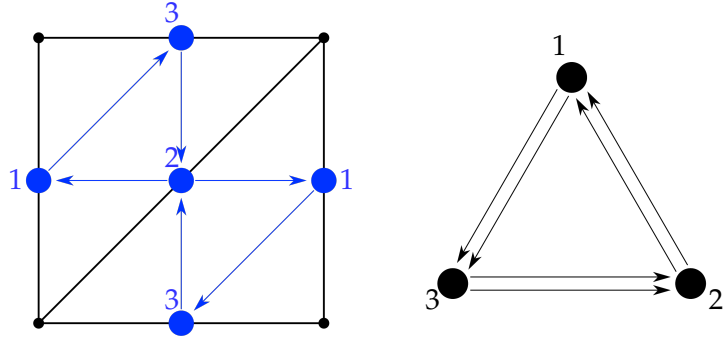
\begin{figure}[h!]
\small
\centering
\begin{overpic}
[width=0.60\textwidth]{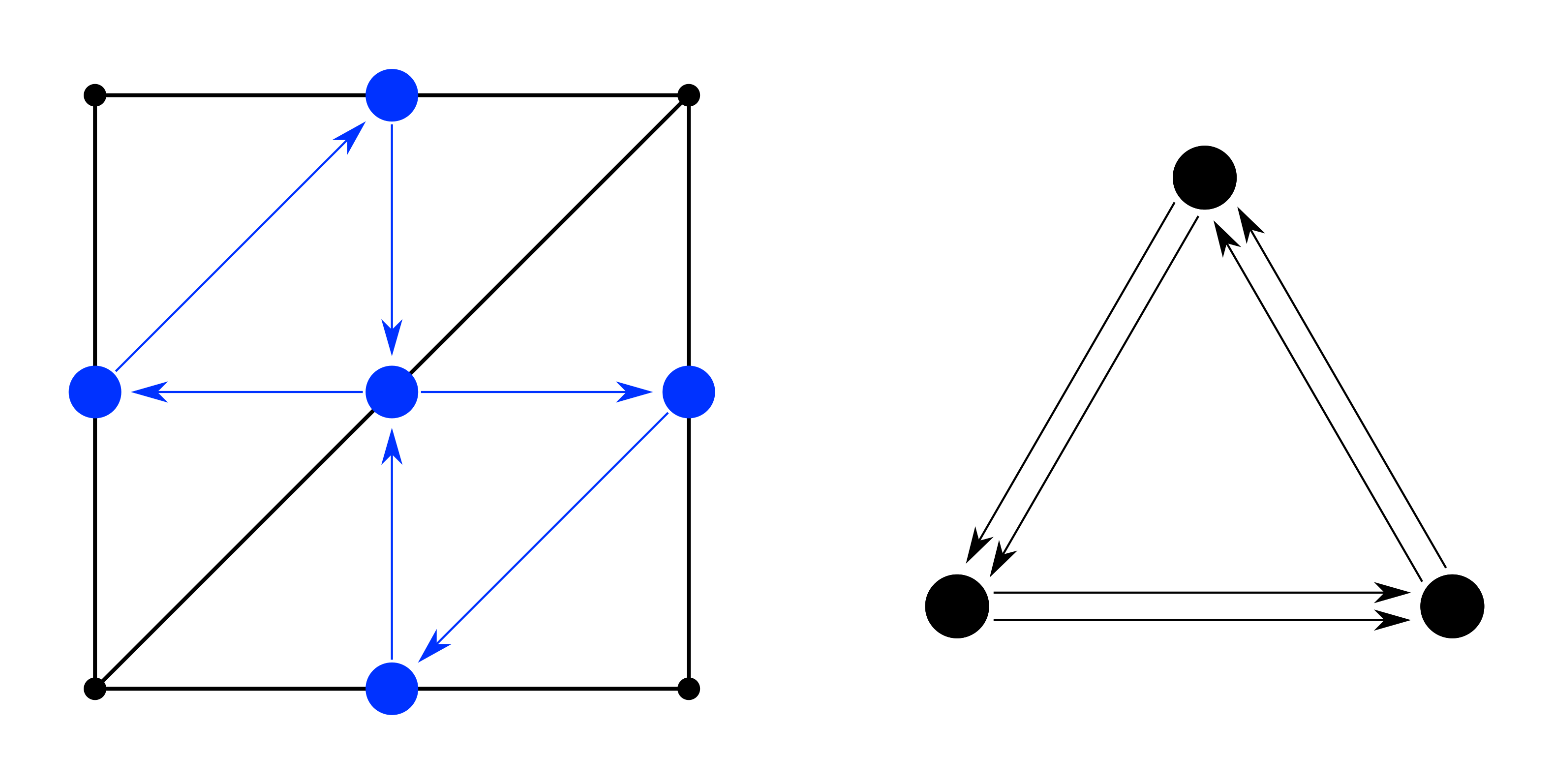}
\put(2,24){\textcolor{Blue}{$1$}}
\put(46.5,24){\textcolor{Blue}{$1$}}
\put(23,27){\textcolor{Blue}{$2$}}
\put(23,46.5){\textcolor{Blue}{$3$}}
\put(23,8){\textcolor{Blue}{$3$}}
\put(75,42){$1$}
\put(96,8){$2$}
\put(58,8){$3$}
\end{overpic}
\caption{\small Triangulation of the once-punctured torus $C$, where opposite edges are identified, and the resulting Markov quiver for the $\N=2^*$ gauge theory. The quiver is obtained by adding a node to the midpoint of each edge and clockwise oriented arrows on the faces of the triangulation.}
\label{triangulationquiver}
\normalsize
\end{figure}

So far we constructed the quiver from the spectrum, but we wanted to find the spectrum in the first place. Luckily there are methods to find the quiver without prior knowledge of the spectrum (cf.~\cite[Section~2.1.2]{alim2014mathcal} for a short overview), and we will therefore consider the quiver as given. For example, given an ideal triangulation $\scrT_\vartheta$ of the ultraviolet curve $C$ as discussed in Section~\ref{BPSgeometry}, we can construct a quiver by assigning a node $i$ to the midpoint of each edge $E_{\gamma_i}$, and arrows between nodes by drawing clockwise oriented arrows into each triangle and counting these (cf.~\cite{bridgeland2015quadratic}). This is illustrated for the $\N=2^*$ theory in Figure~\ref{triangulationquiver}.\footnote{Our basis of charges is slightly different from \cite{alim2013bps,alim2014mathcal}. To align with \cite{longhi2015structure} we choose $\gamma_i$ such that $\langle\gamma_i,\gamma_{i+1}\rangle=2$.} 

\subsection{$\Pi$-Stability and Mutations}
\label{subsec:mutations}
\noindent
With the BPS quiver at hand, we can map the problem of computing the BPS spectrum to a problem in representation theory: stable BPS states of the four-dimensional field theory correspond bijectively to $\Pi$-stable representations of the BPS quiver. Any BPS state \eqref{eq:gammabps} defines a $\C$-vector space of linear representations of the BPS quiver with dimension vector \smash{$\vec n=(n_i)\in\Z_{\geq0}^{2r+n_{\rm f}}$} given by
\begin{align}\label{eq:repspace}
\scrR(\vec n) = \bigoplus_{\bullet_i\rightarrow \bullet_j} \, {\rm Hom}\big(\C^{n_i},\C^{n_j}\big) \ ,
\end{align}
where the direct sum runs through all arrows of the quiver. The complex algebraic gauge group
\begin{align}
\scrG(\vec n) = \Big( \prod_{i=1}^{2r+n_{\rm f}} \, GL(n_i,\C) \Big) \, \Big/ \, \C^\times
\end{align}
acts naturally by basis changes of the vector spaces $\C^{n_i}$ and by conjugating elements of \eqref{eq:repspace} as bifundamental fields. The corresponding gauge orbits are precisely the isomorphism classes of representations of the state with charge $\gamma_\bps$, which are parameterized by the quotient variety $\scrM(\vec n)=\big[\scrR(\vec n)/\scrG(\vec n)\big]$ of complex dimension
\begin{align}\label{eq:expecteddim}
d(\vec n) = 1 + \sum_{\bullet_i\rightarrow \bullet_j} \, n_i\,n_j - \sum_{i=1}^{2r+n_{\rm f}} \, n_i^2 \ .
\end{align}

A homomorphism from a quiver representation $\gamma$ with dimension vector \smash{$\vec m\in\Z_{\geq0}^{2r+n_{\rm f}}$} to $\gamma_\bps$ is a set of linear maps $\sff_i:\C^{m_i}\to\C^{n_i}$ for each node $\bullet_i$ which commute with the linear maps representing the arrows $\bullet_i\longrightarrow\bullet_j$ of the BPS quiver, that is, \smash{$\sff_j\,{\sf a}^{\vec m}_{ji} = {\sf a}_{ji}^{\vec n}\,\sff_i$}. In this way the representations of the BPS quiver form a $\C$-linear abelian category. If the maps $\sff_i:\C^{m_i}\to\C^{n_i}$ are all injective, we say that $\gamma$ is a subrepresentation of the quiver representation $\gamma_\bps$; concretely this means that $m_i\leq n_i$ for each node $\bullet_i$ and the linear maps $\sfa_{ji}^{\vec m}:\C^{m_i}\to\C^{m_j}$ are the restrictions of the linear maps $\sfa^{\vec n}_{ji}:\C^{n_i}\to\C^{n_j}$.

The quiver representation $\gamma_\bps$ is called $\Pi$-stable if all proper subrepresentations $\gamma$ have central charges satisfying the strict inequality
\begin{align}
{\rm arg}\big(Z(\gamma;u)\big) < {\rm arg}\big(Z(\gamma_\bps;u)\big) \ .
\end{align}
This condition ensures that the BPS particle with electromagnetic charge $\gamma_\bps=\gamma+\cdots$ cannot decay into one of its constituent BPS states with charge $\gamma$. The $\Pi$-stable representations are the supersymmetric ground states of the low energy effective $\N=4$ quantum mechanics of a BPS particle based on the BPS quiver, whose interactions are governed by a $\scrG(\vec n)$-invariant holomorphic superpotential $\cW$ which is a formal sum of traces over oriented cycles of the quiver~\cite{Douglas:2000ah,alim2014mathcal}. The extrema of the superpotential give F-flatness conditions $\partial\cW=0$ which are realised as relations for the BPS quiver, that is, linear combinations of oriented paths obtained by composing arrows of the quiver.

The stability condition on quiver representations captures wall-crossing phenomena as well: as we vary over the $u$-plane we may encounter walls of marginal stability, where the arguments of $Z(\gamma;u)$ and $Z(\gamma_\bps;u)$ agree and a previously $\Pi$-stable representation decays. There can also be wall-crossings of the second kind, across the types of walls discussed in Section~\ref{BPSgeometry}, which occur when a BPS ray $Z(\gamma_i;u)$ rotates out of the specified half-plane $\mathbb{H}_\vartheta$. Then the description as a quiver representation breaks down, since there is no longer a positive integral basis. This is fixed by compensating the rotation with a quiver mutation. We will return to this point momentarily.

When the F-term equations are trivially satisfied (for example by setting the vacuum expectation values of some chiral fields to zero), the $\Pi$-stable representations with dimension vector $\vec n$ form an open subset of the moduli space $\scrM(\vec n)$ of the expected complex dimension $d(\vec n)$ given by \eqref{eq:expecteddim}; in general, this dimension is lowered by the number of non-redundant F-flatness constraints. The constraint $d(\vec n)\geq0$ on the dimension vector already imposes constraints on the allowed combinations \eqref{eq:gammabps}, and the resulting BPS state is a particle of spin $\frac12\,\big(d(\vec n)+1\big)$~\cite{alim2014mathcal}. For $SU(2)$ theories ($r=1$) only hypermultiplets (with $d(\vec n)=0$) and vector multiplets (with $d(\vec n)=1$) appear in the BPS spectrum. 

After finding the allowed dimension vectors $\vec n$, one may then determine which representations correspond to  one-particle bound states in the four-dimensional field theory. Their counterparts in quiver theory are Schur representations, that is, representations of the BPS quiver with a one-dimensional space of endomorphisms~\cite{Douglas:2000qw,Fiol:2000wx,Fiol:2006jz}. In particular, Schur representations are indecomposable, and $\Pi$-stable representations are always Schur representations, but the converses are not necessarily true. By focusing on Schur modules we avoid subtleties associated with ``threshold'' bound states, as then the set of integers $(n_i)$ is relatively prime; otherwise, if there is a common divisor $d>1$, then $Z(\gamma_\bps;u)$ aligns with the central charges of the direct sum of $d$ copies of a possibly destabilising proper subrepresentation with dimension vector \smash{$\frac1d\,\vec n$}. Finally, imposing $\Pi$-stability effectively resolves any fixed point singularities and also compactifies the moduli space~$\scrM(\vec n)$, giving a well-defined $L^2$-cohomology which quantizes the moduli space and computes the BPS spectrum.

\begin{example}\label{ex:nodes}
The simplest BPS states are always given by the basis charges $\gamma_i$ themselves, which are trivially $\Pi$-stable everywhere in the Coulomb moduli space $\scrB$ as they have no proper subrepresentations. In this case the dimension vector $\vec n=(0,\dots,0,1,0,\dots,0)$ has all zero entries except for the $i$-th slot where the entry is one,  corresponding to the Schur representation of the BPS quiver which has a single non-trivial one-dimensional vector space $\C$ at the node $\bullet_i$ and all arrow morphisms set to zero. The F-term equations are then always trivially satisfied, the expected dimension \eqref{eq:expecteddim} is equal to zero, and there are no moduli so that the nodes of any BPS quiver each correspond to a BPS hypermultiplet.
\end{example}

\begin{example}\label{ex:AD3quiver}
A BPS state of the AD$_3$ theory is specified by a pair of relatively prime non-negative integers $\vec n=(n_1,n_2)$ corresponding to a representation of the Dynkin quiver of type $A_2$ in Figure~\ref{quiverexamples}. The expected dimension \eqref{eq:expecteddim} of the moduli space $\scrM(n_1,n_2)$ is
\begin{align}
d(n_1,n_2) = 1 + n_2\,n_1 - \big(n_1^2+n_2^2\big) = 1-(n_1-n_2)^2 -n_1\,n_2 \ .
\end{align}
This is non-negative only when $(n_1,n_2)=(1,0)$, $(0,1)$ or $(1,1)$, for which the dimension is zero. All of these determine Schur representations corresponding to the hypermultiplets $\gamma_1$, $\gamma_2$ and $\gamma_1+\gamma_2$ in the BPS spectrum of the AD$_3$ theory (cf. Example~\ref{ex:AD3wallcrossing}). Thanks to the theorems of Gabriel and Kac (see e.g.~\cite[Appendix~A]{Douglas:2000qw} for a concise review), these are the only indecomposable representations of the Dynkin quiver of type $A_2$. A $\Pi$-stability analysis is considered in~\cite{alim2014mathcal}, reproducing the chamber structure of~Example~\ref{ex:AD3wallcrossing}.
\end{example}

\begin{example}\label{ex:Fterm}
For the $\N=2^*$ gauge theory, summing over traces of oriented cycles of the Markov quiver constructed in Figure~\ref{triangulationquiver} gives the superpotential~\cite{alim2013bps,alim2014mathcal}
\begin{align}
\cW = \Tr\,(\sfa_{12}\,\sfa_{23}\,\sfa_{31} + \sfb_{12}\,\sfb_{23}\,\sfb_{31} + \sfa_{12}\,\sfb_{23}\,\sfa_{31}\,\sfb_{12}\,\sfa_{23}\,\sfb_{31}) \ ,
\end{align}
where $(\sfa_{i,i+1},\sfb_{i,i+1}):\bullet_{i+1}\rightrightarrows\bullet_i$ are the pairs of arrows between neighbouring nodes of the Markov quiver. The F-term constraints $\partial\cW=0$ yield six relations which are realised as paths from nodes $\bullet_{i+2}$ to nodes $\bullet_i$ given by
\begin{align}\label{eq:Fterm2*}
\sfa_{i,i+1}\,\sfa_{i+1,i+2} + \sfb_{i,i+1}\,\sfa_{i+1,i+2}\,\sfb_{i+2,i}\,\sfa_{i,i+1}\,\sfb_{i+1,i+2} &= 0 \ , \nonumber\\[4pt]
\sfb_{i,i+1}\,\sfb_{i+1,i+2} + \sfa_{i,i+1}\,\sfb_{i+1,i+2}\,\sfa_{i+2,i}\,\sfb_{i,i+1}\,\sfa_{i+1,i+2} &= 0 \ ,
\end{align}
for $i=1,2,3$ (read cyclically modulo~$3$).  A representation of the Markov quiver is specified by a triple of non-negative integers $\vec n=(n_1,n_2,n_3)$, which we assume are relatively prime.

Let us consider the subclass of representations with $n_i=0$ for a single $i\in\{1,2,3\}$. Then any linear map representing an arrow to or from the node $\bullet_i$ is necessarily zero, and so the F-flatness constraints \eqref{eq:Fterm2*} are trivially satisfied. The expected dimension \eqref{eq:expecteddim} of the moduli space $\scrM(n_{i-1},n_{i+1},0)$ is
\begin{align}
d(n_{i-1},n_{i+1},0) = 1 + 2\,n_{i-1}\,n_{i+1} - \big(n_{i-1}^2+n_{i+1}^2\big) = 1-(n_{i-1}-n_{i+1})^2 \ .
\end{align}
This is non-negative only when either $n_{i-1}=n_{i+1}=1$, where the dimension is one, or when $n_{i+1}=n_{i-1}\pm1$, where the dimension is zero. Both solutions determine Schur representations and, since the resulting quiver is an extended Dynkin quiver of type \smash{$\widehat{A}_1$}, as in Example~\ref{ex:AD3quiver} they exhaust the indecomposable representations with $n_i=0$~\cite{Fiol:2006jz}. The former case corresponds to the charge $\gamma_{i-1}+\gamma_{i+1}$, while the latter case determines two infinite families of charges $(k+1)\,\gamma_{i+1}+k\,\gamma_{i-1}$ and $k\,\gamma_{i+1}+(k+1)\,\gamma_{i-1}$ with $k\geq0$. These are of course the familiar vector multiplet and towers of hypermultiplets in the BPS spectrum of the pure $\N=2$ gauge theory (cf. Example~\ref{ex:puresu2walls}), which we also expect to see in the $\N=2^*$ spectrum: considering bound states of only two of the three hypermultiplets $\gamma_{i-1}$ and $\gamma_{i+1}$ reduces the Markov quiver quantum mechanics to the Kronecker subquiver in Figure~\ref{quiverexamples}, and heuristically corresponds to decoupling the adjoint hypermultiplet in the asymptotically free limit $m\to\infty$.\footnote{This limit can be understood as follows: Fix the central charges of two states $\gamma_{i\pm 1}$ and  allow $Z(\gamma_i)$ to vary as we continuously vary the mass $m$. Then taking the infinite mass limit also sends $|Z(\gamma_i)|=|m-Z(\gamma_{i-1})-Z(\gamma_{i+1})|$ to infinity. The state $\gamma_i$ therefore decouples from the theory. From the perspective of the BPS quiver the node corresponding to $\gamma_i$ is removed and the two nodes that are left form the BPS quiver of the pure $SU(2)$ theory.} Later on we will map out the regions in the Coulomb branch $\scrB_m$ of the $\N=2^*$ theory where these particles are stable BPS states.

However, in contrast to the other two quivers shown in Figure~\ref{quiverexamples}, the Markov quiver is an example of a quiver of ``wild'' representation type~\cite{assem2006wild}: the indecomposable modules occur in families of arbitrary numbers of parameters, and no description of all the isomorphism classes of indecomposable representations is possible, as their classification would contain the classification of indecomposable modules over all finite-dimensional algebras.\footnote{This is in fact true for any connected quiver unless its underlying graph is a simply laced Dynkin diagram of unextended or extended type, in which case there are respectively finitely or infinitely many isomorphism classes of indecomposable representations (see e.g.~\cite{keller2008} for a review). The basic examples are provided respectively by the BPS quivers for the AD$_3$ theory and the pure $SU(2)$ gauge theory.} This complexity makes the determination of a complete list of all BPS states in the spectrum of the $\N=2^*$ theory an extremely arduous task.
\end{example}

In general, $\Pi$-stability of quiver representations is difficult to check even for relatively simple quivers. A simpler and more algorithmic way to compute the BPS spectrum is via the {mutation method}. This consists of elementary changes of quivers that occur when we vary the choice of half-plane $\mathbb{H}_{\vartheta}$. By convention it is rotated clockwise, and therefore at some point the leftmost charge $\gamma$ leaves the half-plane and is replaced by its antiparticle $-\gamma$ which becomes the new rightmost charge. The leftmost and rightmost charges are always basis elements of the positive integral basis for the respective half-planes, but with this replacement the basis is no longer a positive integral basis. We therefore have to perform a basis change, which is called a {mutation on} $\gamma$. A mutation generally produces a different but physically equivalent BPS quiver, which acts as a one-dimensional Seiberg duality on the corresponding $\N=4$ quiver quantum mechanics and preserves the BPS spectrum.

Let $\gamma_i$ be the BPS particle that rotates out of the plane, and let $\gamma_j$ be any other charge with $j\neq i$. Then the mutation is given by 
\begin{align}\label{fliprule}
\gamma_i \longmapsto \gamma_i' = -\gamma_i \qquad \mbox{and} \qquad
\gamma_j \longmapsto \gamma_j' = \gamma_j + \langle\gamma_j,\gamma_i\rangle_+\,\gamma_i \ ,
\end{align}
where $\langle\cdot,\cdot\rangle_+$ is again the positive of part of $\langle\cdot,\cdot\rangle$ as in (\ref{fliprules}). 
We can continue doing this until we have rotated the half-plane by $\pi$. At this stage every BPS particle has been leftmost and has therefore been mutated on. We have thus found the whole spectrum. It is worth noting that the quiver after a rotation by $\pi$ is the same quiver as the original one, but now labelled by the antiparticles $\{-\gamma_i\}$. It can often be helpful to start from that quiver and perform inverse mutations, which use the rule 
\begin{align}\label{unfliprule}
\gamma_i \longmapsto \gamma_i' = -\gamma_i \qquad \mbox{and} \qquad
\gamma_j \longmapsto \gamma_j' = \gamma_j + \langle\gamma_i,\gamma_j\rangle_+\,\gamma_i \ .
\end{align}

In Section~\ref{spectrumonwall} below we will use these methods to construct the spectrum of the $\N{=}2^*$ theory on the walls given in Section~\ref{longhispectrum}. In Section~\ref{subsec:triangBPS} we will then have a look at the ideal triangulations  involved on the ultraviolet curve $C$ to gain an understanding of how the states correspond to cycles on the Seiberg-Witten curve $\Sigma_{u;m}$.

\subsection{Spectrum on the Walls}\label{spectrumonwall}
\begin{figure}[h!]
\small
\centering
\begin{overpic}
[width=0.50\textwidth]{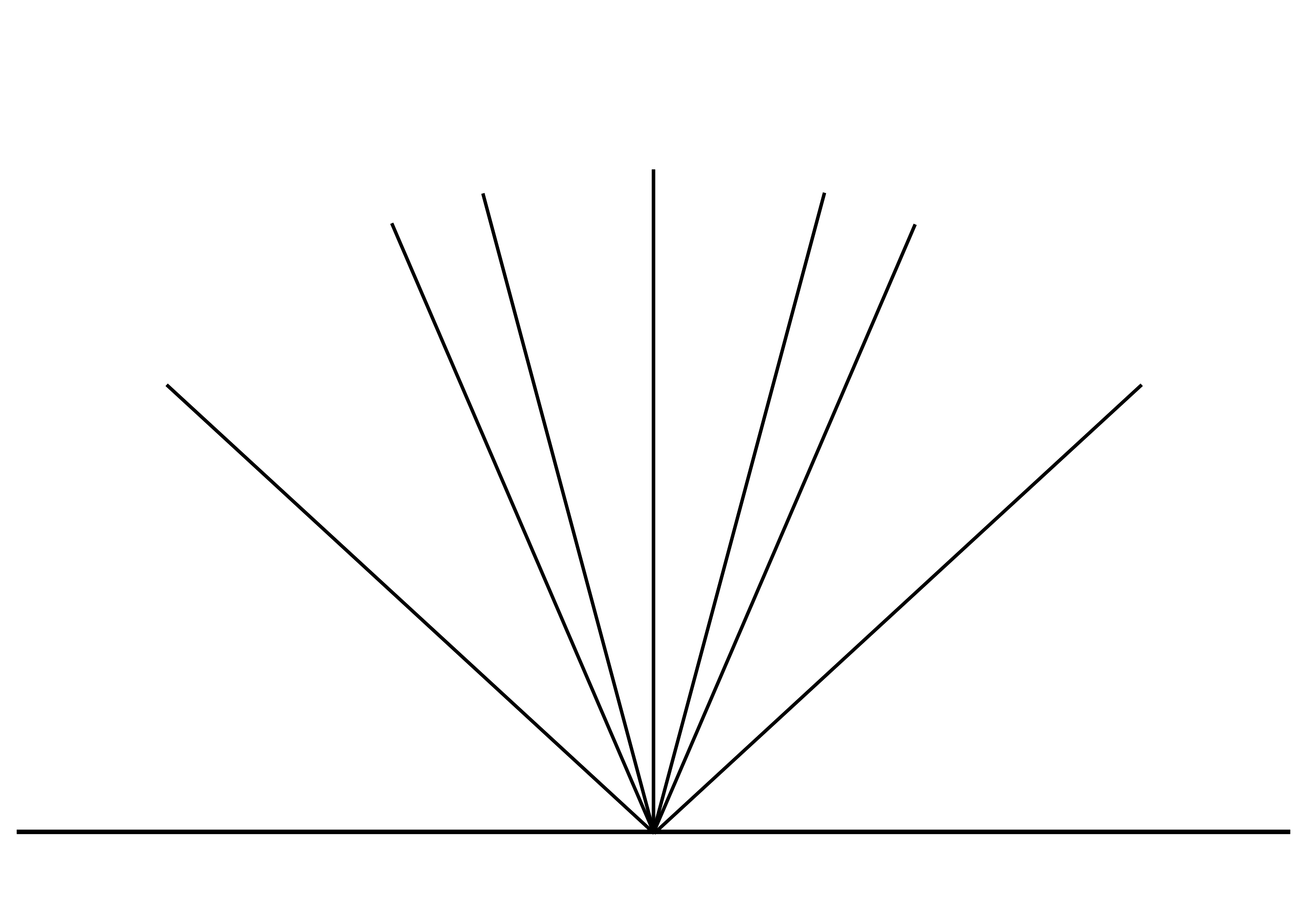}
\put(87,43){$\gamma_1$}
\put(70,53){$2\gamma_1+\gamma_2$}
\put(60,58){$3\gamma_1+2\gamma_2$}
\put(43,60){$\gamma_1+\gamma_2$}
\put(38,65){$\gamma_{\rm f}\pm(\gamma_1+\gamma_2),$}
\put(21,58){$2\gamma_1+3\gamma_2$}
\put(12,53){$\gamma_1+2\gamma_2$}
\put(8,43){$\gamma_2$}
\put(44,35){$\dots$}
\put(51,35){$\dots$}
\end{overpic}
\caption{\small Charge rays $Z(\gamma;u)$ of the BPS spectrum in the upper half-plane for $u\in\scrE_3$. We include the first two members of the infinite families $\gamma_1+k\,(\gamma_1+\gamma_2)$ and $\gamma_2+k\,(\gamma_1+\gamma_2)$.}
\label{gammarays}
\normalsize
\end{figure}
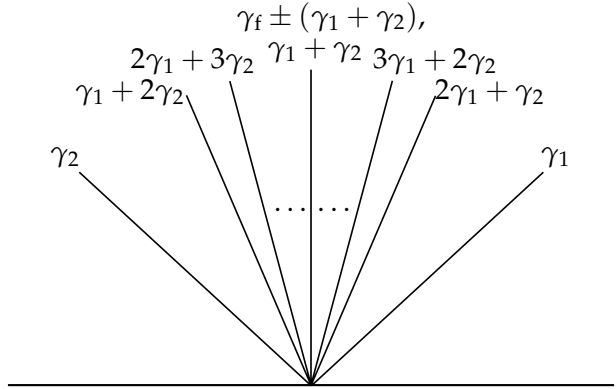

\nid
Following \cite{longhi2015structure} and the discussion in Section~\ref{longhispectrum}, we consider the $\N=2^*$ theory at the wall of marginal stability $\scrE_3$ in the Coulomb branch $\scrB_m$ defined by
\begin{equation}
\scrE_3 = \Big\{u\in\scrB_m \ \Big| \ \frac{Z(\gamma_3;u)}{Z(\gamma_{\rm f};u)}\in\mathbb{R}_{>0} \ \ , \ \ \arg\big( Z(\gamma_1;u)\big)<\arg \big(Z(\gamma_2;u)\big)\Big\} \ ,
\end{equation}
where the spectrum consists of the charges shown in Figure~\ref{gammarays}.
We will construct this spectrum using mutations of BPS quivers, where our starting point is the Markov quiver for the $\N=2^*$ theory constructed in Figure~\ref{triangulationquiver}. It consists of three states $\gamma_i$ with pairing $\langle\gamma_i,\gamma_{i+1}\rangle=2$. We note that the Markov quiver is invariant under a cyclic relabelling $\gamma_i\mapsto\gamma_{i\pm1}$, which generates the action of a $\Z_3$ automorphism of the category of quiver representations on the BPS spectrum. Therefore the analysis on the wall $\scrE_3$ easily translates to the other walls $\scrE_1$ and $\scrE_2$ in \eqref{longhiwall}.

We start the mutation method by mutating on the state $\gamma$ with largest $\arg(Z(\gamma))$, which is $\gamma_2$. Applying the mutation rule (\ref{fliprule}) we get
\begin{align}
\gamma_1\longmapsto\gamma_1' = \gamma_1+2\gamma_2 \ , \quad
\gamma_2\longmapsto\gamma_2' = -\gamma_2 \qquad \mbox{and} \qquad
\gamma_3\longmapsto\gamma_3' = \gamma_3 \ .
\end{align}
The charge $\gamma$ with largest $\arg( Z(\gamma))$ is now $\gamma_1+2\gamma_2$, as expected. We can define a quiver consisting of states \smash{$\gamma_i^{\textrm{\tiny$(k)$}}$}, which is the Markov quiver after $k$ mutations with a renaming to exchange \smash{$\gamma_1^{\textrm{\tiny$(k)$}}$} and \smash{$\gamma_2^{\textrm{\tiny$(k)$}}$} after each mutation in order to preserve the pairing $\langle\gamma^{\textrm{\tiny$(k)$}}_i,\gamma^{\textrm{\tiny$(k)$}}_{i+1}\rangle=2$. The resulting quiver is then labelled by
\begin{align}\label{kflipquiver}
\gamma_1^{\textrm{\tiny$(k)$}} := \gamma_1 - k\,(\gamma_1+\gamma_2) \ , \quad 
\gamma_2^{\textrm{\tiny$(k)$}} := \gamma_2 + k\,(\gamma_1+\gamma_2) \qquad \mbox{and} \qquad
\gamma_3^{\textrm{\tiny$(k)$}} := \gamma_3 \ .
\end{align}
With the renaming of charges after each mutation, $\gamma_2^{\textrm{\tiny$(k)$}}$ always has largest $\arg( Z(\gamma))$ for a given $k$, and since $\langle\gamma_i^{\textrm{\tiny$(k)$}},\gamma_{i+1}^{\textrm{\tiny$(k)$}}\rangle=2$, mutating it gives
\begin{align}
\gamma_1^{\textrm{\tiny$(k)$}}&\longmapsto\gamma_1 - k\,(\gamma_1+\gamma_2)+2\,\big(\gamma_2 + k\,(\gamma_1+\gamma_2)\big)=\gamma_2 + (k+1)\,(\gamma_1+\gamma_2) \ , \nonumber\\[4pt]
\gamma_2^{\textrm{\tiny$(k)$}}&\longmapsto-\gamma_2 - k\,(\gamma_1+\gamma_2)=\gamma_1 - (k+1)\,(\gamma_1+\gamma_2) \ , \nonumber\\[4pt]
\gamma_3^{\textrm{\tiny$(k)$}}&\longmapsto\gamma_3 \ .
\end{align}
Hence $k$ is increased by one and $\gamma_1$ is exchanged with $\gamma_2$; in particular, $\gamma_1^{\textrm{\tiny$(k+1)$}} = -\gamma_2^{\textrm{\tiny$(k)$}}$. The family (\ref{kflipquiver}) contains states of the form $\gamma_2+k\,(\gamma_1+\gamma_2)$, as well as the state $\gamma_3$ which is not mutated on. These assemble for all $k\geq0$ into the same BPS quiver, reflecting the well-known fact that the mutation equivalence class of the Markov quiver consists of a single element.

To understand the family $\gamma_1+k\,(\gamma_1+\gamma_2)$ it is helpful to start from the other direction. The starting point is then the initial quiver but labelled by the antiparticles $\tgamma_i=-\gamma_i$. We can then mutate again on states, but this time using the inverse mutation rule (\ref{unfliprule}).
The calculation then carries through as  before and we can define a similar family of charges $\tgamma_i^{\textrm{\tiny$(k)$}}$ with $\langle\tgamma^{\textrm{\tiny$(k)$}}_i,\tgamma^{\textrm{\tiny$(k)$}}_{i+1}\rangle=2$ by
\begin{align}\label{kunflipquiver}
\tgamma_1^{\textrm{\tiny$(k)$}} &:= -\big(\gamma_1 + k\,(\gamma_1+\gamma_2)\big) \ , \nonumber \\[4pt]
\tgamma_2^{\textrm{\tiny$(k)$}} &:= -\big(\gamma_2 - k\,(\gamma_1+\gamma_2)\big) \ , \nonumber \\[4pt]
\tgamma_3^{\textrm{\tiny$(k)$}} &:= -\gamma_3 \ ,
\end{align}
where $\tgamma_3^{\textrm{\tiny$(k)$}}=\tgamma_3$ does not change under the mutations. In this way we find the states $\gamma_1+k\,(\gamma_1+\gamma_2)$ and mutate on them.

There is one missing piece to the spectrum shown in Figure~\ref{gammarays}. From either direction, the quark given by $\gamma_3$ appeared as a node of the quiver, but we never actually mutated on it. This is because it aligns with the vector multiplet $\gamma_1+\gamma_2$ which is the result of the juggle that relates the $k\to\infty$ limits of (\ref{kflipquiver}) and (\ref{kunflipquiver}). Whereas the particle $\gamma_3$ is guaranteed to be a stable BPS hypermultiplet as it is always a node of the Markov quiver (cf. Example~\ref{ex:nodes}), the stability of $\gamma_1+\gamma_2$ does not follow from the present analysis as it would formally require infinitely many mutations. We must therefore resort directly to representation theory on the accumulation ray. 

The quiver representation $\gamma_1+\gamma_2$ has dimension vector $(1,1,0)$ with two non-trivial linear maps $\C\to\C$ representing the arrows $\bullet_2\rightrightarrows\bullet_1$, which can be characterized by a pair of complex numbers $\vec{\sf a}\in\C^2$; the F-flatness conditions \eqref{eq:Fterm2*} are trivially satisfied (cf.~Example~\ref{ex:Fterm}). For generic \smash{$\vec{\sf a}$}, an easy analysis of commutative diagrams shows that the only proper subrepresentation is $\gamma_1$. On the wall $\scrE_3$, the phase of $Z(\gamma_1)$ is smaller than the phase of $Z(\gamma_2)$, and so $\arg(Z(\gamma_1))<\arg(Z(\gamma_1+\gamma_2))<\arg(Z(\gamma_2))$. Hence $\gamma_1+\gamma_2$ is $\Pi$-stable for generic $\vec{\sf a}$. On the other hand, $\gamma_2$ is a destabilizing proper subrepresentation when \smash{$\vec{\sf a}=\vec0$}, and so the moduli space of $\Pi$-stable representations of $\gamma_1+\gamma_2$ is \smash{$\big(\C^2\setminus\{\vec0\}\big)\big/\,\C^\times \cong \PP^1$}, the spin~$1$ moduli space of a vector multiplet~\cite{alim2014mathcal}.

Finally, while the mutation method accounts for the totality of BPS states on either side of the accumulation ray in the central charge plane containing $\gamma_3$ and $\gamma_1+\gamma_2$, we still need to account for the BPS state $\gamma_1+\gamma_2+\gamma_{\rm f} = 2(\gamma_1+\gamma_2)+\gamma_3$ along the ray in Figure~\ref{gammarays}. The stability analysis is a bit lengthy in this case. Since $\Pi$-stable representations are Schur modules, we begin by considering the moduli of representations of the Markov quiver with dimension vector $(2,2,1)$ that correspond to Schur modules. Such representations are parameterized by row vectors $\sfa_{31},\sfb_{31}:\C^2\to\C$, column vectors $\sfa_{23},\sfb_{23}:\C\to\C^2$, and square matrices $\sfa_{12},\sfb_{12}:\C^2\to\C^2$. 

An endomorphism of a generic representation of this form is given by a complex number $\lambda\in\C^\times$ acting at the node $\bullet_3$, and a pair of matrices $\sfS_1,\sfS_2\in GL(2,\C)$ acting at the nodes $\bullet_1$ and $\bullet_2$ respectively. Commutativity of all relevant diagrams results in six equations
\begin{align}\label{eq:Markovendo}
\sfa_{31}\,\sfS_1 &= \lambda\,\sfa_{31} \ , \quad \sfS_2\,\sfa_{23} = \lambda\,\sfa_{23}  \qquad \mbox{and} \qquad \sfS_1\,\sfa_{12} = \sfa_{12}\,\sfS_2 \ , \nonumber \\[4pt]
\sfb_{31}\,\sfS_1 &= \lambda\,\sfb_{31} \ , \quad \sfS_2\,\sfb_{23} = \lambda\,\sfb_{23}  \qquad \mbox{and} \qquad \sfS_1\,\sfb_{12} = \sfb_{12}\,\sfS_2 \ .
\end{align}
The condition for $\gamma_1+\gamma_2+\gamma_{\rm f}$ to be a Schur representation is that the equations \eqref{eq:Markovendo} have the unique solution $\sfS_1=\sfS_2=\lambda\,{\mathbbm 1}_2$. This rules out representations with $\sfa_{i,i+1}=\sfb_{i,i+1}=0$ for more than one $i\in\{1,2,3\}$. 

Let us next look at what constraints are imposed by $\Pi$-stability of the resulting Schur representations. For this, note that the only potential destabilizing subrepresentations of $\gamma_1+\gamma_2+\gamma_{\rm f}$ on the wall $\scrE_3$ are the constituent BPS states whose central charges lie either on or to the left of the accumulation ray shown in Figure~\ref{gammarays}. These are the states with electromagnetic charges $\gamma_2$, $\gamma_3$, $\gamma_1+\gamma_2$ and $\gamma_1+2\gamma_2$. An easy analysis of commutative diagrams shows that $\gamma_2$ can only be a subrepresentation if $\sfa_{12}=\sfb_{12}=0$, while $\gamma_3$ can only be a subrepresentation if $\sfa_{23}=\sfb_{23}=0$. These subrepresentations can thus be discarded if we require at least one of the arrows of $\bullet_2\rightrightarrows\bullet_1$ and of $\bullet_3\rightrightarrows\bullet_2$ to be non-zero maps.

A generic subrepresentation with dimension vector $(1,1,0)$ can be expressed through a commutative diagram
\begin{equation}\label{eqn:defretpicture1}
\begin{tikzcd}
\C^2 \ar[rr,shift right=-0.5ex,"\sfa_{31}"] \ar[rr,shift right=0.5ex,swap,"\sfb_{31}"]  & & \C \ar[rr,shift right=-0.5ex,"\sfa_{23}"] \ar[rr,shift right=0.5ex,swap,"\sfb_{23}"] & & \C^2 \ar[llll,bend left=-40,shift left=0.5ex,"\sfb_{12}"] \ar[llll,bend left=-40,shift left=-0.5ex,swap,"\sfa_{12}"]  \\ & & & & \\
\C \ar[uu,"\sff_1"] \ar[rr,shift right=-0.5ex,"0"] \ar[rr,shift right=0.5ex,swap,"0"] & & 0 \ar[uu,"0"] \ar[rr,shift right=-0.5ex,"0"] \ar[rr,shift right=0.5ex,swap,"0"] & & \C \ar[uu,"\sff_2"] \ar[llll,bend right=-40,shift left=0.5ex,"\alpha"] \ar[llll,bend right=-40,shift left=-0.5ex,swap,"\beta"] 
\end{tikzcd}
\end{equation}
In this diagram the top row is the original representation $\gamma_1+\gamma_2+\gamma_{\rm f}$, while the bottom row is the subrepresentation $\gamma_1+\gamma_2$ which is characterized by a pair of complex numbers $(\alpha,\beta):\bullet_2\rightrightarrows\bullet_1$ and linear embeddings $\sff_1,\sff_2:\C\hookrightarrow\C^2$, with $0$ denoting the zero map. As we saw above, $\Pi$-stability of the vector multiplet $\gamma_1+\gamma_2$ implies that at least one of $\alpha$ or $\beta$ must be non-zero. It is easy to see that the rightmost inner square always commutes, for any choice of embedding $\sff_2$ and column vectors $\sfa_{23},\sfb_{23}$. On the other hand, commutativity of the leftmost inner square requires ${\rm im}(\sff_1) = \C$ to lie in the kernels of both $\sfa_{31}$ and $\sfb_{31}$, whereas commutativity of the outer square necessitates that ${\rm im}(\sff_2)=\C$ does {not} land in both kernels of $\sfa_{12}$ and $\sfb_{12}$. Hence all such subrepresentations can be discarded if either $\ker(\sfa_{31})\cap\ker(\sfb_{31})=0$ or $\sfa_{12}=\sfb_{12}=0$; however, the latter case was already ruled out above. A similar analysis of the subrepresentations with dimension vector $(1,2,0)$ shows that the condition $\ker(\sfa_{31})\cap\ker(\sfb_{31})=0$ also eliminates the BPS hypermultiplet $\gamma_1+2\gamma_2$ from the possibilities.

Finally, we consider those $\Pi$-stable representations which solve the F-flatness conditions \eqref{eq:Fterm2*}.
The condition $\ker(\sfa_{31})\cap\ker(\sfb_{31})=0$ implies that either $\sfa_{31}=\sfb_{31}=0$, or else the kernels of $\sfa_{31}$ and $\sfb_{31}$ are both one-dimensional subspaces of $\C^2$ which are linearly independent. In the latter case the maps can be brought to the form $
\sfa_{31} = \begin{pmatrix} 1 & 0 \end{pmatrix}$ and $ \sfb_{31} = \begin{pmatrix}
0 & 1 \end{pmatrix}$
by a suitable complex gauge transformation in $\scrG(2,2,1)$. Substituting this into \eqref{eq:Fterm2*} with $i=2$ then gives $\sfa_{23}=\sfb_{23}=0$, which we already ruled out above. Hence 
\begin{align}
\sfa_{31}=\sfb_{31}=0 \ .
\end{align}

The only non-trivial F-term equations \eqref{eq:Fterm2*} that remain are those with $i=1$, which now read as
\begin{align}\label{eq:Ftermleft}
\sfa_{12}\,\sfa_{23} = 0 \qquad \mbox{and} \qquad \sfb_{12}\,\sfb_{23} = 0 \ .
\end{align}
For a generic Schur representation the two rightmost inner maps in the top row of \eqref{eqn:defretpicture1} have linearly independent one-dimensional images in $\C^2$, so again a suitable $\scrG(2,2,1)$-transformation brings them to the form
\begin{align}
\sfa_{23} = \begin{pmatrix}
1 \\ 0
\end{pmatrix} \qquad \mbox{and} \qquad \sfb_{23} = \begin{pmatrix}
0 \\ 1
\end{pmatrix} \ .
\end{align}
From \eqref{eq:Ftermleft} it then follows that the outer maps in the top row of \eqref{eqn:defretpicture1} generically have one-dimensional linearly independent images in $\C^2$, and using the remaining complex gauge symmetry they may be brought into the form
\begin{align}
\sfa_{12} = \begin{pmatrix}
0 & 1 \\ 0 & 0
\end{pmatrix} \qquad \mbox{and} \qquad \sfb_{12} = \begin{pmatrix}
0 & 0 \\ 0 & 1
\end{pmatrix} \ .
\end{align}
Substituting into \eqref{eq:Markovendo} then fixes $\sfS_1=\sfS_2=\lambda\,{\mathbbm 1}_2$, as required. There are no moduli remaining in the generic gauge orbit, so this completes the demonstration that the state $\gamma_1+\gamma_2+\gamma_{\rm f}$ is a $\Pi$-stable hypermultiplet along the accumulation ray for the wall $\scrE_3$.

\subsection{Geometry of BPS States}\label{subsec:triangBPS}
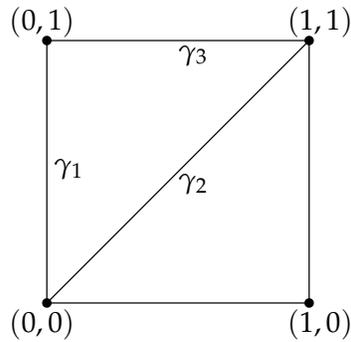
\begin{figure}[h!]
\small
\centering
\begin{overpic}
[width=0.40\textwidth]{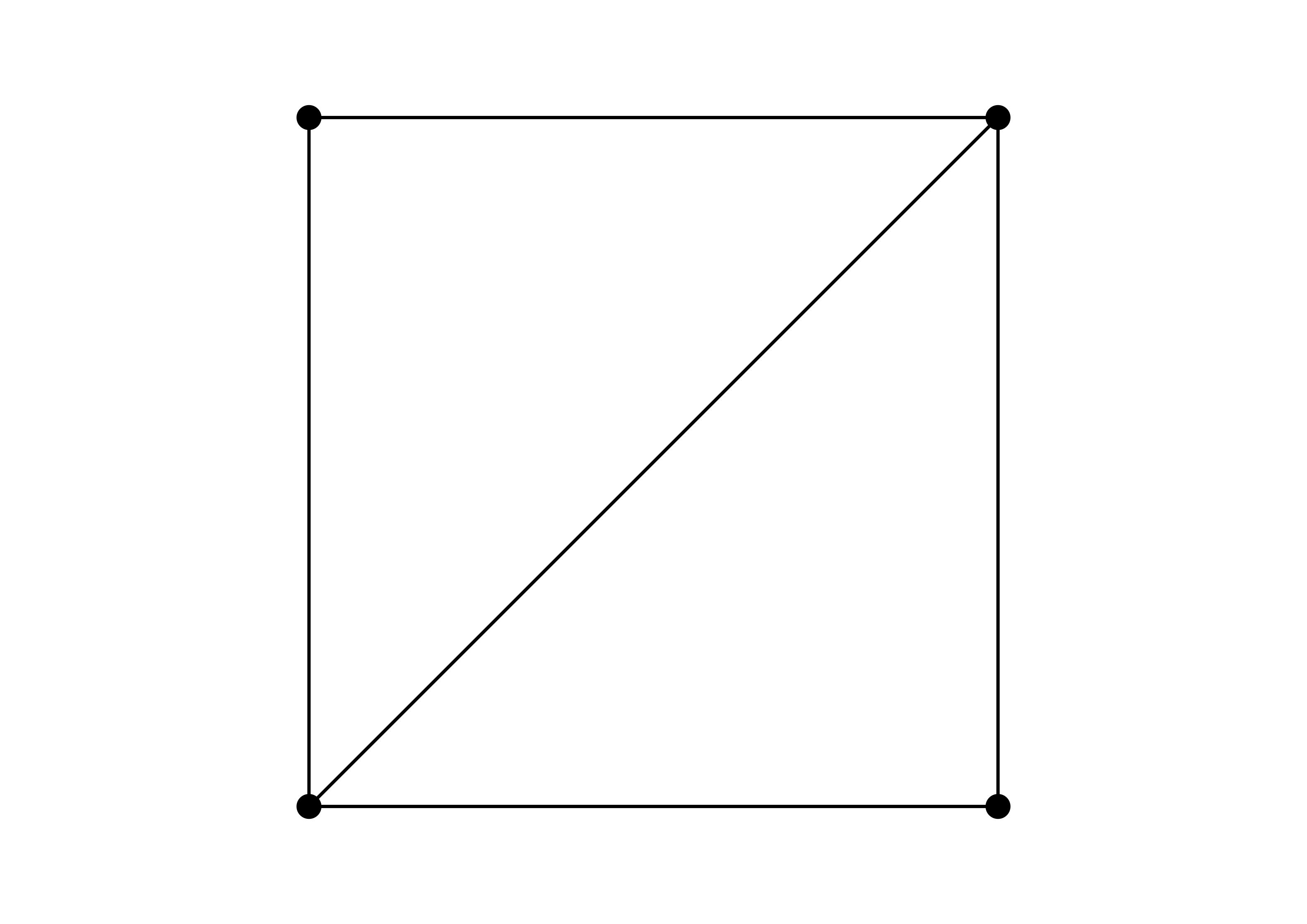}
\put(16,3){$(0,0)$}
\put(72,3){$(1,0)$}
\put(16,64){$(0,1)$}
\put(72,64){$(1,1)$}
\put(50,32){$\gamma_2$}
\put(50,58){$\gamma_3$}
\put(25,35){$\gamma_1$}
\end{overpic}
\caption{\small Triangulation of the once-punctured torus $C$ for $\tau_0=\I$. Edges are labelled by homology elements $\gamma_i$ of the Seiberg-Witten curve $\Sigma_{u;m}$.}
\label{T0}
\normalsize
\end{figure}

\nid
The rules \eqref{fliprule} for quiver mutations are identical to the flip rules \eqref{fliprules} dictating the transformations which the Fock-Goncharov coordinates \smash{$\cX_\gamma^{u,\vartheta}$} undergo at critical phases $\vartheta_{\rm c}$ that correspond to BPS states.
In fact, we can explicitly relate the perspective of BPS quivers to the perspective of ideal triangulations $\scrT_\vartheta$ on $C$ from Section~\ref{BPSgeometry}, which is part of the general correspondence between mutations and flips~\cite{bridgeland2015quadratic}. For the $\N=2^*$ theory, the ultraviolet curve $C$ is the once-punctured torus. We will always draw it as a quadrilateral with the puncture in the corners. We label edges of the triangulation by their corresponding state $\gamma_i$ as shown in Figure~\ref{T0}, with the triangulation considered on the covering space~$\C\setminus(\Z+\tau_0\,\Z) \to C$.

One crucial tool for studying these triangulations systematically is the action of the mapping class group of the once-punctured torus $C$, which is identified as the group $SL(2,\Z)$ implementing the action of S-duality in the $\N=2^*$ gauge theory. Here we will focus on its action on the symplectic generators $(A,B)$ of the first homology of $C$, under the canonical identification $H_1(C,\Z)\cong\Z\oplus\Z$, and use the standard $SL(2,\Z)$ generators
\begin{align}
\sfT:\ (A,B)\longmapsto (A+B, B) \qquad \mbox{and} \qquad
\sfS:\ (A,B)\longmapsto (-B,A) \ .
\end{align}
We also define $\Tdual=\sfT\,\sfS\,\sfT$ which maps $(A,B$) to $(A,A+B)$. These transformations can be used to cyclically permute the edge labels $\gamma_i\mapsto \gamma_{i+1}$ by acting with $\sfS$. We are therefore content again to perform the analysis solely for the wall $\scrE_3$.

\begin{figure}[h!]
\small
\centering
\begin{overpic}
[width=0.80\textwidth]{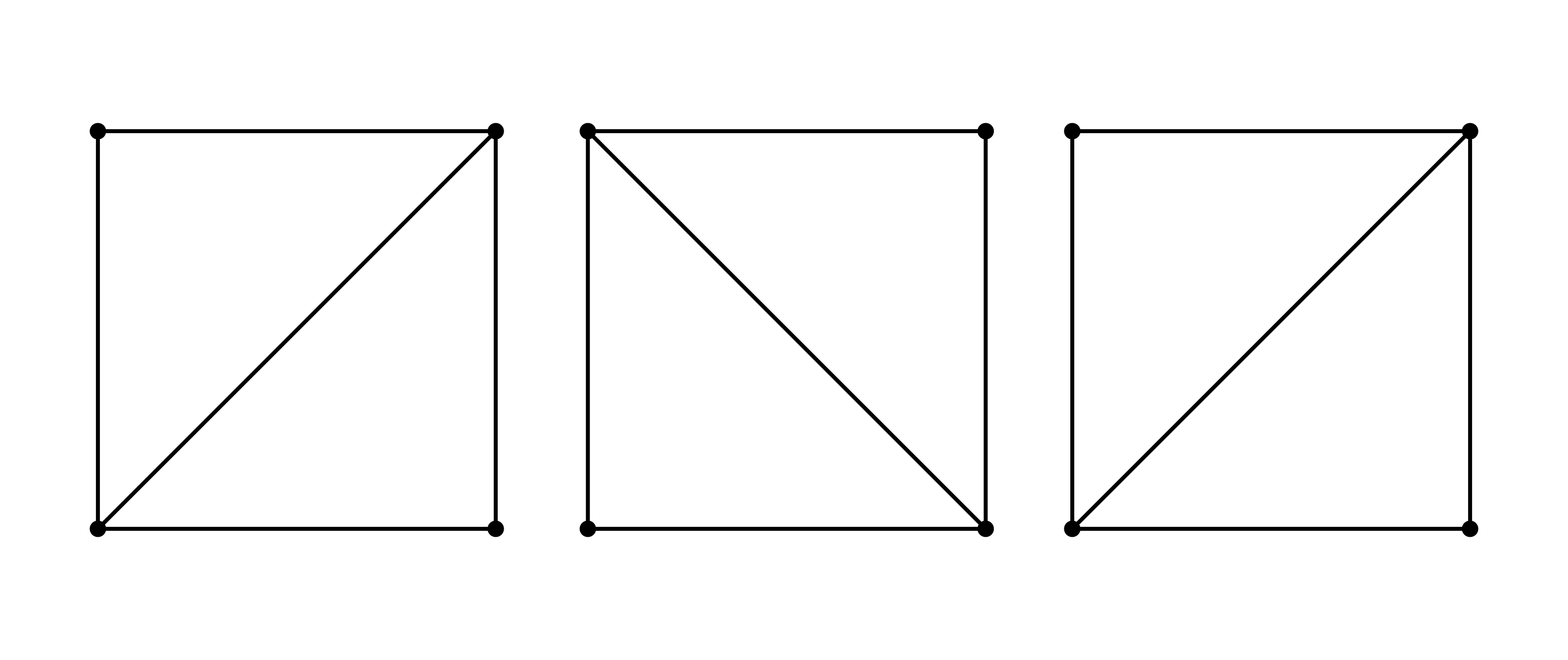}
\put(2,5){$(0,0)$}
\put(26,5){$(1,0)$}
\put(2,35){$(0,1)$}
\put(26,35){$(1,1)$}
\put(35,5){$(0,0)$}
\put(58,5){$(1,0)$}
\put(35,35){$(0,1)$}
\put(58,35){$(1,1)$}
\put(67,5){$(0,0)$}
\put(92,5){$(1,0)$}
\put(67,35){$(-1,1)$}
\put(92,35){$(0,1)$}
\put(33,20){$\mapsto$}
\put(64,20){$\simeq$}
\put(20,19){$\gamma_2$}
\put(47,19){$\gamma_2'$}
\put(82,19){$\gamma_1'$}
\put(17,6){$\gamma_3$}
\put(49,6){$\gamma_3$}
\put(81,6){$\gamma_3$}
\put(7,20){$\gamma_1$}
\put(38,20){$\gamma_1'$}
\put(69,20){$\gamma_2'$}
\end{overpic}
\caption{\small We first flip the edge corresponding to $\gamma_2$ and then act with $\Tdual^{-1}$ to restore the initial triangulation, now with $\gamma_1'=\gamma_2^{\textrm{\tiny (1)}}$ as the diagonal which is to be flipped next.}
\label{Flip1}
\normalsize
\end{figure}
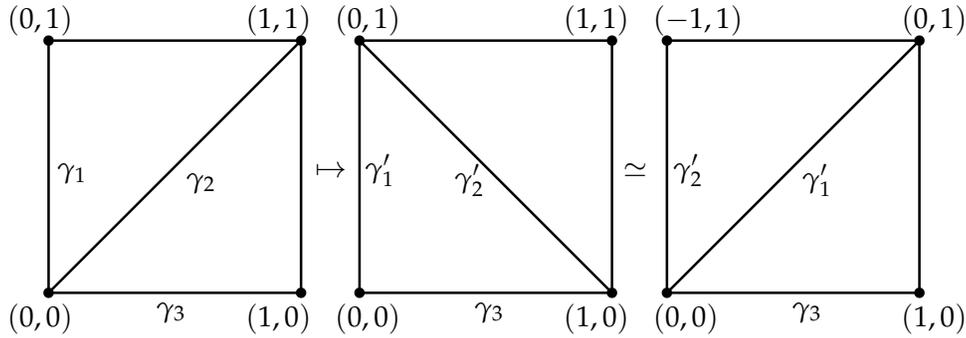

\begin{figure}[h!]
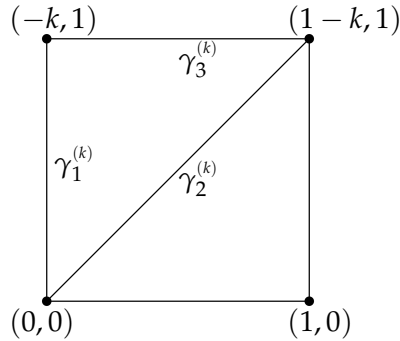

\small
\centering
\begin{overpic}
[width=0.40\textwidth]{TorusT0.png}
\put(16,3){$(0,0)$}
\put(72,3){$(1,0)$}
\put(16,64){$(-k,1)$}
\put(72,64){$(1-k,1)$}
\put(50,32){$\gamma_2^{\textrm{\tiny$(k)$}}$}
\put(50,56){$\gamma_3^{\textrm{\tiny$(k)$}}$}
\put(25,35){$\gamma_1^{\textrm{\tiny$(k)$}}$}
\end{overpic}
\caption{\small Triangulation after $k$ flips of $\varPi_2^{\gamma_1,\gamma_2}$.}
\label{Flipk}
\normalsize
\end{figure}

We will take the triangulation in Figure~\ref{T0} as our starting point and go through a sequence of flips which correspond to the mutations we have done in Section~\ref{spectrumonwall}. These mutations start with the infinite family of BPS states $\varPi_2^{\gamma_1,\gamma_2}$, the first flip of which is the flip of $\gamma_2$. To handle the infinite number of flips in a systematic way, we reparametrize the torus after each flip so that our triangulation has the same form as the one we started with, with the next edge to be flipped being the diagonal from $(0,0)$ to the top right corner. This is done by acting with $\Tdual^{-1}$ as shown in Figure~\ref{Flip1}.
With the renaming $\gamma_2^{\textrm{\tiny (1)}}=\gamma_1'$ and $\gamma_1^{\textrm{\tiny (1)}}=\gamma_2'$ as we did in defining the states in (\ref{kflipquiver}), the resulting triangulation is the one in Figure~\ref{T0} with $\gamma_i$ replaced by $\gamma_i^{\textrm{\tiny (1)}}$. Applying $\Tdual^{-1}$ after every flip, we arrive at the triangulation after $k$ flips which is labelled by (\ref{kflipquiver}) and is shown in Figure~\ref{Flipk}.
We can understand the resulting triangulations geometrically as the edges wrapping the $A$-cycle of $C$ with negative winding number. After an infinite number of windings, there is a {juggle} transformation that relates the infinite negative winding triangulation to infinite positive winding which is the limit of the family of BPS states $\varPi_1^{\gamma_1,\gamma_2}$.

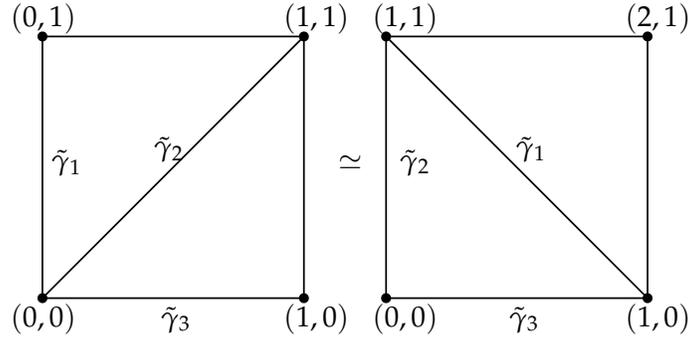
\begin{figure}[h!]
\small
\centering
\begin{overpic}
[width=0.55\textwidth]{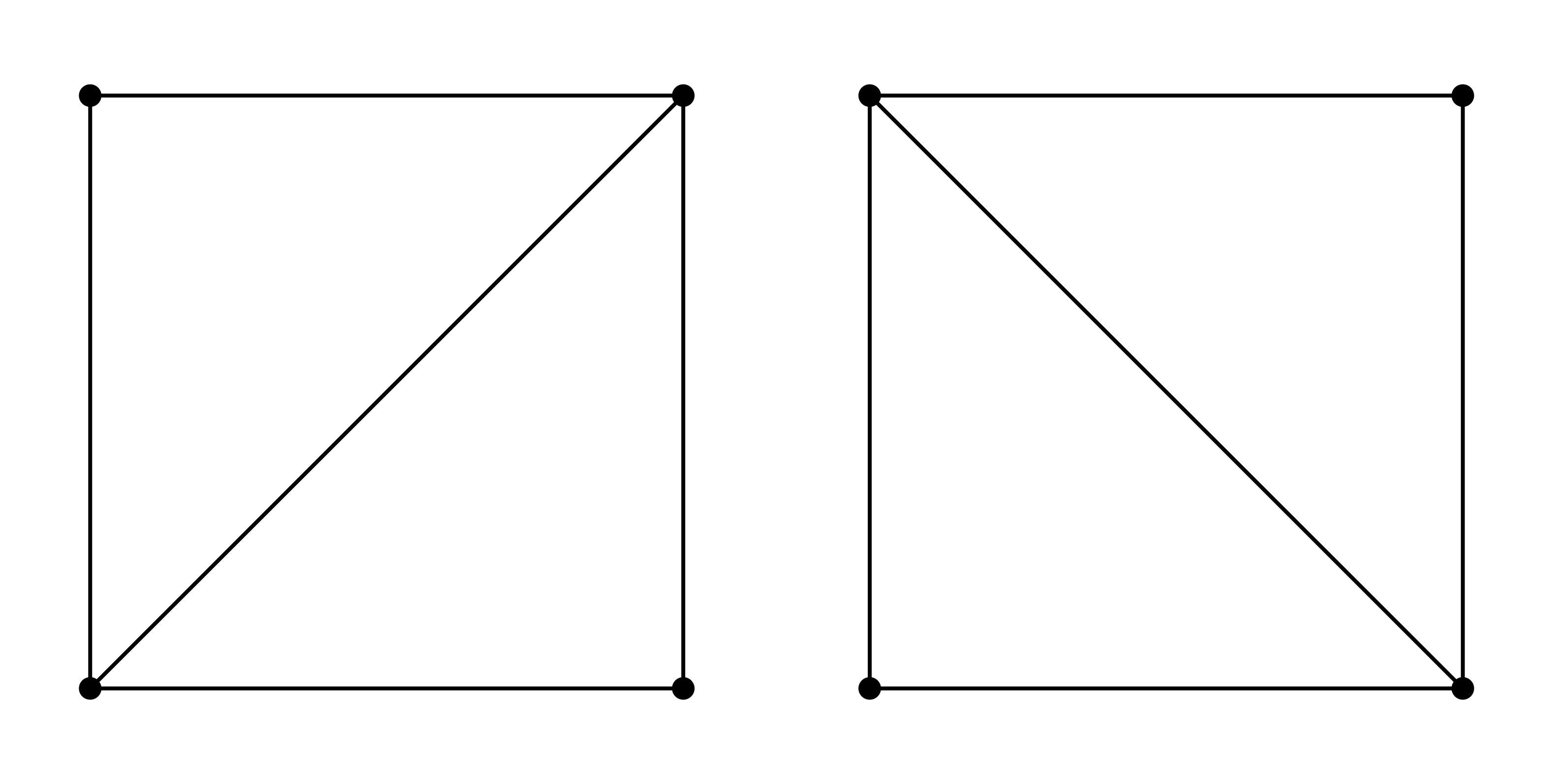}
\put(1,2){$(0,0)$}
\put(41,2){$(1,0)$}
\put(1,46){$(0,1)$}
\put(41,46){$(1,1)$}
\put(54,2){$(0,0)$}
\put(91,2){$(1,0)$}
\put(54,46){$(1,1)$}
\put(91,46){$(2,1)$}
\put(23,2){$\tgamma_3$}
\put(74,2){$\tgamma_3$}
\put(7,25){$\tgamma_1$}
\put(58,25){$\tgamma_2$}
\put(22,27){$\tgamma_2$}
\put(75,27){$\tgamma_1$}
\put(49,25){$\simeq$}
\end{overpic}
\caption{\small Acting on the initial triangulation with $\Tdual$ gives a triangulation that puts $\tgamma_1$ in a good position to be unflipped.}
\label{Unflip0}
\normalsize
\end{figure}

\begin{figure}[h!]
\small
\centering
\begin{overpic}
[width=0.80\textwidth]{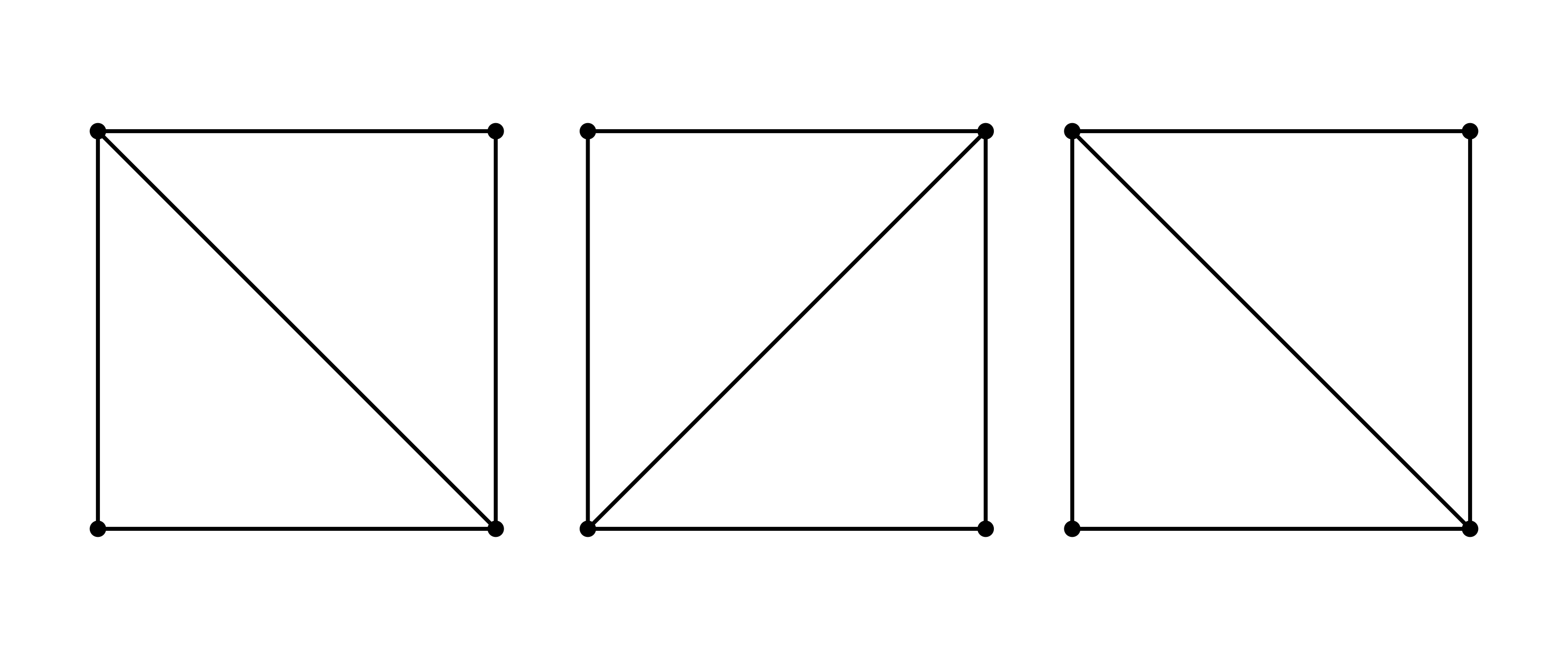}
\put(2,5){$(0,0)$}
\put(26,5){$(1,0)$}
\put(2,35){$(1,1)$}
\put(26,35){$(2,1)$}
\put(35,5){$(0,0)$}
\put(58,5){$(1,0)$}
\put(35,35){$(1,1)$}
\put(58,35){$(2,1)$}
\put(67,5){$(0,0)$}
\put(92,5){$(1,0)$}
\put(67,35){$(2,1)$}
\put(92,35){$(3,1)$}
\put(33,20){$\mapsto$}
\put(64,20){$\simeq$}
\put(19,21){$\tgamma_1$}
\put(47,22){$\tgamma_1'$}
\put(81,21){$\tgamma_2'$}
\put(17,6){$\tgamma_3$}
\put(49,6){$\tgamma_3$}
\put(81,6){$\tgamma_3$}
\put(7,20){$\tgamma_2$}
\put(38,20){$\tgamma_2'$}
\put(69,20){$\tgamma_1'$}
\end{overpic}
\caption{\small We first unflip the edge corresponding to $\tgamma_1$ and then act with $\Tdual$ to restore the initial triangulation, now with $\tgamma_2'=\tgamma_1^{\textrm{\tiny (1)}}$ as the diagonal which is to be unflipped next.}
\label{Unflip1}
\normalsize
\end{figure}

\begin{figure}[h!]
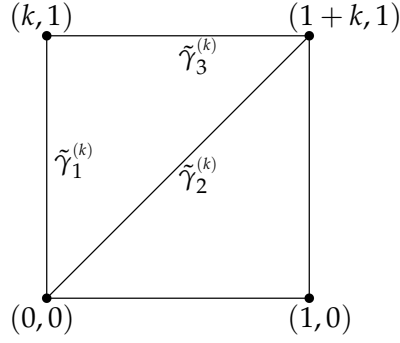

\small
\centering
\begin{overpic}
[width=0.40\textwidth]{TorusT0.png}
\put(16,3){$(0,0)$}
\put(72,3){$(1,0)$}
\put(16,64){$(k,1)$}
\put(72,64){$(1+k,1)$}
\put(50,32){$\tgamma_2^{\textrm{\tiny$(k)$}}$}
\put(50,56){$\tgamma_3^{\textrm{\tiny$(k)$}}$}
\put(25,35){$\tgamma_1^{\textrm{\tiny$(k)$}}$}
\end{overpic}
\caption{\small Triangulation after $k$ unflips of $\varPi_1^{\gamma_1,\gamma_2}$.}
\label{Unflipk}
\normalsize
\end{figure}

For the infinite family $\varPi_1^{\gamma_1,\gamma_2}$ our starting point is again the triangulation in Figure~\ref{T0}, but now with labels replaced by the antiparticles $\gamma_i\mapsto\tgamma_i = -\gamma_i$. We can apply the transformation $\Tdual$ to it in order to get $\tgamma_1$ into a nice position to unflip, as shown in Figure~\ref{Unflip0}.
From there we can proceed to unflip $\tgamma_1$. Similarly to our previous discussion of the family $\varPi_2^{\gamma_1,\gamma_2}$, we can combine every flip with the action of $\Tdual$ and the renaming of charges we did in (\ref{kunflipquiver}). This is shown for the first unflip in Figure~\ref{Unflip1}.
Finally, acting with $\Tdual^{-1}$ restores our original homology basis $(A,B)$  and yields the triangulation after $k$ unflips shown in Figure~\ref{Unflipk}. 
As expected, we now obtain $k$ windings of the $A$-cycle of $C$ in the positive direction.

\begin{figure}[h!]
\small
\centering
\begin{overpic}
[width=0.40\textwidth]{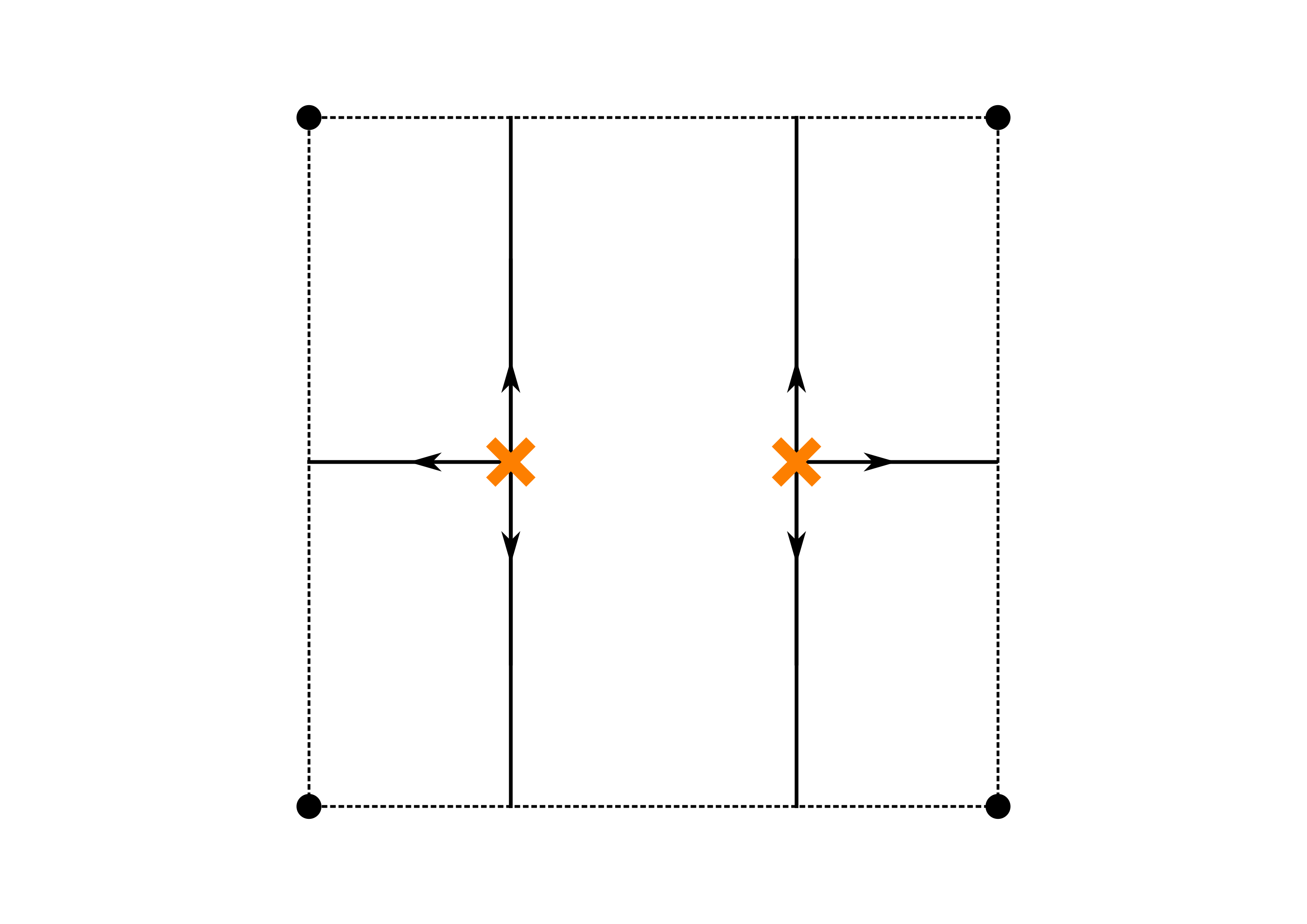}
\put(16,3){$(0,0)$}
\put(72,3){$(1,0)$}
\put(16,64){$(0,1)$}
\put(72,64){$(1,1)$}
\end{overpic}
\caption{\small Topology of the Stokes graph corresponding to the accumulation ray. There are two double walls which lift to $\gamma_1+\gamma_2$ and one that lifts to $\gamma_3$.}
\label{torusjuggle}
\normalsize
\end{figure}
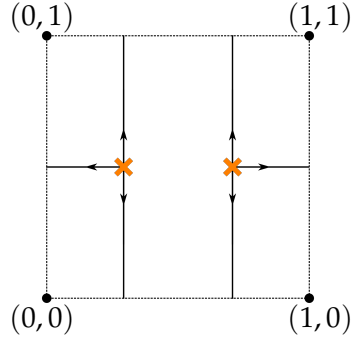

We observe that the triangulations for this spectrum are very similar to the ones we have seen for the pure $SU(2)$ gauge theory in Figures~\ref{puresu2flips} and~\ref{juggletriangulations}, which is not surprising as the BPS spectra of the two theories are very similar. The only difference occurs at the accumulation ray in the central charge plane, which for the pure $SU(2)$ theory contains only the state $\gamma_1+\gamma_2$, while the $\N=2^*$ theory at the wall $\scrE_3$ also contains the states $\gamma_3$ and $\gammavf$ as part of this accumulation ray.
On the accumulation ray itself we cannot draw a triangulation, but we may instead use the Stokes graph which consists of double walls that can be lifted to cycles on $\Sigma_{u;m}$ which support the BPS states. The situation is analogous to the pure $SU(2)$ theory which has double walls wrapping a cylinder as shown in Figure~\ref{juggletriangulations}. The Stokes graph for our case is shown in Figure~\ref{torusjuggle} and similarly has two double walls winding the $A$-cycle which correspond to the vector state $\gamma_1+\gamma_2$. 
There is furthermore a double wall winding the $B$-cycle that lifts to $\gamma_3$, and the combination of walls supports $2\,(\gamma_1+\gamma_2)+\gamma_3=\gammavf$. 

The network also includes the flavour state $\gamma_{\rm f}=\gamma_1+\gamma_2+\gamma_3$ which corresponds to a closed finite curve around the puncture and similarly bounds a family of closed curves around the puncture (cf. \cite{gaiotto2013wall}). From this geometric perspective, the mass $m$ of the $\N=2^*$ adjoint hypermultiplet is one of the eigenvalues of the monodromy around the puncture $z=0$ on $C$.

\section{\tops{$\boldsymbol{\N=2^*}$ Spectrum Away from the Walls}{BPS Spectrum Away from the Walls}}\label{perturbedspectrum}
\nid
So far we have restricted our analysis of the BPS spectrum of the rank one $\N=2^*$ gauge theory to points in the Coulomb branch $\scrB_m$ which lie on the wall of marginal stability $\scrE_3$, where the central charge of $\gamma_3$ aligns with $m=Z(\gamma_{\rm f})$. Following the prescription in~\cite{longhi2015structure}, we now perturb away from this wall.

\subsection{Perturbing the Spectrum}\label{subsec:perturbing}
\begin{figure}[h!]
\small
\centering
\begin{overpic}
[width=0.50\textwidth]{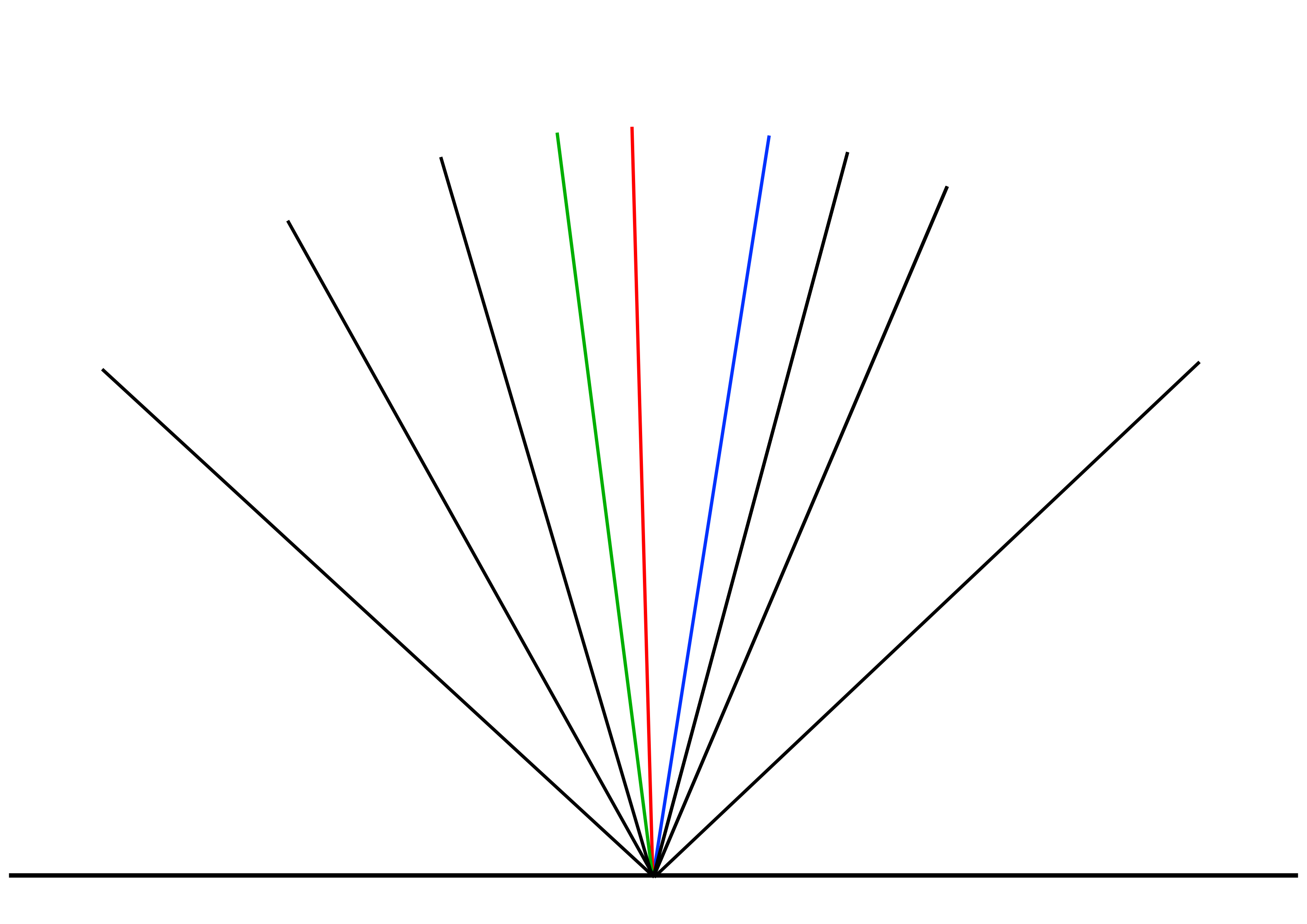}
\put(37.5,50){$\dots$}
\put(27,59.5){\textcolor{Green}{$\gamma_1+\gamma_2$}}
\put(39,63){\textcolor{Red}{$\gammavf$}}
\put(59.5,59.5){\textcolor{Blue}{$\gamma_3$}}
\put(44.5,50){\textcolor{Red}{$...$}}
\put(51,50){\textcolor{Blue}{$\dots$}}
\put(58,50){$...$}
\put(92,44){$\gamma_1$}
\put(5,44){$\gamma_2$}
\end{overpic}
\caption{\small Charge rays $Z(\gamma;u)$ of the $\N=2^*$ spectrum after the perturbation (\ref{perturbspectrum}). The rays of $\gamma_1+\gamma_2$, $\gamma_3$ and $\gammavf$ become separated, and the perturbations of $\gamma_3$ and $\gammavf$ lead to new behaviour.}
\label{gammaraysp}
\normalsize
\end{figure}
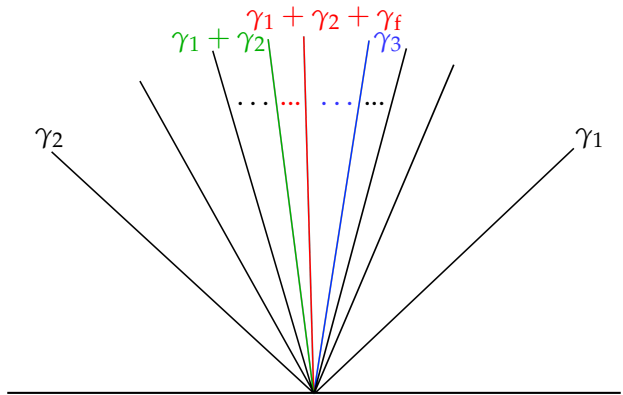

\nid
For $u\in\scrE_3$, we keep the central charge $Z_2=Z(\gamma_2;u)$ constant and perturb $Z_1=Z(\gamma_1;u)$ by
\begin{equation}\label{perturbspectrum}
Z_1\longmapsto Z_1+\delta \, Z_2 \ ,
\end{equation}
where $\delta\in\mathbb{R}_{>0}$ is a small parameter. This perturbs the phase of $Z_1$ by a small positive amount, and thus closes the wedge between $Z_2$ and $Z_1$ as $Z_1$ rotates anticlockwise. Since $Z_{\rm f}=Z(\gamma_1+\gamma_2+\gamma_3)=m$ is constant on the moduli space, this change of $Z_1$ while keeping $Z_2$ constant perturbs $Z_3=Z(\gamma_3;u)$ as well by
\begin{equation}
Z_3\longmapsto Z_3-\delta \, Z_2 \ .
\end{equation}
Hence $Z_3$ rotates slightly clockwise. In a similar manner the central charges $Z(\gamma_1+\gamma_2;u)$ and $Z(\gamma_1+\gamma_2+\gamma_{\rm f}\,;u)$ are changed by $\delta \, Z_2$, and rotate slightly anticlockwise. We should however keep in mind that these two vectors have different magnitudes, so adding a small piece affects them differently. Therefore the central charge $Z(\gamma_1+\gamma_2;u)$ rotates further than $Z(\gamma_1+\gamma_2+\gamma_{\rm f}\,;u)$ and the two charges become separated. We should also note that the bound states $\gamma_i + n\,(\gamma_1+\gamma_2)$ involve multiple copies of $\gamma_1$ and are affected as well, rotating slightly anticlockwise. This movement of charges is shown in Figure~\ref{gammaraysp}.

If we consider the spectrum generator (\ref{Spectrum}), we now have to move $\cK_{\gamma_3}$ and $\cK_{\gamma_1+\gamma_2+\gamma_{\rm f}}$ through parts of the family $\varPi_1$. We therefore expect the form
\begin{equation}\label{PertSpec}
\bbS = \varPi_2^{\gamma_1,\gamma_2}\,\cK^{-2}_{\gamma_1+\gamma_2}\cdots\cK_{\gamma_1+\gamma_2+\gamma_{\rm f}}\cdots\cK_{\gamma_3} \ \prod_{0\leq k<N}^\curvearrowleft \, \cK_{(k+1)\,\gamma_1+k\,\gamma_2} \ ,
\end{equation}
where $\varPi_1^{\gamma_1,\gamma_2}$ is now truncated after the $N$-th flip. What do these permutations do to the spectrum and what happens in the dotted areas? We first examine $\cK_{\gamma_3}$ which is commuted through all $\cK_{\gamma_1+k\,(\gamma_1+\gamma_2)}$ for $k\ge N$. The electric-magnetic pairing gives
\begin{equation}
\langle\gamma_3,\gamma_1+k\,(\gamma_1+\gamma_2)\rangle=\langle\gamma_3,\gamma_1\rangle=2 \ ,
\end{equation}
so we can use the corresponding Kontsevich-Soibelman wall-crossing formula to find the resulting bound states between $\gamma_3$ and $\gamma_1+k\,(\gamma_1+\gamma_2)=-\tgamma_1^{\textrm{\tiny$(k)$}}$ through
\begin{equation}\label{primaryWCF}
\cK_{\gamma_3}\,\cK_{-\tgamma_1^{\textrm{\tiny$(k)$}}}= \varPi_2^{\gamma_3,-\tgamma_1^{\textrm{\tiny$(k)$}}}  \, \cK^{-2}_{-\tgamma_1^{\textrm{\tiny$(k)$}}+\gamma_3} \, \varPi_1^{\gamma_3,-\tgamma_1^{\textrm{\tiny$(k)$}}} \ .
\end{equation}
It follows that the permutation of $\cK_{\gamma_3}$ results in an infinite family of vector multiplets $\gamma_1 + k\,(\gamma_1+\gamma_2) + \gamma_3$ with $k\ge N$, together with the usual two families of hypermultiplet bound states. Since the states $\gamma_1+k\,(\gamma_1+\gamma_2)$ accumulate near $\gamma_1+\gamma_2$, the same is true for at least some of these new bound states, and since $\arg (Z_1+Z_2)>\arg(Z_1+Z_2+Z_{\rm f})$, the resulting symplectomorphisms will also permute with $\cK_{\gammavf}$.

Before permuting with $\cK_{\gammavf}$, there is however more non-trivial behaviour due to the permutation of $\cK_{\gamma_3}$. The bound states discussed above lie between the respective charges $-\tgamma_1^{\textrm{\tiny$(k)$}}$ and $\gamma_3$. When $\gamma_3$ interacts with other states \smash{$-\tgamma_1^{\textrm{\tiny $(k')$}}$}, the same can happen for the bound states, which can then lead to the formation of secondary bound states. We will discuss this in detail in Section~\ref{SecondaryStates} below, but first we will have a closer look at the states found in (\ref{primaryWCF}).

\subsection{Primary Bound States from Quiver Mutations}\label{PrimaryStates}
\nid
We can make the computation of \eqref{primaryWCF} very explicit and derive the BPS quivers involved at every step. Starting with unflips from $\gamma_1$, we follow the previously discovered family $\{\tilde\gamma_i^{\textrm{\tiny$(k)$}}\}$ up to $k=N$ which gives the charges after the $N$-th unflip. The quiver at that point is labelled by the charges
\begin{align}
\tgamma_1^{\textrm{\tiny $(N)$}} &= -\big(\gamma_1+N\,(\gamma_1+\gamma_2)\big)  \ , \nonumber \\[4pt]
\tgamma_2^{\textrm{\tiny $(N)$}} &= -\big(\gamma_2-N\,(\gamma_1+\gamma_2)\big) \ , \nonumber \\[4pt]
\tgamma_3^{\textrm{\tiny $(N)$}} &= -\gamma_3 \ ,
\end{align}
and now instead of unflipping $\tgamma_1^{\textrm{\tiny $(N)$}}$, we unflip $\tgamma_3=\tgamma_3^{\textrm{\tiny $(N)$}}$ to thus get
\begin{align}
\bar{\gamma}_1^{\textrm{\tiny $(N)$}} \longmapsto \tgamma_1^{\textrm{\tiny $(N)$}}+2\tgamma_3 \ , \quad
\bar{\gamma}_2^{\textrm{\tiny $(N)$}} \longmapsto \tgamma_2^{\textrm{\tiny $(N)$}} \qquad \mbox{and} \qquad
\bar{\gamma}_3^{\textrm{\tiny $(N)$}} \longmapsto -\tgamma_3 \ .
\end{align}
This starts the sequence of unflips leading to the vector multiplet $\tgamma_1^{\textrm{\tiny $(N)$}}+\tgamma_3$. We can encode it in a family, where we again combine every unflip with a renaming of charges as done previously. This gives
\begin{align}\label{tgammakl}
\tgamma_1^{\textrm{\tiny $(N|l)$}} &= \tgamma_1^{\textrm{\tiny $(N)$}}-l\,\big(\tgamma_1^{\textrm{\tiny $(N)$}}+\tgamma_3\big) \ , \nonumber \\[4pt]
\tgamma_2^{\textrm{\tiny $(N|l)$}} &= \tgamma_2^{\textrm{\tiny $(N)$}} \ , \nonumber \\[4pt]
\tgamma_3^{\textrm{\tiny $(N|l)$}} &= \tgamma_3+l\,\big(\tgamma_1^{\textrm{\tiny $(N)$}}+\tgamma_3\big) \ ,
\end{align}
where $l\geq0$ and $\langle\tgamma_i^{\textrm{\tiny $(N|l)$}},\tgamma_{i+1}^{\textrm{\tiny $(N|l)$}}\rangle=2$. Unflipping $\tgamma_3^{\textrm{\tiny $(N|l)$}}$ gives the next collection of charges at step $l+1$, similarly to the hypermultiplet families considered earlier.

For the other side of the accumulation ray in the central charge plane, we would like to start coming from the left. In many theories with a vector multiplet, there is only one accumulation ray and one can just start from the quiver with charges $\{-\gamma_i\}$. This is not possible here as we would then have to first work through the ray at $\gamma_1+\gamma_2$ and all the states that were generated by the symplectomorphism $\cK_{\gamma_1+\gamma_2+\gamma_{\rm f}}$.

There is however another way to find the quiver on the other side. Consider the spectrum generator \eqref{PertSpec} when we truncate $\varPi_1^{\gamma_1,\gamma_2}$ at $N+1$ instead of $N$, that is
\begin{equation}
\bbS = \cdots\cK_{-(\tgamma_1^{\textrm{\tiny $(N+1)$}}+2\tgamma_3)}\,\cK_{-\tgamma_3}\,\cK_{-\tgamma_1^{\textrm{\tiny $(N)$}}} \cdots \ .
\end{equation}
From here we can move $\cK_{-\tgamma_3}$ further to the right without changing the family $\tgamma_i^{\textrm{\tiny $(N+1|l)$}}$.\footnote{As we will see in Section~\ref{SecondaryStates}, moving $\gamma_3$ will often affect the families it previously interacted with in non-trivial ways. Moving it individually however is still a valid working assumption, and such scenarios do occur in our numerical examination in Section~\ref{Pythonspectrum1} for example.}
The resulting permutation gives
\begin{equation}\label{g13lhs}
\bbS = \cdots\cK_{-(\tgamma_1^{\textrm{\tiny $(N+1)$}}+2\tgamma_3)}\,\varPi_2^{-\tgamma_3,-\tgamma_1^{\textrm{\tiny $(N)$}}}\,\cK_{-\tgamma_3-\tgamma_1^{\textrm{\tiny $(N)$}}}^{-2}\cdots\cK_{-\tgamma_3}\cdots \ ,
\end{equation}
and we can now use this to read off our quiver. We start to the right of \smash{$\cK_{-(\tgamma_1^{\textrm{\tiny $(N+1)$}}+2\tgamma_3)}$}, where the quiver is labelled by charges \smash{$\tgamma^{\textrm{\tiny $(N+1|1)$}}_i$} as defined in (\ref{tgammakl}).

We therefore start from the charges\footnote{Note that the definition of the charges $\{\hgamma_i^{\textrm{\tiny $(N)$}}\}$ uses a cyclic relabelling $i\mapsto i-1$ for later convenience.}
\begin{align}\label{gammahat}
\hgamma_3^{\textrm{\tiny $(N)$}}& := \tgamma_1^{\textrm{\tiny $(N+1|1)$}} = -\tgamma_3 \ , \nonumber\\[4pt]
\hgamma_1^{\textrm{\tiny $(N)$}}& := \tgamma_2^{\textrm{\tiny $(N+1|1)$}} = -\big(\gamma_2-(N+1)\,(\gamma_1+\gamma_2)\big)=-\tgamma_1^{\textrm{\tiny $(N)$}} \ , \nonumber\\[4pt]
\hgamma_2^{\textrm{\tiny $(N)$}}& := \tgamma_3^{\textrm{\tiny $(N+1|1)$}}=\tgamma_1^{\textrm{\tiny $(N+1)$}}+2\tgamma_3 \ .
\end{align}
From this quiver we could continue to the family $\tgamma_i^{\textrm{\tiny $(N+1|l)$}}$ by unflipping $\tgamma_3^{\textrm{\tiny $(N+1|1)$}}$, but we are instead interested in the other direction by flipping the charge \smash{$-\tgamma_1^{\textrm{\tiny $(N)$}}=\hgamma_1^{\textrm{\tiny $(N)$}}$}. This flip yields
\begin{align}
\hgamma_1^{\textrm{\tiny $(N)$}}& \longmapsto -\hgamma_1^{\textrm{\tiny $(N)$}} = \tgamma_1^{\textrm{\tiny $(N)$}} \ , \nonumber\\[4pt]
\hgamma_2^{\textrm{\tiny $(N)$}}& \longmapsto \hgamma_2^{\textrm{\tiny $(N)$}} = \tgamma_1^{\textrm{\tiny $(N+1)$}}+2\tgamma_3  \ , \nonumber\\[4pt]
\hgamma_3^{\textrm{\tiny $(N)$}}& \longmapsto \hgamma_3^{\textrm{\tiny $(N)$}} + 2 \hgamma_1^{\textrm{\tiny $(N)$}} = -\big(\tgamma_3^{\textrm{\tiny $(N)$}} + 2 \tgamma_1^{\textrm{\tiny $(N)$}}\big) \ .
\end{align}
This is exactly the start of the winding sequence of $\tgamma_1^{\textrm{\tiny $(N)$}}$ and $\tgamma_3$ that we expected. We therefore obtain the family of charges
\begin{align}\label{gammahats}
\hgamma_1^{\textrm{\tiny $(N|l)$}}& := \hgamma_1^{\textrm{\tiny $(N)$}}+ l\, \big(\hgamma_1^{\textrm{\tiny $(N)$}}+\hgamma_3^{\textrm{\tiny $(N)$}}\big) = -\big(\tgamma_1^{\textrm{\tiny $(N)$}} + l \, (\tgamma_1^{\textrm{\tiny $(N)$}} + \tgamma_3)\big) \ , \nonumber\\[4pt]
\hgamma_2^{\textrm{\tiny $(N|l)$}}& := \hgamma_2^{\textrm{\tiny $(N)$}} = \tgamma_1^{\textrm{\tiny $(N+1)$}}+2\tgamma_3 \ , \nonumber\\[4pt]
\hgamma_3^{\textrm{\tiny $(N|l)$}}& := \hgamma_3^{\textrm{\tiny $(N)$}}- l\, \big(\hgamma_1^{\textrm{\tiny $(N)$}}+\hgamma_3^{\textrm{\tiny $(N)$}}\big) = -\big(\tgamma_3 - l \, (\tgamma_1^{\textrm{\tiny $(N)$}} + \tgamma_3)\big) \ .
\end{align}

If we compare the families in (\ref{tgammakl}) and (\ref{gammahats}), we notice that the state that is not involved in the winding has changed from $\tgamma_2^{\textrm{\tiny $(N)$}}$ to $
\hgamma_2^{\textrm{\tiny $(N)$}} = \tgamma_1^{\textrm{\tiny $(N+1)$}}+2\tgamma_3.$
The charge $\hgamma_2^{\textrm{\tiny $(N)$}}$ can be brought into a suggestive form that relates it to \smash{$\tgamma_2^{\textrm{\tiny $(N)$}}$}, by rewriting it as
\begin{equation}\label{eq:gamma2juggle}
\hgamma_2^{\textrm{\tiny $(N)$}}= -\big(\gamma_1+(N+1)\,(\gamma_1+\gamma_2)+2\gamma_3\big)= \tgamma_1^{\textrm{\tiny $(N-1)$}}-2\gamma_{\rm f} =  -\big( \tgamma_2^{\textrm{\tiny $(N)$}}+2\gamma_{\rm f} \big) \ .
\end{equation}
In the form \eqref{eq:gamma2juggle}, this suggests that $\hgamma_2^{\textrm{\tiny $(N)$}}$ is the change that the charge $\tgamma_2^{\textrm{\tiny $(N)$}}$ undergoes after the juggle.

The same arguments can be applied to $\tgamma_i^{\textrm{\tiny$(k)$}}$ for any $k\ge N$. One obtains families of quiver charges \smash{$\tgamma_i^{\textrm{\tiny $(k|l)$}}$} defined as in (\ref{tgammakl}) and \smash{$\hgamma_i^{\textrm{\tiny $(k|l)$}}$} defined as in (\ref{gammahat}); from an algebraic perspective, these families include at least some of the indecomposable `band representations' of the Markov quiver, see~\cite{Derksen2008}. These yield bound states $l \, \tgamma^{\textrm{\tiny$(k)$}}_1 + (l+1)\,\tgamma_3$ and $(l+1)\, \tgamma^{\textrm{\tiny$(k)$}}_1 + l\,\tgamma_3$ respectively. The bound state $\tgamma_i^{\textrm{\tiny $(k|0)$}}=\tgamma_3$ is only included in the first such families at $k=N$, and in the following we will consider only the states $\tgamma_i^{\textrm{\tiny $(k|l)$}}$ with $l\ge1$.
We expect this behaviour to persist at least until the symplectomorphism $\cK_{\gammavf}$ is reached. The different families labelled by $(k|l)$ can moreover overlap and form further bound states, which are discussed in Section~\ref{SecondaryStates} below.

\begin{notation}\label{not:simplifty}
For the the remainder of this section we will simplify our notation considerably. For the BPS quiver method it was very important to keep track of signs of charges, but for the spectrum itself this is unnecessary clutter. We therefore define
\begin{equation}\label{tgammasimple}
\tgamma^{\textrm{\tiny$(k)$}} := \gamma_1 + k\,(\gamma_1+\gamma_2) = -\tgamma_1^{\textrm{\tiny$(k)$}} \ ,
\end{equation}
and for the $(k|l)$-families we define
\begin{align}\label{gammaklsimple}
\tgamma^{\textrm{\tiny $(k|l)$}} & := \gamma_3 + l\, \big(\gamma_3+\tgamma^{\textrm{\tiny$(k)$}}\big) \ , \nonumber \\[4pt]
\tgamma^{\textrm{\tiny $(k|\infty)$}} & := \gamma_3 + \tgamma^{\textrm{\tiny$(k)$}} \ , \nonumber \\[4pt]
\tgamma^{\textrm{\tiny $(k|-l)$}} & := \tgamma^{\textrm{\tiny$(k)$}} + l \, \big(\gamma_3+\tgamma^{\textrm{\tiny$(k)$}}\big) \ .
\end{align}
For $l=0$ these give $\tgamma^{\textrm{\tiny $(k|0)$}}=\gamma_3$, while $\tgamma^{\textrm{\tiny $(k|-0)$}}=\tgamma^{\textrm{\tiny$(k)$}}$, but we will not usually use the notation~$\tgamma^{\textrm{\tiny $(k|\pm\,0)$}}$. 
\end{notation}

\subsection{Overlapping Families and Secondary States}\label{SecondaryStates}
\nid
The $(k|l)$-families discussed in Section~\ref{PrimaryStates} are bound states of charges $\tgamma^{\textrm{\tiny$(k)$}}$ and $\gamma_3$, and their central charges therefore lie in between the two corresponding rays. Since $\gamma_3$ interacts with various states $\tgamma^{\textrm{\tiny$(k)$}}$, the different $(k|l)$-families can overlap and this overlapping can lead to secondary BPS states.

As the simplest case, consider the charges $\tgamma^{\textrm{\tiny $(k+1|1)$}}$ and $\tgamma^{\textrm{\tiny$(k)$}}$. The two copies of $\gamma_3$ in $\tgamma^{\textrm{\tiny $(k+1|1)$}}$ reduce the argument of its central charge and it can therefore end up to the right of $\tgamma^{\textrm{\tiny$(k)$}}$. Since $\langle \tgamma^{\textrm{\tiny $(k+1|1)$}},\tgamma^{\textrm{\tiny$(k)$}}\rangle = 2$, the Kontsevich-Soibelman wall-crossing formula gives another vector multiplet \smash{$\tgamma^{\textrm{\tiny $(k+1|1)$}}+\tgamma^{\textrm{\tiny$(k)$}}$} accompanied by the usual families of hypermultiplets. 

Another simple permutation that can occur is when $\tgamma^{\textrm{\tiny $(k+1|1)$}}$ interacts with $\tgamma^{\textrm{\tiny $(k|-1)$}}$. The electric-magnetic pairing is again two, so we once more obtain a vector multiplet accompanied by two families of hypermultiplets.

The overlap of families is in general not small and one should consider all the pairings that can occur, as listed in Table~\ref{klpairings}.
\begin{table}[h!]
\centering
\begin{tabular}{c|c c c}
$\langle\cdot,\cdot\rangle$ & $\tgamma^{\textrm{\tiny $(k|l')$}}$ & $\tgamma^{\textrm{\tiny $(k|\infty)$}}$ & $\tgamma^{\textrm{\tiny $(k|-l')$}}$ \\ \hline
$\tgamma^{\textrm{\tiny $(k+1|l)$}}$ & $2\,\big(1+l'\,(1-l)\big)$ & $2\,(1-l)$ & $2\,\big(l'\,(1-l)-l\big)$ \\
$\tgamma^{\textrm{\tiny $(k+1|\infty)$}}$ & $-2\,l'$ & $-2$  & $-2\,(1+l')$ \\
$\tgamma^{\textrm{\tiny $(k+1|-l)$}}$ & $-2\,\big(l'\,(l+2)+1\big)$ & $-2\,(l+2)$ & $-2\,\big(l'\,(l+2)+l-1\big)$ \\
\end{tabular}
\caption{\small Intersection pairings for different states at level $k+1$ with other states at level $k$. Here $l,l'\ge0$ except for the first row, where $l\ge1$, and the third column, where $l'\ge1$.}
\label{klpairings}
\end{table}
Most of these pairings are never positive. If we ignore negative pairings, the only bound states that occur are those with pairings
\begin{equation}
\langle \tgamma^{\textrm{\tiny $(k+1|l)$}}, \tgamma^{\textrm{\tiny $(k|-l')$}} \rangle = 2\,\big(1+l'\,(1-l)\big) \ ,
\end{equation}
which is two if and only if $l=1$ or $l'=0$. Therefore $\tgamma^{\textrm{\tiny $(k+1|1)$}}$ will bind with every state $\tgamma^{\textrm{\tiny $(k|-l')$}}$, and every state $\tgamma^{\textrm{\tiny $(k+1|l)$}}$ will bind with $\tgamma^{\textrm{\tiny $(k|-0)$}}=\tgamma^{\textrm{\tiny$(k)$}}$. We also see that it is possible for (arbitrary large) negative pairings to occur, which should not happen in $SU(2)$ theories~\cite{galakhov2013wild,longhi2015structure}. When exploring the spectrum numerically in Section~\ref{Pythonspectrum}, we do indeed find some cases of wall-crossings with pairing $-2$. We interpret these in  terms of the reverse orderings of the pairing two wall-crossing formula in Section~\ref{missingstates}. 
It is furthermore possible for non-neighbouring families to overlap. The resulting pairings are never positive and again large negative pairings can occur.

Although our analysis stops here, nothing (in principle at least) prevents one from going further in the study of higher bound states.
The resulting secondary families can again overlap and therefore produce tertiary BPS states. This process can go on indefinitely in principle, but we expect it to terminate at some point. This expectation is supported by numerical analysis of the $\N=2^*$ spectrum that we present in Section~\ref{Pythonspectrum} below. 

\begin{notation}\label{not:level}
To keep track of the different kinds of secondary states, we introduce the \textit{level} of a state $\gamma$, and denote it by $\mathsf{level}(\gamma)$. It is zero for our initial spectrum of BPS states $\gamma\in\{\gamma_1,\gamma_2,\gamma_3,\gammavf\}$, one for $\gamma=\tgamma^{\textrm{\tiny$(k)$}}$, and it is added according to the rule
\begin{equation}\label{lvldef}
\mathsf{level}(\gamma+\gamma') = \text{max}\big(\mathsf{level}(\gamma),\mathsf{level}(\gamma')\big)+1 \ .
\end{equation}
\end{notation}

\subsection{Numerical Analysis of the Spectrum}\label{Pythonspectrum}
\nid
We can further investigate the $\N=2^*$ spectrum numerically using a Python program and the Kontsevich-Soibelman wall-crossing formula. A detailed explanation of the algorithm~\cite{Rueter} is presented in Appendix~\ref{PythonAlgorithm}. In this section we will present our findings and discuss their physical relevance. One crucial aspect of the algorithm is that we try to break down the perturbation into small steps that lead to the permutations involving two factors $\cK_{\gamma}\,\cK_{\gamma'}$. This does however limit the computation in some regards, and in particular makes it impossible to explore parts of the spectrum that involve the reverse ordered $\langle\gamma,\gamma'\rangle=2$ wall-crossing formula discussed in Section~\ref{missingstates}.

In the following we will always set $Z_1=1 + \I\,$, $Z_2 = -1 + \I\,$, and take $m$ to be purely imaginary.\footnote{It is convenient to place the adjoint mass $m\in \I\,\mathbb{R}$ so that the upper half-plane $\mathbb{H}_0\hookrightarrow\mathbb{C}$ is centred on it. While physically $m>0$ is of course preferable, at the level of BPS quivers it makes no difference up to a simple rotation in $\mathbb{C}$.} The parameters we consider are then $|m|$ and the perturbation $\delta>0$. Since our calculations involve various infinite families of states, we further introduce some cutoffs. The states $\tgamma^{\textrm{\tiny$(k)$}}$ are only constructed up to $k\le\mathsf{kcut}$, and similarly we only construct states $\tgamma^{\textrm{\tiny $(k|\pm\,l)$}}$ up to $l\le\mathsf{subcut}$, which is also the cutoff we use for any other secondary families. In the same way we also cut off the level defined in (\ref{lvldef}) and construct only states $\gamma$ with $\mathsf{level}(\gamma)\le\mathsf{lvlcut}$.

\subsubsection{\tops{Analysis at Small Mass and Large Perturbation}{Python Exploration at small m and large delta}}\label{Pythonspectrum1}
\nid
Our main computation is at a relatively small choice of $m$ at $|m|=3$ and a perturbation of $\delta\approx0.15$, which is large in the sense that it allows $\gamma_3$ to interact with all bound states except $\tgamma^{\textrm{\tiny (1)}}=2\gamma_1+\gamma_2$.\footnote{We actually choose $\delta=0.98\cdot0.15$ as the computation turns out to be slightly simpler then.} With this choice of $m$, the remaining initial central charges are $Z_3=\I$ and $Z(\gammavf\,;u)=5\,\I\,$. Although it may seem odd to choose a relatively large perturbation instead of a small one, we should keep in mind that in any case both $\gamma_3$ and $\gammavf$ will permute past infinitely many states which are accumulated near $\gamma_1+\gamma_2$. We therefore choose the computationally easier route with $\gamma_3$ being close to $\gamma_1$, where there are fewer states $\tgamma^{\textrm{\tiny$(k)$}}$ to consider. We also focus on the wedge of states between $\gamma_3$ and $\gammavf$. Given that $\gammavf$ has pairing $-2$ with all states $\tgamma^{\textrm{\tiny$(k)$}}$, its wall-crossing is much more difficult to handle and we will briefly discuss this issue in Section~\ref{gammavfcrossing}. It is furthermore difficult to push numerically to much higher $k$, in order to include the states behind $\gammavf$, as the behaviour gets more complicated with the inclusion of larger~$k$.

For our first explorations of the $\N=2^*$ spectrum, we truncate any bound state families to their first members by $\textsf{subcut}=1$. Any further bound states and secondary families are likely to lie between these states, so this approximation gives a good indication of the qualitative behaviour of these families, and in particular of the overlap between families discussed in Section~\ref{SecondaryStates}. We will have a brief look at a calculation with $\textsf{subcut}=4$ later, but we have to reduce $\textsf{kcut}$ considerably for it.

\paragraph{States up to level $\boldsymbol{2}$ and $\boldsymbol{k=14}$.} We can compute states to the lowest level, by setting the cutoff $\textsf{lvlcut}=2$, which results in the bound states $\tgamma^{\textrm{\tiny $(k|\pm\,l)$}}$ as defined in (\ref{gammaklsimple}). The wedge between $\gamma_3$ and $\gammavf$ includes the states $\left\{\tgamma^{\textrm{\tiny$(k)$}}\right\}_{1<k<12}$, but taking into account the resulting bound states with $\gamma_3$, we also encounter the states $\tgamma^{\textrm{\tiny $(k|1)$}}$ for $k=12,13,14$. Hence if we wish to compute the entire spectrum between $\gamma_3$ and $\gammavf$, we need to set $\textsf{kcut}\geq14$. We can then compute the spectrum of states up to level $2$ completely with the above choice of parameters.

Already with this level cutoff we encounter some non-trivial and unexpected behaviour in the form of wall-crossings involving negative pairings
\begin{equation}\label{deepfamilyoverlap}
\langle\tgamma^{\textrm{\tiny $(k|1)$}}, \tgamma^{\textrm{\tiny $(k-2|1)$}}\rangle = -2 \ ,
\end{equation}
which occur for states with $k=10,11,12,13,14$. For all but the $k=14$ state, these are between $\gamma_3$ and $\gammavf$. As mentioned earlier, wall-crossings with negative pairing should not occur, and in Section~\ref{neghypercrossing} we will interpret it instead in terms of the reverse ordering of a wall-crossing formula with pairing two.

We also encounter vectors interacting with other states, for example $\gamma_3+\tgamma^{\textrm{\tiny$(k)$}}$ with $\tgamma^{\textrm{\tiny $(k-1)$}}$ for $k\ge 7$. For the parts of the spectrum that end up between $\gamma_3$ and $\gamma_1+\gamma_2+\gamma_{\rm f}$ these interactions have zero pairing, which is expected since otherwise the families accumulating near the vector multiplet would lead to arbitrarily large positive or negative pairings because these bound states involve arbitrarily many copies of the vector state. However, the interactions of vector states with zero pairing still lead to complicated behaviour, as the families of states that accompany them will have non-zero pairing. In particular, the families of bound states in front of the vector states can be written as
\begin{equation}
\tgamma^{\textrm{\tiny $(k|+l)$}}=\gamma_3 + l \,(\gamma_3+\tgamma^{\textrm{\tiny$(k)$}})
\end{equation}
and $\langle\gamma_3,\tgamma^{\textrm{\tiny$(k)$}}\rangle=2$ for all $k$. Hence each of the bound states accompanying $\gamma_3+\tgamma^{\textrm{\tiny$(k)$}}$ will form secondary bound states with $\tgamma^{\textrm{\tiny $(k-1)$}}$ as they interact with it. This gives rise to an infinite number of vector states each accompanied by the usual families of hypermultiplets.

We do find further vector states interacting with pairing $-2$ with another state for $k=13,14$, but these interactions occur in the portion behind $\gammavf$.

\begin{figure}[h!]
\centering
\begin{overpic}
[width=\textwidth]{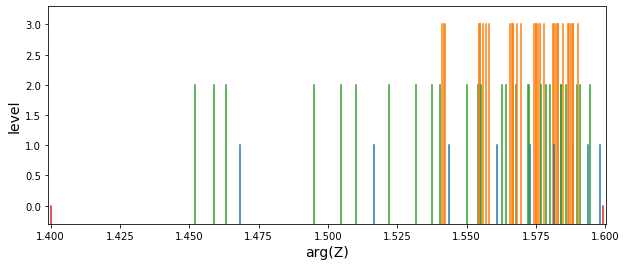}
\end{overpic}
\caption{\small States between $\gamma_3$ (leftmost) and $\gammavf$ (rightmost) for $\textsf{kcut}=10$, $\textsf{subcut}=1$, and $\textsf{lvlcut}=3$. We plot the phase of the states on the horizontal axis against their level on the vertical axis and also colour them by level.}
\label{SecondaryFamily}
\end{figure}

\paragraph{States up to level $\boldsymbol{3}$ and $\boldsymbol{k=10}$.} We now increase the level cutoff to $3$ to allow for the first kind of secondary states. We can compute the spectrum up to $k=10$. Figure~\ref{SecondaryFamily} shows the resulting spectrum, where we plot the level of the states against their phase for the states between $\gamma_3$ and $\gammavf$. 
The states at level $0$ are just $\gamma_3$ and $\gammavf$ which are the leftmost and rightmost state in the plot respectively. The states at level $1$ are then the states $\tgamma^{\textrm{\tiny $(k)$}}$ and the states at level $2$ are the primary bound states $\tgamma^{\textrm{\tiny $(k|l)$}}$.

At level $3$ there are different kinds of secondary states, which result from the primary states $\tgamma^{\textrm{\tiny $(k|l)$}}$ moving past either other primary families $\tgamma^{\textrm{\tiny $(k'|l')$}}$ or past states $\tgamma^{\textrm{\tiny $(k')$}}$. The leftmost family of secondary states is a result of the bound state $\tgamma^{\textrm{\tiny $(5|1)$}}$ interacting with $\tgamma^{\textrm{\tiny $(4)$}}$. There are furthermore bound states formed by $\tgamma^{\textrm{\tiny $(k|1)$}}$ and $\tgamma^{\textrm{\tiny $(k-1|-1)$}}$ for $k\ge6$. These are the two simplest kinds of secondary bound states that we anticipated in Section~\ref{SecondaryStates}. 

For $k=9,10$ the resulting secondary states are also involved in interactions with negative pairing $-2$ as they interact with~$\tgamma^{\textrm{\tiny $(k-2)$}}$.
For example, the vector state $\tgamma^{\textrm{\tiny $(10|1)$}}+\tgamma^{\textrm{\tiny $(9|-1)$}}$ is involved in a wall-crossing of the form
\begin{equation}\label{negpairvector}
\langle\tgamma^{\textrm{\tiny $(k|1)$}}+\tgamma^{\textrm{\tiny $(k-1|-1)$}},\tgamma^{\textrm{\tiny $(k-2)$}}\rangle=-2
\end{equation}
as it interacts with $\tgamma^{\textrm{\tiny $(8)$}}$. Naively this indicates that the family of bound states that accumulate near the state (\ref{negpairvector}) will interact with $\tgamma^{\textrm{\tiny $(8)$}}$ with arbitrarily large negative pairings. In Section~\ref{negvectorcrossing} we will discuss how the vector $\tgamma^{\textrm{\tiny $(10|1)$}}+\tgamma^{\textrm{\tiny $(9|-1)$}}$ and its bound states can be interpreted as bound states of $\tgamma^{\textrm{\tiny $(8)$}}$.

\begin{figure}[h!]
\centering
\begin{overpic}
[width=\textwidth]{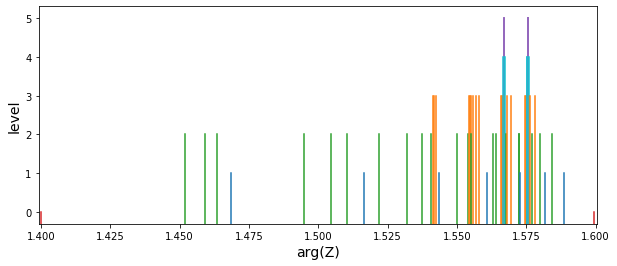}
\end{overpic}
\caption{\small States between $\gamma_3$ (leftmost) and $\gammavf$ (rightmost) for $\textsf{kcut}=8$, $\textsf{subcut}=1$, and $\textsf{lvlcut}=5$.}
\label{SecondaryFamilies}
\end{figure}

\paragraph{States up to level $\boldsymbol{5}$ and $\boldsymbol{k=8}$.} Pushing the parameters of the computation further is not attainable without tradeoffs as the spectrum gets very complicated and ensuring the criteria outlined in Appendix~\ref{pertcriteria} gets much harder. We can push the level cutoff up to $5$ by going to $\textsf{kcut}=8$, where we find two regions where higher level states accumulate up to level $5$ as shown in Figure~\ref{SecondaryFamilies}.

\begin{figure}[h!]
\centering
\begin{overpic}
[width=\textwidth]{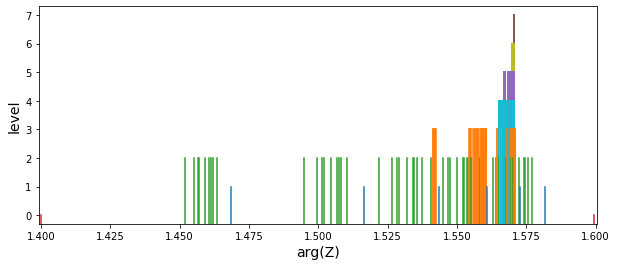}
\end{overpic}
\caption{\small States between $\gamma_3$ (leftmost) and $\gammavf$ (rightmost) for $\textsf{kcut}=7$, $\textsf{subcut}=4$, and $\textsf{lvlcut}=8$. The maximum level reached is $7$ so there are no further states for this choice of cutoffs.}
\label{MoreSecondaryFamilies}
\end{figure}

\paragraph{Spectrum for increased \textsf{subcut} up to $\boldsymbol{k=7}$.} If we wish to increase the number $\textsf{subcut}$ of states computed per family of hypermultiplets, we again need to reduce $\textsf{kcut}$ considerably. We can for example reach $\textsf{subcut}=4$ by restricting $\textsf{kcut}$ to $7$. In that case we can compute all resulting states, finding a maximum level of $7$ and the spectrum shown in Figure~\ref{MoreSecondaryFamilies}.
In this case there is no unexpected behaviour in the form of negative pairings or permutations of factors $\cK_\gamma^{-2}$ with non-zero pairing. These mainly arise at higher $k$, where even the first families of bound states have deeper overlap and pairings such as (\ref{deepfamilyoverlap}) can occur.

\paragraph{Summary.} Examining the spectrum between $\gamma_3$ and $\gammavf$ at $|m|=3$ with a large perturbation, we can confirm the existence of some of the secondary states anticipated in Section~\ref{SecondaryStates} and observe that the process of forming secondary states continues to a high level as seen in Figures~\ref{SecondaryFamilies} and~\ref{MoreSecondaryFamilies}. We also find what naively looks like wall-crossings with negative pairings such as (\ref{deepfamilyoverlap}), and even states leading to permutations of $\cK_\gamma^{-2}$ factors with non-zero pairing such as (\ref{negpairvector}). We will discuss these wall-crossings in more detail in Section~\ref{missingstates}, where we will seek to explain them in terms of the reverse ordering of the primitive pairing two wall-crossing formula.

\subsubsection{\tops{Varying the Mass and Perturbation}{Varying $m$ and $\delta$}}\label{mdeltavary}
\nid
In Section~\ref{Pythonspectrum1} we explored the spectrum at a specific choice of the mass parameter $m$ and of the perturbation away from the wall $\scrE_3$ which is parametrised by $\delta$ as given in (\ref{perturbspectrum}). We can also vary these two parameters. Generally speaking, increasing the mass $m$ makes $\gamma_3$ and $\gammavf$ stay closer together during the perturbation, so there are fewer states $\tgamma^{\textrm{\tiny$(k)$}}$ between them. Furthermore, both are affected less by the perturbation as a whole and to have $\gamma_3$ interacting with the same number of states $\tgamma^{\textrm{\tiny$(k)$}}$ would require a larger choice of $\delta$. On the other hand, even with fewer states $\tgamma^{\textrm{\tiny$(k)$}}$ between them, the bound states that result from interactions of $\gamma_3$ now stay closer to $\gamma_3$, which leads to overlapping bound state families, and therefore to secondary and higher states as we saw in Section~\ref{Pythonspectrum1}.

We can also vary $\delta$ and in particular reduce it since we used a large perturbation in Section~\ref{Pythonspectrum1}. Since the states $\tgamma^{\textrm{\tiny$(k)$}}$ are packed more closely together as we approach the accumulation ray in the central charge plane, reducing $\delta$ increases the number of states $\tgamma^{\textrm{\tiny$(k)$}}$ that fit between $\gamma_3$ and $\gammavf$. On the other hand, the states with large $k$ are now heavier and as a result  their bound states with $\gamma_3$ are closer to $\tgamma^{\textrm{\tiny$(k)$}}$, so there is less overlap of bound state families, and therefore reduced possibility of secondary and higher states appearing in the BPS spectrum. 

For any choice of $m$ and $\delta$, most of the complicated behaviour seems to occur close to $\gammavf$. Numerically it is also not possible to push $k$ far beyond the value needed to generate all the states $\tgamma^{\textrm{\tiny$(k)$}}$ that are between $\gamma_3$ and $\gammavf$. There seems to consistently be difficult behaviour of $\tgamma^{\textrm{\tiny$(k)$}}$ and their first bound states in the region just behind $\gammavf$. Our findings in Section~\ref{gammavfcrossing} illuminate this somewhat, as the wall-crossing of $\gammavf$ itself should remove various states from the spectrum and without handling it correctly there will be further complicated behaviour in this region. For the cases at large $m$ it is generally harder to even compute states up to $\gammavf$ at low level cutoffs, since bound states of $\gamma_3$ and $\tgamma^{\textrm{\tiny$(k)$}}$ for higher $k$ enter our region of  interest, which leads to more secondary bound states and generally more complicated behaviour.

We computed the spectrum for mass values $|m|\in\{3,8,50,1000\}$ and for each case used multiple different perturbations, which need to be somewhat tailored to the mass in question. For each of those mass values, we encounter negative pairings of the type (\ref{deepfamilyoverlap}) and vectors of the form $\gamma_3+\tgamma^{\textrm{\tiny$(k)$}}$ wall-crossing with $\tgamma^{\textrm{\tiny$(k-1)$}}$ with pairing zero as seen in the $|m|=3, \delta\approx 0.15$ case. We can also again find vector states involved in wall-crossings with negative pairing, but unlike our previous discussion, these are now high level states.

\begin{figure}[h!]
\centering
\begin{overpic}
[width=\textwidth]{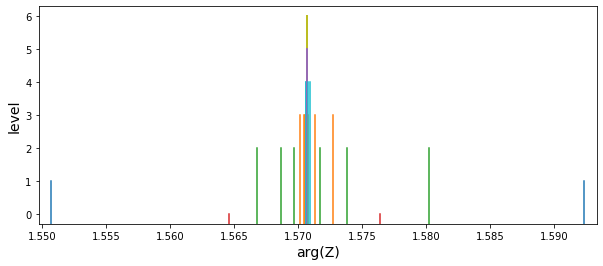}
\end{overpic}
\caption{\small BPS spectrum between $\tgamma^{\textrm{\tiny $(2)$}}$ (leftmost) and $\tgamma^{\textrm{\tiny $(3)$}}$ (rightmost) for $m=50\,\I$, $\delta\approx0.3$, $\textsf{kcut}=4$, $\textsf{subcut}=1$, and $\textsf{lvlcut}=7$. There are high level secondary states between $\gamma_3$ and $\gammavf$ (shown in red), which are the result of overlapping bound state families.}
\label{spectrumverylargemass}
\end{figure}

As an extreme example of the spectrum at a large perturbation and larger mass, consider $|m|=50$ and $\delta\approx0.3$. In this case, $\gamma_3$ and $\gammavf$ are not separated by a state $\tgamma^{\textrm{\tiny $(k)$}}$, and both sit in between $\tgamma^{\textrm{\tiny $(2)$}}$ and $\tgamma^{\textrm{\tiny $(3)$}}$. Still we can find secondary states up to level $6$, which are the result of bound states $\tgamma^{\textrm{\tiny $(k|l)$}}$ staying close to $\gamma_3$ for $k=3,4$. Figure~\ref{spectrumverylargemass} shows the resulting spectrum between $\tgamma^{\textrm{\tiny $(2)$}}$ and $\tgamma^{\textrm{\tiny $(3)$}}$. 
This highlights the importance of secondary states, which form a rich spectrum even when the primary bound states are seemingly trivial.

Overall the experimentation with different mass and perturbation sizes confirms that the qualitative findings in Section~\ref{Pythonspectrum1} are robust and unlikely to be coincidental for the chosen parameters there.

\subsection{Reorganising the Negative Pairing Wall-Crossings}\label{missingstates}
\nid
We have encountered wall-crossings  of the form $\cK_{\gamma}\,\cK_{\gamma'}=\cK_{\gamma'}\cdots\cK_{\gamma}$ with negative pairings $\langle\gamma,\gamma'\rangle=-2$. As explained earlier, such wall-crossings should not occur in rank one $\N=2$ theories as they violate the no-exotics theorem. Furthermore, Denef's equations (\ref{denefeqn}) show that there should be further symplectomorphism factors separating $\cK_{\gamma}\,\cK_{\gamma'}$, which therefore prevent the primitive wall-crossing with negative pairing.

A natural way to make sense of these wall-crossings is as the reverse ordering of the wall-crossing formula for $\langle\gamma',\gamma\rangle=2$, namely
\begin{equation}\label{inverseKSWCF}
\varPi_2^{\gamma',\gamma} \, \cK^{-2}_{\gamma+\gamma'} \, \varPi_1^{\gamma',\gamma}=\cK_{\gamma'}\,\cK_{\gamma} \ .
\end{equation}
This requires the appropriate bound states of $\gamma$ and $\gamma'$ to appear as BPS particles as well. Indeed, following the Denef equations (\ref{denefeqn}), these states should be BPS particles in the theory when coming close to the wall $\scrW(\gamma,\gamma')$ from the side where $\arg (Z(\gamma'))>\arg (Z(\gamma))$.

This leads us to propose a variety of states which are in the BPS spectrum, that were missed in our earlier analysis. In this section we show how at least some of these states are certainly in the spectrum and can be identified with other bound states resulting from earlier wall-crossings. The states that we manage to match up are for the most part from the $(k|l)$-families of states with $l>1$, which explains why they were missed in the numerical explorations of Section~\ref{Pythonspectrum}.

We thereby identify a few of the seemingly missing states by explicitly matching up secondary and tertiary bound states. The existence of the remaining states in the spectrum  is however highly non-trivial and poses interesting constraints on the BPS spectrum. For example, in Section~\ref{neghypercrossing} we will see that it relates the behaviour of high level secondary states at $k'<k$ to the spectrum in between $\tgamma^{\textrm{\tiny $(k|1)$}}$ and $ \tgamma^{\textrm{\tiny $(k-2|-1)$}}$ near the wall $\scrW(\tgamma^{\textrm{\tiny $(k|1)$}}, \tgamma^{\textrm{\tiny $(k-2|-1)$}})\subset\scrB_m$. It would be interesting to gain further insight into this and identify the pattern of required states in more detail. We should emphasize that this behaviour is highly unusual for rank one $\N=2$ field theories at least: bound states are formed as hypermultiplets accompanying a vector multiplet which at some point on the Coulomb branch leave the spectrum at an entirely different wall-crossing, where they are reorganised as hypermultiplets accompanying different vector bound states of different constituents. 
 
We perform this analysis for the wall-crossings with the pairings (\ref{deepfamilyoverlap}) which involve two hypermultiplets, as well as the wall-crossings involving vector multiplets with pairings (\ref{negpairvector}). We further apply a similar analysis to $\gammavf$ which undergoes wall-crossings with $\langle\gammavf,\tgamma^{\textrm{\tiny $(k)$}}\rangle=-2$, and likewise can be interpreted as a reversion of the primitive pairing two Kontsevich-Soibelman wall-crossing formula.

\subsubsection{Wall-Crossings of Hypermultiplets}\label{neghypercrossing}
\nid
We will first discuss the case of a hypermultiplet wall-crossing with negative pairing which we encountered in our Python exploration in (\ref{deepfamilyoverlap}). For this first case the pairing in question is
\begin{equation}
\langle\tgamma^{\textrm{\tiny $(k|1)$}}, \tgamma^{\textrm{\tiny $(k-2|-1)$}}\rangle = -2 \ .
\end{equation} 
The expected bound states from the Denef formula are the vector multiplet $\tgamma^{\textrm{\tiny $(k|1)$}}+\tgamma^{\textrm{\tiny $(k-2|-1)$}}$ as well as the hypermultiplets $(n+1)\,\tgamma^{\textrm{\tiny $(k|1)$}}+n\,\tgamma^{\textrm{\tiny $(k-2|-1)$}}$ and $n\,\tgamma^{\textrm{\tiny $(k|1)$}}+(n+1)\,\tgamma^{\textrm{\tiny $(k-2|-1)$}}$ for $n\geq0$. 

Writing out the vector multiplet more explicitly in terms of our original basis $\gamma_i$, it is given by
\begin{equation}\label{missingvector1}
\tgamma^{\textrm{\tiny $(k|1)$}}+\tgamma^{\textrm{\tiny $(k-2|-1)$}} = 3\,\gamma_1 + (3\,k-4)\,(\gamma_1 + \gamma_2) + 3\,\gamma_3 \ .
\end{equation}
Because it has the same amount of charges $\gamma_1$ and $\gamma_3$, this state could result for example from a wall-crossing with $\langle\tgamma^{\textrm{\tiny $(k'+1|l)$}}, \tgamma^{\textrm{\tiny $(k')$}}\rangle=2$ which leads to the vector multiplet
\begin{equation}\label{lowerkvector}
\tgamma^{\textrm{\tiny $(k'+1|l)$}} + \tgamma^{\textrm{\tiny $(k')$}} = (l+1)\,\gamma_1+(l\,k'+l+k')\,(\gamma_1+\gamma_2)+(l+1)\,\gamma_3 \ ,
\end{equation}
which is indeed the same as (\ref{missingvector1}) for $l=2$ and $k'=k-2$. Note that this state would be missing in the corresponding numerical calculation, since it is built from an $l=2$ state, which lies beyond the cutoff of bound states $\textsf{lvlcut}=1$. This suggests that previously created bound states at lower $k$ can be reinterpreted as bound states between the states involved in (\ref{deepfamilyoverlap}). 

We can similarly look for the needed hypermultiplets, starting with
\begin{align}\label{missinghypers1}
& \tgamma^{\textrm{\tiny $(k|1)$}}+n\, \big(\tgamma^{\textrm{\tiny $(k|1)$}}+\tgamma^{\textrm{\tiny $(k-2|-1)$}}\big) \nonumber \\[4pt] & \hspace{4cm} =(3\,n+1)\,\gamma_1+(3\,n\,k+k-4\,n)\,(\gamma_1+\gamma_2)+(3\,n+2)\,\gamma_3 \ .
\end{align}
We again consider bound states arising from $\langle\tgamma^{\textrm{\tiny $(k'|l)$}}, \tgamma^{\textrm{\tiny $(k'-1)$}}\rangle=2$, but this time allow for arbitrary $l$. The hypermultiplets of the form $(n'+1)\,\tgamma^{\textrm{\tiny $(k'+1|l)$}}+n'\,\tgamma^{\textrm{\tiny $(k')$}}$ are given by
\begin{equation}\label{missinghypercandidates1}
(n'\,l+l+n')\,\gamma_1+\big((n'+1)\,l\,k' + (n'+1)\,l + n\,k'\big)\,(\gamma_1+\gamma_2)+(n'+1)\,(l+1)\,\gamma_3 \ .
\end{equation}
Equating (\ref{missinghypers1}) and (\ref{missinghypercandidates1}) gives
\begin{align}
l=\frac{3\,n+1-n'}{n'+1} \qquad \mbox{and} \qquad
k-k'=\frac{7\,n+1-n'}{3\,n+1} \ ,
\end{align}
which need to have simultaneous integer solutions with $k-k'\ge0$ and $l\ge0$ for the states to match. In fact, we should really be interested in $l>0$ as $\tgamma^{\textrm{\tiny $(k'+1|0)$}}=\gamma_3$ which leads to trivial solutions due to notational coincidence. Similarly $k-k'=1$ gives rise to a trivial solution for $n=0$. The non-trivial solutions then correspond to $k-k'=2$ which is solved either by: (a) $(n',l)=(0,4)$ giving the $n=1$ hypermultiplet of (\ref{missinghypers1}), not even as a secondary bound state but as $\tgamma^{\textrm{\tiny $(k-1|4)$}}$; or by (b) $(n',l)=(1,3)$ giving the $n=2$ hypermultiplet of (\ref{missinghypers1}), this time as a secondary bound state.

We can similarly look for the hypermultiplets 
\begin{align}\label{missinghypers2}
&\tgamma^{\textrm{\tiny $(k-2|-1)$}}+n\, \big(\tgamma^{\textrm{\tiny $(k|1)$}}+\tgamma^{\textrm{\tiny $(k-2|-1)$}}\big)\nonumber\\[4pt]
& \hspace{2cm} =(3\,n+2)\,\gamma_1+\big(3\,n\,k+2\,k-4\,(n+1)\big)\,(\gamma_1+\gamma_2)+(3\,n+1)\,\gamma_3 \ ,
\end{align}
and compare them to the other set of hypermultiplet bound states $n'\,\tgamma^{\textrm{\tiny $(k'+1|l)$}}+(n'+1)\,\tgamma^{\textrm{\tiny $(k')$}}$ which are given by
\begin{equation}\label{missinghypercandidates2}
(n'\,l+n'+1)\,\gamma_1+\big(n'\,l\,(k'+1)+(n'+1)\,k'\big)\,(\gamma_1+\gamma_2)+n'\,(l+1)\,\gamma_3 \ .
\end{equation}
This matches the hypermultiplet (\ref{missinghypers2}) with $n=1$, by taking $l=1$, $n'=2$ and $k'=k-2$. There are no other non-trivial solutions.

So far we have found the vector bound state of the BPS states $\tgamma^{\textrm{\tiny $(k|1)$}}$ and $ \tgamma^{\textrm{\tiny $(k-2|-1)$}}$ involved in the wall-crossing with negative pairing (\ref{deepfamilyoverlap}), which is the same state as the vector multiplet $\tgamma^{\textrm{\tiny $(k-1|2)$}} + \tgamma^{\textrm{\tiny $(k-2)$}}$. We similarly found two of the hypermultiplets of the family $(n+1)\,\tgamma^{\textrm{\tiny $(k|1)$}}+n\, \tgamma^{\textrm{\tiny $(k-2|-1)$}}$, namely the $n=1$ case in which it agrees with $\tgamma^{\textrm{\tiny $(k-1|4)$}}$ and the $n=2$ case in which it agrees with the secondary bound state $2\,\tgamma^{\textrm{\tiny $(k-1|3)$}}+\tgamma^{\textrm{\tiny $(k-2)$}}$. We further found the first hypermultiplet of the other family $n\,\tgamma^{\textrm{\tiny $(k|1)$}}+(n+1)\, \tgamma^{\textrm{\tiny $(k-2|-1)$}}$ as the secondary bound state $2\,\tgamma^{\textrm{\tiny $(k-1|1)$}}+3\,\tgamma^{\textrm{\tiny $(k-2)$}}$.

We should also check that, whenever the wall-crossing with $\langle\tgamma^{\textrm{\tiny $(k|1)$}}, \tgamma^{\textrm{\tiny $(k-2|-1)$}}\rangle=-2$ occurs, the secondary bound states from $\tgamma^{\textrm{\tiny $(k-1|l)$}}$ and $\tgamma^{\textrm{\tiny $(k-2)$}}$ are indeed  in the spectrum: Since
\begin{align}
\arg \big(Z(\tgamma^{\textrm{\tiny $(k-1|1)$}})\big) &< \arg \big(Z(\tgamma^{\textrm{\tiny $(k|1)$}})\big) \nonumber \ , \\[4pt]
\arg\big( Z(\tgamma^{\textrm{\tiny $(k-2|-1)$}})\big) &< \arg \big(Z(\tgamma^{\textrm{\tiny $(k-2|l)$}})\big) \ ,
\end{align}
it follows that when $\arg\big( Z(\tgamma^{\textrm{\tiny $(k-2|-1)$}})\big)=\arg \big(Z(\tgamma^{\textrm{\tiny $(k|1)$}})\big)$ we have
\begin{equation}
\arg \big(Z(\tgamma^{\textrm{\tiny $(k-1|1)$}})\big)<\arg \big(Z(\tgamma^{\textrm{\tiny $(k-2|l)$}})\big) \ .
\end{equation}
This indicates that the wall-crossing with pairing $\langle\tgamma^{\textrm{\tiny $(k-1|1)$}},\tgamma^{\textrm{\tiny $(k-2|l)$}}\rangle=2$ has indeed taken place, giving rise to the required bound states.

\subsubsection{Wall-Crossings of Vector Multiplets}\label{negvectorcrossing}
\nid
In a similar vein we can look at the vector multiplets involved in the wall-crossings associated to (\ref{negpairvector}) which involves the pairing
\begin{equation}
\langle\tgamma^{\textrm{\tiny $(k|1)$}}+\tgamma^{\textrm{\tiny $(k-1|-1)$}},\tgamma^{\textrm{\tiny $(k-2)$}}\rangle=-2 \ .
\end{equation}
In this case we are looking to reinterpret the vector multiplet $\tgamma^{\textrm{\tiny $(k|1)$}}+\tgamma^{\textrm{\tiny $(k-1|-1)$}}$ as a vector bound state of $\tgamma^{\textrm{\tiny $(k-2)$}}$ which can be found by subtracting one state from the other. The result gives rise to the decomposition
\begin{equation}
\tgamma^{\textrm{\tiny $(k|1)$}}+\tgamma^{\textrm{\tiny $(k-1|-1)$}} = \tgamma^{\textrm{\tiny $(k-2)$}} + \tgamma^{\textrm{\tiny $(k|2)$}} \ .
\end{equation}
This is now a state that is certainly in the BPS spectrum and that we indeed expect to miss with our Python exploration, as states with $l>1$ are not included in the numerical calculations that find (\ref{negpairvector}).

We can also identify one of the hypermultiplets required for this wall-crossing, namely the state $2\,\tgamma^{\textrm{\tiny $(k-2)$}} + \tgamma^{\textrm{\tiny $(k|2)$}}$ which agrees with $\tgamma^{\textrm{\tiny $(k-1|-3)$}}$. The remaining hypermultiplets do not correspond to members of the $(k|l)$-families of states and would need to be found in higher level secondary bound states.

\subsubsection{\tops{Wall-Crossings of $\gammavf$}{wall-crossing for gamma1 + gamma2 + gammaf}}\label{gammavfcrossing}
\nid
We can similarly examine the wall-crossings of $\gammavf$ as it interacts with the states $\tgamma^{\textrm{\tiny $(k)$}}$. As before this involves a negative pairing $\langle\gammavf,\tgamma^{\textrm{\tiny $(k)$}}\rangle=-2$ and again we can search for appropriate bound states to reorganise the BPS spectrum. The corresponding vector state is
\begin{equation}
\gammavf + \tgamma^{\textrm{\tiny $(k)$}} = \gamma_1+(k+2)\,(\gamma_1+\gamma_2)+\gamma_3 \ ,
\end{equation}
which is just the vector state formed by $\tgamma^{\textrm{\tiny $(k+2)$}}$ with $\gamma_3$. Looking further for the corresponding hypermultiplets, we can start with the states $\tgamma^{\textrm{\tiny $(k|l)$}}$ and compare them to the bound states of $\tgamma^{\textrm{\tiny $(k'|l)$}}$ with $ \tgamma^{\textrm{\tiny $(k'-1)$}}$. We summarise our findings in Table~\ref{gammavfhyperstable}.
\begin{table}[h!]
\centering
\begin{tabular}{c c| c} 
Expected bound state of $\gammavf$, $\tgamma^{\textrm{\tiny $(k)$}}$ & & Corresponding bound state of $\tgamma^{\textrm{\tiny $(k')$}}$\\
\hline
\hline
$\gammavf + \tgamma^{\textrm{\tiny $(k)$}}$ & & $\tgamma^{\textrm{\tiny $(k+2)$}}+\gamma_3$\\
\hline
$(n+1)\,(\gammavf) + n\, \tgamma^{\textrm{\tiny $(k)$}} $& $n=1$ & $\tgamma^{\textrm{\tiny $(k+4|1)$}}$\\
& $n=2$ & $\tgamma^{\textrm{\tiny $(k+3|2)$}}$\\
& $n=3$ & $2\,\tgamma^{\textrm{\tiny $(k+3|1)$}}+\tgamma^{\textrm{\tiny $(k+2)$}}$\\
\hline
$n\,(\gammavf) + (n+1)\, \tgamma^{\textrm{\tiny $(k)$}}$ & $n=1$ & $\tgamma^{\textrm{\tiny $(k+1|-1)$}}$\\
 & $n\ge 1$ & $2\,\tgamma^{\textrm{\tiny $(k+1)$}}+\tgamma^{\textrm{\tiny $(k+2|n-1)$}}$\\
\end{tabular}
\caption{\small Bound states between $\gammavf$ and $\tgamma^{\textrm{\tiny $(k)$}}$, and their reinterpretation in terms of primary and secondary states resulting from $\gamma_3$ interacting with $\tgamma^{\textrm{\tiny $(k')$}}$ for $k'>k$.} 
\label{gammavfhyperstable}
\end{table}
In this case we actually find one of the hypermultiplet families completely, namely $n\,(\gammavf) + (n+1)\, \tgamma^{\textrm{\tiny $(k)$}}$ which can be identified with $2\,\tgamma^{\textrm{\tiny $(k+1)$}}+\tgamma^{\textrm{\tiny $(k+2|n-1)$}}$ through
\begin{align}\label{gammavfhyperfamily}
2\,\tgamma^{\textrm{\tiny $(k+1)$}}+\tgamma^{\textrm{\tiny $(k+2|n-1)$}} 
=\tgamma^{\textrm{\tiny $(k)$}} + n\,(\gammavf + \tgamma^{\textrm{\tiny $(k)$}}) \ .
\end{align}

It would be interesting to find a more systematic way of identifying these different bound states. So far we have just compared some $(k|l|n)$-families of states to the missing bound states to find ad hoc agreements for particular states in all cases except (\ref{gammavfhyperfamily}). This procedure is limited in scope, as the potential ways that higher level secondary states can form grows quickly, so there are many candidate bound states to compare to each bound state that is needed for the reverse Kontsevich-Soibelman wall-crossing formula (\ref{inverseKSWCF}). Understanding this process in more detail would lead to a better understanding of the $\N=2^*$ spectrum in general, as it also relates very different states in the spectrum and in particular states at largely separated levels.

\section*{Acknowledgments}
\noindent
We are grateful to Lotte Hollands for helpful discussions. The work of PR was supported by a James Watt Scholarship from Heriot-Watt University. The work of RJS was supported by
the Consolidated Grant ST/P000363/1 ``Particle Theory at the Higgs Centre''
from the Science and Technology Facilities Council.

\appendix

\section{\tops{Python Algorithm for Exploring the $\boldsymbol{\N=2^*}$ Spectrum}{Python Algorithm for Exploring the N=2* Spectrum}}\label{PythonAlgorithm}
\nid
To explore the complicated BPS spectrum of the rank one $\N=2^*$ gauge theory away from the wall of marginal stability $\scrE_3$, we use an algorithm implemented in Python~\cite{Rueter} that perturbs the spectrum step by step, performs the necessary permutations of symplectomorphisms $\cK_\gamma$ in the Kontsevich-Soibelman wall-crossing formula, and then adds in any new resulting bound states.

\subsection{Implementation of the Algorithm}\label{app:description}
\nid
We start by initialising the spectrum on the wall as given in Section \ref{spectrumonwall}, where we specify as input the central charges $Z_1$ and $Z_2$ as well as the mass $m$, which is used to compute the central charge $Z_3$ via the relation $Z_1+Z_2+Z_3=m$. The spectrum contains infinite families of states so we also take a cutoff $\textsf{kcut}$ for the first families of states. Then the spectrum on the wall consists of $\gamma_1, \gamma_2, \gamma_3$, the vector multiplet $\gamma_1+\gamma_2$ and the hypermultiplet $\gammavf$, as well as $\tgamma^{\textrm{\tiny$(k)$}}$ for $k\le\textsf{kcut}$. In a similar manner we put a second cutoff $\textsf{subcut}$, which is used for the bound state families that result from the perturbation of the spectrum.

We add small perturbations of the central charges as described in Section~\ref{subsec:perturbing} by
\begin{equation}
Z_1 \longmapsto Z_1 + \delta \, Z_2 \ ,
\end{equation}
where central charges such as $Z_3$ that include copies of $Z_1$ are also affected, which we keep track of by a perturbation weight assigned to every state. An electromagnetic charge $\gamma=n_1\,\gamma_1 + n_2\,\gamma_2 + n_3\,\gamma_3$ will have $(n_1-n_3)\,\delta \, Z_2$ added to its central charge by the perturbation. We perform this perturbation in small steps $\epsilon\ll\delta$ and adjust the magnitude of $\epsilon$ as we go.

Ideally we would choose $\epsilon$ to be so small as to only allow two states $\gamma$ and $\gamma'$ to interact, so that we can apply the Kontsevich-Soibelman wall-crossing formula to them and add the bound states to the BPS spectrum. It turns out that this criterion for perturbations is too restrictive and we instead allow for perturbations where only two states form bound states. We will return to this issue in more detail in Appendix~\ref{pertcriteria} below.

The full algorithm then consists of the following steps:
\begin{itemize}
\item[$\diamond$] Perturb the spectrum slightly by $\epsilon$.
\item[$\diamond$] Check that the perturbation fulfills the criteria outlined in Appendix~\ref{pertcriteria} below. If it does not, undo the perturbation and try again with adjusted step size.\footnote{Usually this means reducing the step size, but in some cases we also increase it instead.}
\item[$\diamond$] Permute the symplectomorphisms and add resulting bound states to the BPS spectrum.
\item[$\diamond$] Repeat the procedure until the desired perturbation $\delta$ is reached.
\end{itemize}
The bound states constructed in this way will always be a vector multiplet together with the usual infinite families of hypermultiplets, which are truncated to $\textsf{subcut}$.

\subsection{Permutation Criteria}\label{pertcriteria}
\nid
In the algorithm outlined in Appendix~\ref{app:description}, we would ideally take small enough steps $\epsilon$, so that we only need to permute two terms of the Kontsevich-Soibelman wall-crossing formula. In practice, however, it turns out that this is too restrictive and leads to infinite loops in the algorithm.

We instead implement permutation criteria which ensure that any occurring permutations can be broken down into simple permutations of neighbouring factors, and that the result does not depend on the order of these simple permutations. First of all, any perturbation that leads to at most one permutation is allowed. For larger permutations we encode them in the form of blocks of integers $[p_1,p_2,\dots]$ which represent the ways in which the respective factors are moved. For example, in the block
\begin{equation}\label{eq:blockseg}
[2,-1,-1] \ ,
\end{equation}
the leftmost factor is moved two steps to the right, while the other two factors move one step each to the left. 

Such blocks can be decomposed into pairwise permutations as follows. Find the first negative entry, permute the factor one step to the left, adding in any resulting bound states, and then repeat this procedure until completion. Most of the time we encounter the simple case where one state interacts with one or more other states which do not change their respective ordering. The corresponding blocks are of the form
\begin{equation}\label{niceblocks}
[n,-1,-1,\dots,-1]\qquad \text{or}\qquad [1,1,\dots,1,-n] \ .
\end{equation}
These decompose easily and uniquely into pairwise permutations. 

Issues arise in particular blocks that contain zeroes. For example, consider the block
\begin{equation}\label{zeroblock}
[2,0,-2] \ .
\end{equation}
One interpretation of this block is that there are three ordered states $(\gamma,\gamma',\gamma'')$ that at some point all align and then afterwards are ordered as $(\gamma'',\gamma',\gamma)$. Alternatively, if we tried to break it down into single interactions of neighbouring states, there are two routes, starting with either $\gamma$ or $\gamma''$ and exchanging its position with $\gamma'$. The decomposition into such permutations is therefore not unique and if bound states result from any such permutations, the resulting BPS spectra will differ as well. We therefore only allow such blocks if at most one of the intersection pairings $\langle\cdot,\cdot\rangle$ between the charges involved is non-zero.

This subtle point aside, we can restrict what we call the \emph{permutation length} $\textsf{permlength}$ of a block, which is defined as
\begin{equation}
\textsf{permlength}\big([p_1,p_2,\dots]\big) = \text{max}(|p_1|,|p_2|,\dots) \ .
\end{equation}
The only $\textsf{permlength}=1$ block is $[1,-1]$. For $\textsf{permlength}=2$ there are considerably more possibilities including problematic blocks such as (\ref{zeroblock}). For $\textsf{permlength}\geq3$ the number of possibilities increases drastically. There are even $\textsf{permlength}=3$ blocks which do not contain any zeroes, but decompose into blocks which do contain zeroes while following the decomposition procedure, for example
\begin{equation}\label{bad3block}
[3,1,-1,-3]\,\longmapsto\,[3,0,0,-3] \ .
\end{equation}
We therefore check throughout the permutation process that blocks do not violate the criterion of either having no zeroes or containing at most one non-zero pairing. For almost all of our calculations it is however possible to restrict to $\textsf{permlength}=2$, which is well under control.

Another concern for pertubations is when charges align. It can happen that two or more charges end up with the same phase, either through numerical precision or accidental non-generic angle choice, or through complicated wall-crossings such as the ones anticipated for the blocks \eqref{zeroblock}. We usually forbid perturbations which lead to alignment of charges and try to increase the step size to push past them. Performing consistency checks on spectra which can be computed with this turned off or on indicates that it is still a fairly safe criterion to drop, but we still try to keep it.

Summarising, the usual criteria we enforce on perturbations entail the following:
\begin{enumerate}
\item No symplectomorphism is moved past more than two other symplectomorphisms.
\item Permutations are either of the simple form (\ref{niceblocks}) or involve at most one non-zero pairing $\langle\cdot,\cdot\rangle$.
\item No states have coinciding phases.
\end{enumerate}
If these criteria fail the perturbation is undone and a smaller one is performed, unless the failure was the alignment of charges, in which case a larger perturbation is attempted to push past this alignment. 

For the computations discussed in Section~\ref{Pythonspectrum1} we can usually keep these criteria, but Criterion~1 can be relaxed to $\textsf{permlength}\le3$ and Criterion~3 can be dropped:
\begin{itemize}
\item[$\diamond$] For $\textsf{lvlcut}=1$  and $\textsf{subcut}=1$, we can either compute up to $k=13$ keeping all criteria, or up to $k=14$ if we relax either Criteria~$1$ or~$3$, obtaining the same spectrum in both cases.\footnote{The central charges of the two obtained spectra do not quite agree due to numerical errors.}
\item[$\diamond$] For $\textsf{lvlcut}=5$ and $\textsf{subcut}=1$, we can compute up to $k=8$ keeping all the criteria. We can also relax either Criteria~$1$ or~$3$, or both, and in each case arrive at the same result. 
\item[$\diamond$] For $\textsf{lvlcut}=8$ and $\textsf{subcut}=1$, we can compute up to $k=7$ keeping all criteria. Again relaxing either Criteria~$1$ or~$3$, or both, yields the same spectrum. 
\end{itemize}

In the numerical experiments of Section~\ref{mdeltavary} with varying values of $m$ and $\delta$, we can again usually keep all criteria intact, but occasionally have to abandon one of them. It seems that for higher mass or smaller perturbations, Criterion~$3$ can be usually dropped without problems, while allowing larger permutation lengths $\textsf{permlength}$ leads to issues in the forms of permutations such as (\ref{bad3block}), which decompose into blocks containing zeroes.

As a final remark, we would like to point out that the reverse pairing two wall-crossings investigated in Section~\ref{missingstates} involve permutations where multiple states collapse to one ray on the wall and almost all of the pairings between these states are non-zero. It therefore cannot be broken down into single permutations as done in this appendix and violates the permutation criteria. In particular, if the computation includes a wall-crossing between mutually non-local particles with $\langle\gamma,\gamma'\rangle=-2$, and also one bound state of $\gamma$ and $\gamma'$, it will lead to a block containing a zero as in (\ref{zeroblock}). On the other hand, if it contains multiple bound states it will likewise lead to blocks such as (\ref{bad3block}) as $\gamma$ and $\gamma'$ swap positions, whereas their bound states remain between them. It is therefore not surprising that the algorithm is unable to compute the spectrum for $\textsf{lcut}>1$, for example, unless $\textsf{kcut}$ is set very low, since some of the $l=2$ states are involved as bound states in such reversed wall-crossings.

\printbibliography



\end{document}